\documentclass{aa}

\usepackage{graphicx}
\usepackage{txfonts}
\usepackage[colorlinks,allcolors=blue]{hyperref}
\usepackage[dvipsnames]{xcolor}
\usepackage[final]{changes}
\usepackage{placeins}
\setaddedmarkup{\textcolor{OliveGreen}{#1}}
\setdeletedmarkup{\textcolor{OliveGreen}{\sout{#1}}}
\usepackage[switch]{lineno}
\linenumbers
\begin{document}

   \title{A catalogue of TeV pulsar environments}

   \author{T.Wach
          \inst{1,3}
          \and
          T. Linden
          \inst{2, 3}
          \and
          A. M.W. Mitchell
          \inst{3}
          \and
          S. T. Spencer
          \inst{3,4}\thanks{Now at Cherenkov Telescope Array Observatory, Science Data Management Centre (SDMC), Platanenallee 6, 15738 Zeuthen, Germany}
          }

      \institute{
      Max-Planck-Institut f\"ur Kernphysik, Saupfercheckweg 1, 69117 Heidelberg, Germany\\
      \email{tina.wach@mpi-hd.mpg.de}
      \and
      Stockholm University and The Oskar Klein Centre for Cosmoparticle Physics, Alba Nova, 10691 Stockholm, Sweden
      \and
      Friedrich-Alexander-Universit\"at Erlangen-N\"urnberg, Erlangen Centre for Astroparticle Physics, Nikolaus-Fiebiger-Str. 2, 91058 Erlangen, Germany
      \and
      Department of Physics, Clarendon Laboratory, Parks Road, Oxford, OX1 3PU, United Kingdom
             }

   \date{}

  \abstract
   {Pulsars and their environments represent a major class of Galactic $\gamma$-ray sources. Their complex evolution, shaped by the interactions of the pulsar outflow with both the supernova remnant (SNR) and the surrounding interstellar medium (ISM), produces diverse morphological and spectral characteristics observable from radio to PeV energies.} 
   {This work aims to collect and homogenise data from all major operating TeV observatories, presenting the first comprehensive catalogue of TeV $\gamma$-ray properties of pulsar environments, and facilitating systematic studies of the evolution of pulsar wind nebulae and pulsar halos.}
   {The catalogue was created from information regarding all $\gamma$-ray sources that have been classified as pulsar-associated sources in published results from H.E.S.S., MAGIC, VERITAS, HAWC, and LHAASO. For each source, the observed $\gamma$-ray properties were cross-matched with pulsar properties from the ATNF catalogue and Fermi-\emph{LAT} pulsar catalogue. This information was then used to predict the surface brightness for powerful pulsars around which no nebula has been detected yet.}
   {The final catalogue comprises all currently known TeV sources associated with pulsars, as seen by different instruments, spanning all evolutionary stages. The sample consists of 128 $\gamma$-ray sources, connected to 66 different pulsars. It reflects that the TeV-detected population is dominated by young and energetic pulsars located near the Galactic plane, but includes a growing number of middle-aged systems detected as extended halos. Only a weak correlation is found between TeV luminosity and pulsar characteristic age, indicating that TeV evolution is driven by environmental and transport effects rather than by spin-down age alone. Additionally, five pulsars that should host a PWN detectable by CTAO are identified as prime targets for future observation to improve our understanding of properties inhibiting the formation of a TeV nebula. }
   {This publicly available catalogue provides a uniform foundation for future population studies and for constraining models of particle transport and energy losses in pulsar environments.}


   \maketitle
   \nolinenumbers

\section{Introduction}

The current generation of ground-based gamma-ray detectors has observed a wealth of both galactic and extragalactic GeV-PeV $\gamma$-ray sources over the past two decades. Interestingly, a large fraction of the low-latitude sources identified by the Galactic Plane Survey (HGPS) conducted by the High Energy Stereoscopic System (H.E.S.S.)~\citep{HGPS}, the third source catalogue of the High Altitude Water Cherenkov (HAWC) observatory \citep{hawc}, and the Large High Altitude Air Shower Observatory (LHAASO) catalogue~\citep{cao2023lhaaso} are associated with pulsars and their environments. However, to date, there has been no comprehensive catalogue of pulsar-associated sources that have been compiled across all currently operational TeV $\gamma$-ray instruments.

Cataloguing pulsar-associated sources is complicated by the complex and often poorly constrained evolution of these systems over their lifetime. Young (non-recycled) pulsars are formed in the supernova explosion of massive stars, producing an expanding supernova remnant (SNR) that can be observed from MeV to PeV energies through non-thermal emission generated by accelerated particles interacting with the surrounding medium. Embedded within the SNR, a pulsar wind nebula (PWN) develops, mainly powered by the injection of relativistic electron–positron pairs from the pulsar wind, which radiate via synchrotron and inverse-Compton processes \citep{pwn_theory, PWN_hadrons}.

As the SNR evolves, the reverse shock, generated as the forward shock sweeps up the interstellar medium (ISM), propagates inwards and interacts with the PWN, compressing, distorting, and potentially fragmenting it \citep{pwn_theory}. This interaction can lead to strong asymmetries and offset morphologies, many of which are observed at TeV energies.
While this general picture is well established, the detailed evolution of PWNe beyond the youngest systems remains uncertain, particularly once interactions with the SNR reverse shock become important. It also remains unclear how long PWNe remain dynamically coherent after this phase, how their magnetic and particle content evolves, and under what conditions efficient particle escape occurs \citep{pwn_reverb}.

In some systems, relativistic leptons are thought to escape efficient confinement and propagate diffusively away from the pulsar, producing extended, low surface-brightness $\gamma$-ray emission via inverse-Compton scattering \citep{Aharonian:1995zz}. Such structures, commonly referred to as TeV halos \citep{Linden:2017vvb, halofrac}, were first identified around the middle-aged pulsars Geminga and Monogem \citep{Linden:2017vvb, HAWC:2017kbo}. Early interpretations often associated halo formation primarily with older pulsars and weakened magnetic confinement; however, more recent observations indicate that extended TeV emission is not restricted to a narrow evolutionary stage \citep{hess_source3}. This means that pulsar environments, especially disrupted systems, can exhibit characteristics of both PWNe and pulsar halos simultaneously, making a distinction between the two phases unclear and open to interpretation.

While detailed studies of individual systems have provided critical insights into specific evolutionary stages, many of the outstanding questions regarding PWNe and TeV halos are inherently population-level problems, requiring uniform measurements of source extension, morphology, spectral properties, and pulsar parameters across a wide range of ages and environments. At present, such studies are hindered by the heterogeneous nature of observations across different TeV instruments, each with distinct energy coverage, angular resolution, and sensitivity to extended emission. 

This work aims to mitigate the heterogenous nature of the observational data by providing a comprehensive catalogue collecting properties of the TeV sources around pulsars from all operational TeV $\gamma$-ray instruments. While this cannot address the different systematic uncertainties of distinct detector types, a sample like this can be used to gauge the scale of relevant systematic uncertainties and provides a database against which theoretical models can be tested, giving a comprehensive overview regarding how TeV emission connects to the pulsar properties regardless of the evolutionary state of the system.

\section{Data collection}

Constructing a meaningful catalogue of pulsar environments across all evolutionary stages requires the combination of data from different detector types. Imaging Atmospheric Cherenkov Telescopes (IACTs) are particularly powerful for resolving the detailed morphological structure of the $\gamma$-ray emission around pulsars due to their angular resolution of around $0.06^\circ$ \citep{IACT_perf}. However, their degree-scale field of view (FoV) makes it difficult to detect very extended structures, such as the halos observed around older pulsars (see for example \citep{hess_source2, hess_source8}). Wide-field particle detector (PD) arrays, on the other hand, are well-suited to capture these extended regions, albeit at the cost of reduced angular resolution and point source sensitivity \citep{hawc_perf, wcda_perf}.

To minimise this observational bias, this work made use of data from most operational TeV instruments. Specifically, the data originates from air-Cherenkov telescopes including the H.E.S.S. \citep{hess}, the Major Atmospheric Gamma Imaging Cherenkov (MAGIC) telescopes \citep{magiccrab}, the Very Energetic Radiation Imaging Telescope Array System (VERITAS) \citep{Abeysekara_2018}, as well as PD arrays including the High Altitude Water Cherenkov observatory (HAWC) \citep{2hwc_catalog} and LHAASO \citep{cao2023lhaaso}. This catalogue includes all sources that the respective collaborations, or independent multiwavelength studies, have identified as being associated with a pulsar or classified as a PWN. Systems containing pulsar binaries, as well as composite sources for which the TeV emission is not clearly associated with the pulsar or its nebula are excluded.

The properties included in this catalogue are the morphology of the $\gamma$-ray emission, $1\,\sigma$ containment radius, eccentricity, and position angle, as well as the type and properties of the spectral model. From these, the physical extension of the $\gamma$-ray emission and offset between the pulsar position given by the Australia Telescope National Facility (ATNF) catalogue \citep{ATNF} and the centre of the $\gamma$-ray emission are added to the catalogue data. The spectral properties of the respective sources are used to compute the $\gamma$-ray luminosity $L$ assuming isotropic emission:
\begin{equation}
\label{eq:lum}
    L_\gamma = 4 \pi d^2 F_E ,
\end{equation}
where $F_E$ is the integrated energy flux in units of \(\mathrm{erg\,cm^{-2}\,s^{-1}}\) and $d$ the pulsar distance given by the ATNF catalogue. $F_E$ is obtained by integrating the differential photon flux $\phi(E)$ over a symmetric energy interval centred on the instrument-specific reference energy $E_0$:
\begin{equation}
    F_E = \int_{E_{\min}}^{E_{\max}} E \, \phi(E) \, dE ,
\end{equation}
with $E_{\min} = 10^{-0.5} \times E_0$  and $E_{\max} =  10^{0.5} \times E_0$, corresponding to an energy range spanning one decade around $E_0$. 

The $\gamma$-ray surface brightness is derived from the luminosity $L$ and the physical radius $R$ of the source:
\begin{equation}
\label{eq:surf}
    S = \frac{L}{4 \pi R^2}.
\end{equation}
The available $\gamma$-ray properties are summarised in Table \ref{tab:params}.

\begin{table}
\caption{Summary of $\gamma$-ray properties included in this catalogue.}
\centering
\small
\begin{tabular}{ll}
\hline\hline
Parameter & Description \\ \hline
RA & Right Ascension ($^\circ$) \\
Dec & Declination ($^\circ$) \\
$r_{39}$ & $1\,\sigma$ correlation radius ($^\circ$) \\
$e$ & eccentricity \\
$\varphi$ & Position Angle ($^\circ$) \\
$TS$ & Test statistic of model \\
$E_0$ & Reference Energy (TeV) \\
$N_0$ & Energy flux [$10^{-12} \text{TeV}^{-1}\text{cm}^{-2}\text{s}^{-1}$] \\
$\Gamma$ & Spectral Index \\
$\beta$ & curvature \\
$E_\text{cut}$ & Cutoff Energy (TeV) \\
Offset & Between pulsar and emission centre (pc) \\
Extension & Physical extension (pc) \\
$L_\gamma$ & $\gamma$-ray Luminosity (erg/s) \\
$S$ & Surface brightness (erg/s/pc$^2$)  \\
\hline
\end{tabular}
\label{tab:params}%
\tablefoot{Statistical uncertainties on the reported $\gamma$-ray properties are included where available in the corresponding publications.}
\end{table}

The pulsar properties are queried from the ATNF \citep{ATNF}, using all available columns. Additionally, all available information from the Fermi Third Pulsar Catalog is included into this catalogue (3PC,~\cite{Fermi-LAT:2023zzt}).

\subsection{Instruments}

\paragraph{H.E.S.S. data}

H.E.S.S. is an IACT array located in the Khomas Highlands in Namibia ($16.500^\circ$E, $23.271^\circ$S) at $1800\,$m above sea level. The array consists of five Cherenkov telescopes observing the $\gamma$-ray sky between $0.1\,$TeV to several tens of TeV with an angular resolution of $~0.06^\circ$ \citep{hess}. The H.E.S.S. Galactic Plane Survey (HGPS) \citep{HGPS}, which combines observations from the commissioning phase through 2013, provides a survey of the inner Galaxy, including the morphology, spectral properties, and source associations of the detected $\gamma$-ray emission. For this catalogue, we select all $\gamma$-ray sources listed in the HGPS as associated with known pulsars.

Following the publication of the HGPS, improved reconstruction \citep{impact, runwise} and background estimation techniques \citep{bkg1_hess, bkg2_hess} enhanced the sensitivity to extended and low-surface-brightness emission, refining the analyses of several regions. When available, we include results from these dedicated post-HGPS analyses for individual $\gamma$-ray sources. Information about these updated sources is given in Section \ref{sec:single_sources}. In total, this catalogue includes 36 $\gamma$-ray sources associated with 34 pulsars, detected by H.E.S.S.

\paragraph{MAGIC data}

MAGIC is an IACT array located at the Observatorio El Roque de los Muchachos on La Palma ($17.890^\circ$W and $28.762^\circ$N), at a height of $2200\,$m above sea level. The array consists of two Cherenkov Telescopes and is sensitive to detecting $\gamma$-rays between $\sim 50\,$GeV to several tens of TeV with an angular resolution of approximately $0.07^\circ$ at $1\,$TeV and has been fully operational since 2004 \citep{magiccrab}. At the time of writing, the MAGIC Collaboration has reported the detection of five $\gamma$-ray sources with a firm association to a pulsar, with no catalogue or survey being available.

\paragraph{VERITAS data}

VERITAS is an array of four Cherenkov Telescopes located at the Fred Lawrence Whipple Observatory in Arizona ($110.952^\circ$W, $31.675^\circ$N) at a height of $1270\,$m above sea level. Its commissioning was finalised in 2007. It monitors the $\gamma$-ray sky from approximately $85$\,GeV up to more than $30\,$TeV with an angular resolution better than $0.1^\circ$ \citep{Abeysekara_2018}. VERITAS released a catalogue of the $\gamma$-ray sky in \citet{vts_cat}, VTSCat, combining information from individual studies that have been published since the commissioning of the array. This catalogue includes seven sources detected by VERITAS that have been associated with pulsars.

\paragraph{HAWC data}

The HAWC Observatory is located in the Mexican State of Puebla ($97.310^\circ$W, $18.995^\circ$N) at a height of $4100\,$m above sea level, and has been fully operational since 2015. The array consists of 300 water Cherenkov detectors and is sensitive to $\gamma$-rays from $\sim 100\,$GeV to $\sim 100\,$TeV with an angular resolution of $\sim 0.1^\circ$ above $10\,$TeV \citep{hawc}.

Results from the most recent systematic $\gamma$-ray source search by the HAWC Collaboration were published in their third source catalogue (3HWC) \citep{3hwc}. In this work, all 3HWC sources that are associated with a pulsar, identified as a PWN or a pulsar halo, were included. That dataset, including pulsar associated sources detected after the release of the 3HWC, adds up to 27 $\gamma$-ray sources associated with 20 different pulsars.

Additionally, the 3HWC catalogue provides a selection of $\gamma$-ray sources that could potentially be connected to pulsar halos. These 12 candidates are extended and located within a radius of $1.0^\circ$ of a pulsar with a characteristic age between $100\,$kyr and $400\,$kyr, and a spin-down flux at least $1\%$ of the spin-down flux of the Geminga pulsar. Since the association for all sources in this list has only been made through spatial proximity, these sources are not included in this catalogue.

\paragraph{LHAASO data}
\label{sec:lhaaso}

LHAASO is a particle detector located in Sichuan, China ($100.139^\circ$E, $29.358^\circ$N) at a height of $4410\,$m. The array is fully operational since 2021. The facility comprises several components, of which the Water Cherenkov Detector Array (WCDA) and the Kilometer Square Array (KM2A) are the primary $\gamma$-ray instruments. The WCDA is sensitive to $\gamma$-rays from $\sim 0.3\,$TeV to $\sim 30\,$TeV with an angular resolution of $\sim 0.3^\circ$ at $1\,$TeV, while KM2A operates from $\sim 10\,$TeV to beyond $1\,$PeV with an angular resolution of $\sim 0.5^\circ$ at $20\,$TeV \citep{1LHAASO}.

The first LHAASO catalogue of $\gamma$-ray sources (1LHAASO) \citep{1LHAASO} lists 37 sources that are likely to be associated with a pulsar.
This catalogue adopted all the elements from the list of pulsar-associated sources presented in the 1LHAASO catalogue.

The spectra reported in the 1LHAASO catalogue are provided separately for the WCDA and KM2A sub-arrays, which cover different, only partially overlapping energy ranges. 
For sources detected by both instruments, the independently reported power-law fits can therefore exhibit different spectral indices, particularly when the KM2A spectrum is significantly softer than the WCDA measurement. Such behaviour is expected if the underlying $\gamma$-ray spectrum deviates from a simple power law over the full LHAASO energy range.
To obtain a homogeneous spectral representation for these sources, an exponentially cut-off power-law (ECPL) model was fitted to synthetic spectral energy distribution (SED) points derived from the published WCDA and KM2A spectral models reported in \citet{1LHAASO}. The synthetic SED points were generated by evaluating the published power-law spectra over the corresponding instrument energy ranges and propagating the reported statistical uncertainties on the flux normalisation and spectral index.
The ECPL model is defined as:
\begin{equation}
    \frac{\mathrm{d}N}{\mathrm{d}E}
=
N_0
\left(
\frac{E}{E_{\mathrm{ref}}}
\right)^{-\Gamma}
\exp\left(-\frac{E}{E_{\mathrm{cut}}}\right),
\end{equation}
with a reference energy of $E_{\mathrm{ref}} = 8.3~\mathrm{TeV}$, corresponding to the logarithmic centre of the WCDA and KM2A reference energies. During the fit, increased weight was assigned to spectral points within one decade around $E_{\mathrm{ref}}$ to ensure that the derived integrated energy fluxes are primarily constrained by the energy range contributing most strongly to the luminosity estimate.

The resulting ECPL parameters were used to compute integrated energy fluxes and the corresponding adapted $\gamma$-ray luminosities ($L_{\mathrm{ad}}$) and surface brightnesses ($S_{\mathrm{ad}}$), which are included in the catalogue in the columns `Adapted Luminosity [erg/s]' and `Adapted Surface\_brightness [erg/s/pc2]'. The adapted normalisation $N_0$ refers to the ECPL normalisation evaluated at $E_{\mathrm{ref}} = 8.3~\mathrm{TeV}$. For sources detected by only one of the two LHAASO sub-arrays, the published spectral parameters were retained unchanged and no adapted luminosity or surface brightness was derived.

\subsection{Sources with parameters taken from individual studies}
\label{sec:single_sources}

Below, we summarise the sources for which the $\gamma$-ray properties included in this catalogue were taken from dedicated individual studies rather than from catalogues.

\paragraph{PSR~J0007$+$7303}
is part of the composite CTA~1 (G119.5$+$10.2). $\gamma$-ray emission from the region was first reported by VERITAS in \citet{veritas6}. The morphology of the TeV emission does not show any correlation with the radio shell and its centre is located close to the position of the radio-quiet pulsar PSR~J0007$+$7303 ($\dot{E} = 4.5 \times 10^{35}\,$erg/s) and positionally coincident with a compact X-ray PWN with a jet-like structure, making an association with the pulsar likely. Emission is also observed by LHAASO and identified as 1LHAASO~J0007+7303u \citep{1LHAASO}.

\paragraph{PSR~J0205$+$6449}
is a rather young pulsar ($\tau_c = 5.3\,$kyr), powering PWN 3C58, observed in TeV $\gamma$-ray by MAGIC \citet{magic_3C58}. The nebula is classified as a PWN due to its flat radio spectrum, centrally filled morphology, and strong X-ray synchrotron emission surrounding the pulsar. The pulsar itself has a spin-down power of $\dot{E} = 2.7\times10^{37}\,$erg/s, making it one of the most energetic in the Galaxy.

\paragraph{PSR~J0359$+$5414}
is a radio-quiet pulsar ($\dot{E} = 1.3 \times 10^{36}\,$erg/s) around which the HAWC Collaboration has discovered significant $\gamma$-ray emission \citep{hawc_source1}. The detected emission is spatially coincident with an X-ray PWN detected by Chandra \citep{hawc_source1_xray} and shows similar characteristics to other pulsar-associated sources. A second pulsar in the region, PSR~B0355$+$54 is disfavoured as the origin of the emission due to a spatial offset and tension between upper limits set by VERITAS on the tail emission from PSR~B0355$+$54.

\paragraph{PSR~J0534$+$220}
is the most powerful known pulsar in the Milky Way ($\dot{E} = 4.5 \times 10^{38}\,$erg/s) and powers the Crab Nebula. For this catalogue, the $\gamma$-ray properties of the Crab Nebula as seen by HAWC and LHAASO are taken from their respective catalogues. Observations by MAGIC are taken from \citet{magic_crab} and the properties observed by VERITAS from \citet{vertias_source3}. The properties observed by H.E.S.S. are taken from \citet{hess_crab2024}.

\paragraph{PSR~J0537$-$6910} is located in the Large Magellanic Cloud at a distance of $49.6\,$kpc. Its one of the most powerful pulsars known with a spin-down power of $\dot{E} = 4.9 \times 10^{38}\,\text{erg/s}$, powering the PWN N$157\,$B. \citet{hess_source1} details the properties of the of the $\gamma$-ray emission observed by H.E.S.S.. A spatially coincident X-ray PWN was detected by Chandra \citep{hess_source1_xray}.

\paragraph{PSR~J0633$+$1746}
also referred to as the Geminga Pulsar, is the first pulsar found to host a TeV halo. Emission around the middle-aged pulsar ($\dot{E} = 3.2\times 10^{34}\,$erg/s, $\tau_c = 342\,$kyr) was first discovered by the particle detector MILAGRO \citep{MILAGRO}, and later confirmed with HAWC \citep{hawc_geminga}. \citet{geminga_hawc_new} details an updated analysis of the HAWC data on the pulsar halo, does, however, use a custom diffusion model to describe the emission. Since this approach cannot be consistently incorporated into the homogeneous catalogue framework adopted here, we instead retain the five emission components (3HWC~J0630+186, 3HWC~J0631+169, 3HWC~J0633+191, 3HWC~J0634+165, and 3HWC~J0634+180) associated with the pulsar in the 3HWC catalogue. LHAASO also reported the detection of the TeV halo with the source name 1LHAASO~J0634+1741u.
The H.E.S.S. Collaboration was also able to detect extended $\gamma$-ray emission around the pulsar \citep{hess_source2}. However, due to the limited FoV of the detector,
\citet{hess_source2} was not able to constrain the morphology of the $\gamma$-ray emission and a spectrum was extracted from a region of only $1.0^\circ$ centred on the pulsar position.

\paragraph{PSR~J1016$-$5857} is believed to power the $\gamma$-ray source HESS~J1018$-$589B detected by H.E.S.S. \citep{hess_source7}. While both the pulsar and SNR G284.3$-$1.8 are viable particle accelerators, the pulsar seems to be the most likely origin of the emission due to a combination of the positional coincidence between the centre of the extended $\gamma$-ray emission and the pulsar, its high spin-down luminosity ($\dot{E} = 2.6\times 10^{36}\,$erg/s) and the detection of a spatially coincident X-ray Nebula by Chandra \citep{hess_source7_xray}.

\paragraph{PSR~J1057$-$5226}
is another middle-aged pulsar ($\dot{E} = 3.0\times 10^{34}\,$erg/s, $\tau_c = 535\,$kyr) around which extended $\gamma$-ray emission has recently been detected. The emission detected by H.E.S.S. was found to be marginally offset from the pulsar position. However, no other known accelerator or extended region of dense material is found in the vicinity \citep{hess_source8}. Therefore, it is likely that the $\gamma$-ray emission is caused by relativistic leptons diffusing into the ISM.

\paragraph{PSR~J1747$-$2809}
is part of the composite system G0.9$+$0.1 located close to the Galactic centre. $\gamma$-ray emission from the system was first discovered by H.E.S.S. and identified as HESS~J1747$-$281 \citep{veritas7_hess}. 
Emission was also detected by VERITAS and identified as VER~J1747$-$281 \citep{veritas7}. Both observations suggest the origin of the $\gamma$-ray emission to be the PWN in the core of the remnant powered by PSR~J1747$-$2809 ($\dot{E} = 4.3\times 10^{37}\,$erg/s), due primarily to its position and the apparent lack of hard X-ray emission in the shell of the SNR.

\paragraph{PSR~J1809$-$1917} is believed to power the $\gamma$-ray source HESS~J1809$-$193 first discovered by H.E.S.S. \citep{hess_survey1}. A dedicated analysis of this region revealed that the TeV emission is best described by a compact component centred on the pulsar, surrounded by an extended, bright halo that also encompasses the nearby, more energetic PSR~J1811$-$1925 \citep{hess_source4}. Despite PSR~J1811$-$1925 exhibiting a higher spin-down power ($\dot{E} \sim 6.4\times10^{36}\,\mathrm{erg\,s^{-1}}$) compared to PSR~J1809$-$1917 ($\dot{E} \sim 1.8\times10^{36}\,\mathrm{erg\,s^{-1}}$), the spatial coincidence of the compact component with PSR~J1809$-$1917, together with the detection of a prominent X-ray PWN around it, strongly suggests that this pulsar is the primary driver of the compact $\gamma$-ray emission. While a contribution from PSR~J1811$-$1925 to the extended emission cannot be excluded, the uniformity of the $\gamma$-ray emission suggests that the most viable scenario is particle escape from the PWN around PSR~J1809$-$1917, and both components are added to the catalogue. Significant $\gamma$-ray emission around this pulsar is also observed with HAWC \citep{3hwc}, as the point source 3HWC~J1809$-$190, and by LHAASO \citep{1LHAASO} as the extended source 1LHAASO~J1809$-$1918u. MAGIC reports the detection of a $\gamma$-ray source around the pulsar, does not, however, provide spectral parameters \citep{magic_1809}. The MAGIC detection is, therefore, not added to the catalogue.

\paragraph{PSR~J1813$-$1749}
is a very young ($\tau_c = 5.58\,$kyr) and powerful pulsar with a spin-down power of $\dot{E} = 5.6 \times 10^{37}\,$erg/s. $\gamma$-ray emission around this pulsar has been observed by H.E.S.S. \citep{hess_source3}, MAGIC \citep{magic3}, HAWC \citep{3hwc} and LHAASO \citep{1LHAASO}. IACTs had previously found this $\gamma$-ray emission to be compact emission centred on the pulsar, while PDs found a significantly more extended source. However, a recent study of the region by H.E.S.S. \citep{hess_source3}, revealed that the emission can best be described by two components, a compact source and an extended source with a lower surface brightness. In this study, the compact emission was interpreted as a PWN, while 
the extended component likely results from the escape of relativistic leptons from the nebula into the ISM. 
Both components were added to this catalogue.

\paragraph{PSR~J1826$-$1334} is a middle-aged pulsar ($\tau_c = 21.4\,$kyr, $\dot{E} = 2.8 \times 10^{36}\,$erg/s) that powers a prototypical, spatially extended TeV PWN. The $\gamma$-ray source HESS~J1825$-$137 was discovered during the H.E.S.S. inner galaxy survey \citep{hess_survey1}, and was studied in detail in 2019 \citep{hess_source6}. Multiwavelength studies show that the TeV emission is consistent with an extended PWN that has been shaped by cooling and particle transport as well as by interactions with the local environment. The emission is also spatially coincident with a diffuse X-ray nebula \citep{veritas_source2_xray1, veritas_source2_xray2}.
VERITAS~\citet{vertias_source2}, HAWC (3HWC~J1825$-$138, \citet{J1826_HAWC}), and LHAASO (1LHAASO~J1825-1337u) have also detected emission in this region, and the properties of each source have been added to this catalogue.

\paragraph{PSR~J1826$-$1256}
is a powerful pulsar ($\dot{E} = 3.6 \times 10^{36}\,$erg/s) located near PSR~J1826$-$1334. In the HGPS, H.E.S.S. published emission around the pulsar for the first time, that was spatially separated from the emission around PSR~J1826$-$1334 and identified as HESS~J1826$-$130. H.E.S.S. then published a dedicated study of the region in 2020 \citep{hess_source5}. While an association between the TeV emission and several nearby SNRs cannot be excluded, \citet{hess_source5} notes that these SNRs are likely too old to accelerate particles up to the necessary energies. A more viable association is the energetic pulsar powering the Eel Nebula \citep{hess_source5_xray}. 
HAWC also observed $\gamma$-ray emission in this region, HAWC~J1826$-$128 in \citet{J1826_HAWC} and LHAASO reported the source as 1LHAASO~J1825$-$1256u.

\paragraph{PSR~J1838$-$0655}
is positionally coincident with the $\gamma$-ray emission first discovered by H.E.S.S. as HESS~J1837$-$069 during the first inner galaxy survey \citep{hess_survey1}. The $\gamma$-ray emission properties in this catalogue are adapted from the HGPS \citep{HGPS}. The emission was also observed by MAGIC \citep{magic5}. The energetic pulsar ($\dot{E} = 5.5 \times 10^{36}\,$erg/s) also powers an X-ray Nebula observed by INTEGRAL, the Advanced Satellite for Cosmology and Astrophysics (ASCA) and Chandra \citet{magic5_pulsar}. The positional coincidence of the pulsar, X-ray nebula and $\gamma$-ray emission favour a pulsar origin.

\begin{figure*}
\begin{minipage}[b]{.49\textwidth}
\includegraphics[width=0.99\textwidth]{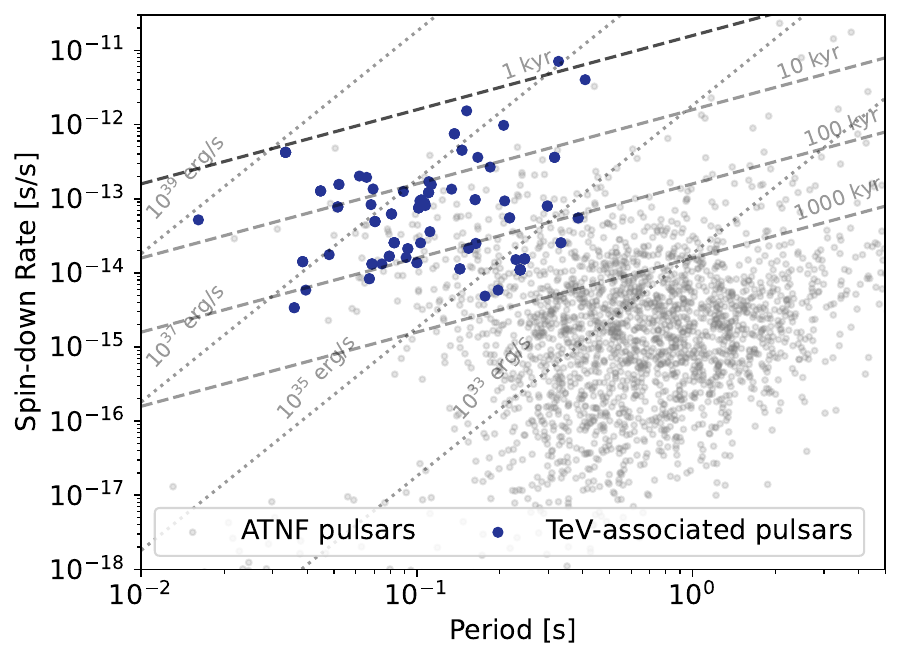}
\end{minipage}\qquad
\begin{minipage}[b]{.49\textwidth}
\includegraphics[width=0.99\textwidth]{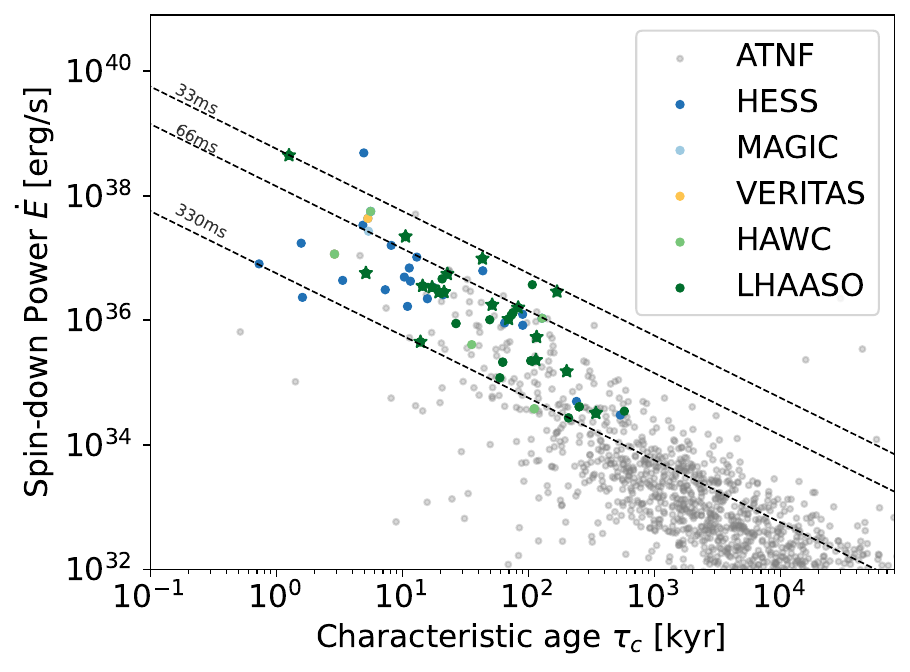}
\end{minipage}
\caption{Left: P-Pdot diagram for Galactic pulsars. The 66 pulsars connected to TeV $\gamma$-ray emission included in this catalogue are shown in blue, while pulsars with only low-energy emission are shown in grey. Right: The spin-down power $\dot{E}$ for TeV detected and non-detected pulsars as a function of the characteristic age, $\tau_c$. The sources detected up to ultra high energies ($>100\,$TeV) are indicated by star markers.}
\label{fig:population}%
\end{figure*}

\paragraph{PSR~J1856$+$0245}
powers $\gamma$-ray emission first discovered by H.E.S.S. and identified as HESS~J1857+026 \citep{magic4_hess}. The existence of extended $\gamma$-ray emission in the region was later confirmed by MAGIC \citep{magic4_magic}. Below $1\,$TeV, both IACT facilities agree that the emission can be described with a symmetric Gaussian model. However, above $1\,$TeV \citet{magic4_magic} claims the detection of two distinct components, one associated with PSR~J1856$+$0245 ($\dot{E} = 4.6 \times 10^{36}\,$erg/s) and the other one associated with a molecular cloud in the region. Since \citet{magic4_magic} was not able to derive separate spectra for each component due to their spatial overlap, the parameters of the single component fit are adopted for this catalogue, which also agrees with later observations by HAWC \citep{magic4_hawc}, LHAASO (1LHAASO~J1857+0245, \citep{1LHAASO}), as well as multiwavelength data \citep{magic4_radio}.

\paragraph{PSR~J1907$+$0602}
powers $\gamma$-ray emission that was first discovered by MILAGRO as MGRO~J$1908+06$ \citep{veritas8_milagro} and later confirmed by H.E.S.S., VERITAS and HAWC \citep{vertias_source4}. The $\gamma$-ray emission is spatially coincident with the SNR G40.5$-$0.5 and the pulsar PSR~J1907$+$0602 ($\dot{E} = 2.8 \times 10^{36}\,$erg/s). However, \citet{vertias_source3_hess} notes that the extension of the $\gamma$-ray emission exceeds the boundary of the radio SNR, making an association with the pulsar far more likely.

\paragraph{PSR~J1928$+$1746}
powers the TeV source 3HWC~J1928$+$178 discovered by HAWC \citep{hawc_source2}. The TeV emission is spatially coincident with the pulsar and the variable X-ray source CXO~J192812.0$+$174712. Since no variability has been detected in the TeV regime, an association with the latter is, however, unlikely. \citet{hawc_source2} find that the most likely origin of the $\gamma$-ray emission is PSR~J1928$+$1746 with a spin-down power of $\dot{E} = 1.6 \times 10^{36}\,$erg/s.
The emission is also observed by LHAASO and identified as 1LHAASO J1928+1746u \citep{1LHAASO}.

\paragraph{PSR~J1930$+$1852}
is a young and energetic pulsar  ($\dot{E} = 1.2\times 10^{37}\,$erg/s, $\tau_c = 2.9\,$kyr) discovered in the SNR G~54.1+0.3. The pulsar is located at the centre of the PWN, also known as the Bull’s Eye Nebula, that displays a centrally peaked radio morphology \citep{veritas5_mw}. $\gamma$-ray emission from this system was detected with H.E.S.S. (HESS J1930+188, \citet{HGPS}), VERITAS (VER~J1930+188, \citet{Abeysekara_2018}), HAWC (3HWC~J1930+188, \citet{3hwc}) and LHAASO (1LHAASO~J1929+1846u, \citet{1LHAASO}). 

\paragraph{PSR~J1932$+$1916}
is a radio-quiet pulsar spatially coincident with the TeV source HAWC~J1932$+$192. X-ray observations of the region with Suzaku and Swift revealed the presence of two emission components, that could be associated with the pulsar ($\dot{E} = 4.1 \times 10^{35}\,$erg/s) and its nebula \citep{hawc_source2_xray}. Moreover, the $\gamma$-ray emission is found to be energetically consistent with a PWN scenario.

\paragraph{PSR~J2016$+$3711}
is a powerful pulsar ($\dot{E} = 2.2\times 10^{37}\,$erg/s, $\tau_c = 11.1\,$kyr) around which MILAGRO detected extended $\gamma$-ray emission \citep{veritas8_milagro}. VERITAS resolved MGRO~J2019$+$37 into two sources, a compact source VER~J2016$+$371 and an extended source VER~J2019$+$368 \citep{veritas8}. VER~J2016$+$371 is positionally coincident with the centre of the composite SNR CTB~87, and multi-wavelength studies support the interpretation of the observed $\gamma$-ray emission as an evolved PWN rather than a young SNR, given its centrally‐filled morphology, lack of a prominent shell, and radio/X-ray spectral properties \citep{veritas8_mw}.

\paragraph{PSR~J2021$+$3651}
is an energetic pulsar ($\dot{E} = 3.4\times 10^{36}\,$erg/s, $\tau_c = 17.2\,$kyr) positionally coincident with the star-forming region Sh~2$-$104. $\gamma$-ray emission in the region was first discovered by VERITAS as VER~J2019$+$368 \citep{veritas8}. The identification of the origin of the emission as either PWN or connected to the star-forming region is more complex due to the spatial proximity of both objects. However, comparisons of the morphology of the TeV emission and the X-ray and radio nebulas, as well as model calculations of the synchrotron radiation, favour the leptonic origin of the source \citep{veritas8, veritas8_xray}. The emission is also observed by HAWC (3HWC J2019+367 \citep{3hwc}) and by LHAASO (1LHAASO~J2020+3649u \citep{1LHAASO})

\subsection{Pulsar properties}

This catalogue contains  $\gamma$-ray sources that have been determined to likely be powered by a pulsar. In cases where more than one pulsar is enclosed in a $\gamma$-ray source region, the pulsar identified as the most likely origin in the respective studies is assumed to be the primary particle accelerator in the region. In addition to the $\gamma$-ray properties, this work also includes information about the respective pulsars sourced from the ATNF catalogue \citep{ATNF} and the Fermi-\emph{LAT} third pulsar catalogue \citep{Fermi-LAT:2023zzt} for all pulsars associated to TeV $\gamma$-ray emission.

\section{Characteristics of the population}

$\gamma$-ray PWNe are preferentially detected around pulsars with high spin-down power, indicating that TeV observations primarily probe systems with strong ongoing particle injection. Pulsars with large values of $\dot{E}$ power bright and relatively compact nebulae that are detectable with current instruments, while older, less energetic pulsars tend to produce fainter and more extended nebulae. As a result, existing TeV surveys are biased towards young, high-$\dot{E}$ systems, a trend reflected in this catalogue, where most detected PWNe are associated with powerful pulsars (Figure~\ref{fig:population}).

\begin{figure*}
\begin{minipage}[b]{.49\textwidth}
\includegraphics[width=0.9\textwidth]{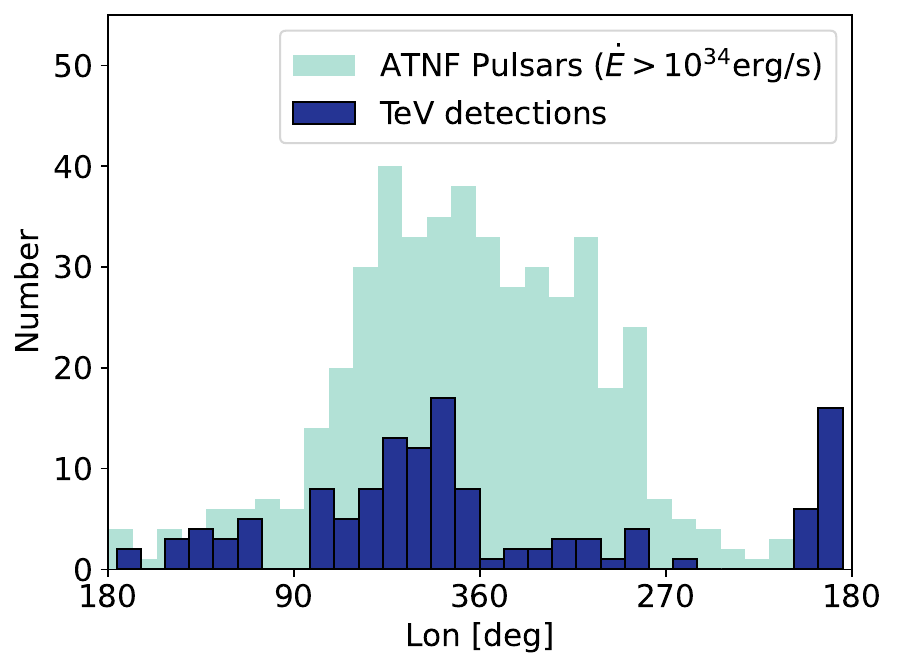}
\end{minipage}\qquad
\begin{minipage}[b]{.49\textwidth}
\includegraphics[width=0.9\textwidth]{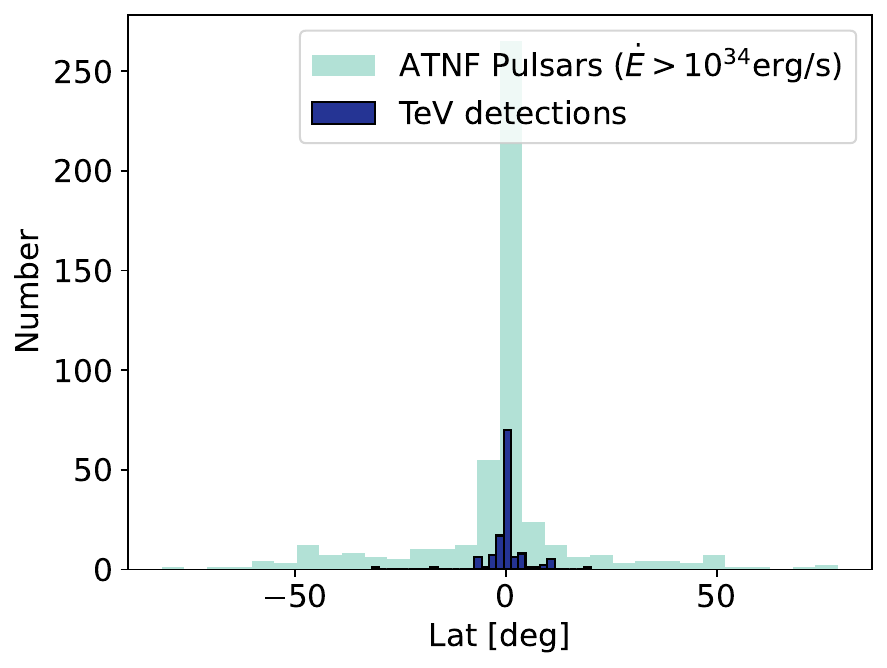}
\end{minipage}
\caption{Galactic position of ATNF pulsars with $\dot{E} > 10^{34}\,$erg/s, compared to the sub-population of pulsars with detected TeV nebulae or halos.}
\label{fig:position}
\end{figure*}

The spatial distribution of TeV PWNe provides additional insight into these selection effects. Most sources are concentrated near the Galactic plane (right panel of Figure~\ref{fig:position}), reflecting both observational and astrophysical factors. Observationally, Cherenkov instruments achieve deeper exposure in the crowded regions of the plane, while astrophysically, pulsar birth is closely linked to massive-star formation in the Galactic disk. This correspondence is apparent when comparing the latitude distribution of catalogue sources with that of ATNF pulsars \citep{ATNF} with $\dot{E} > 10^{34}\,$erg/s.

Although both distributions peak at $b = 0^\circ$, the latitude distribution of pulsars associated with TeV emission is narrower. This may indicate that TeV-detectable PWNe preferentially trace younger, more energetic systems that have not yet migrated far from their birth sites, but observational biases likely also play a role. Older pulsars are expected to reach larger scale heights due to natal kicks, while their nebulae fade and expand, reducing TeV surface brightness. Although this effect is partly offset by reduced diffuse background and source confusion at higher latitudes, the more limited IACT coverage away from the plane reduces TeV sensitivity. Consequently, the observed latitude distribution likely represents a biased subset of the underlying pulsar population, highlighting the importance of accounting for spatial selection effects when interpreting population trends.

In contrast, the longitude distribution shows a different behaviour (left panel of Figure~\ref{fig:position}). While the ATNF pulsar population is denser in the Inner Galaxy, no corresponding enhancement is observed for TeV PWNe. This discrepancy is likely driven by observational limitations, including increased source confusion, higher diffuse backgrounds, and larger average distances towards the inner Galaxy, that reduce TeV detectability and may lead to an under-representation of central PWNe.

\begin{figure}
\centering
\includegraphics[width=0.45\textwidth]{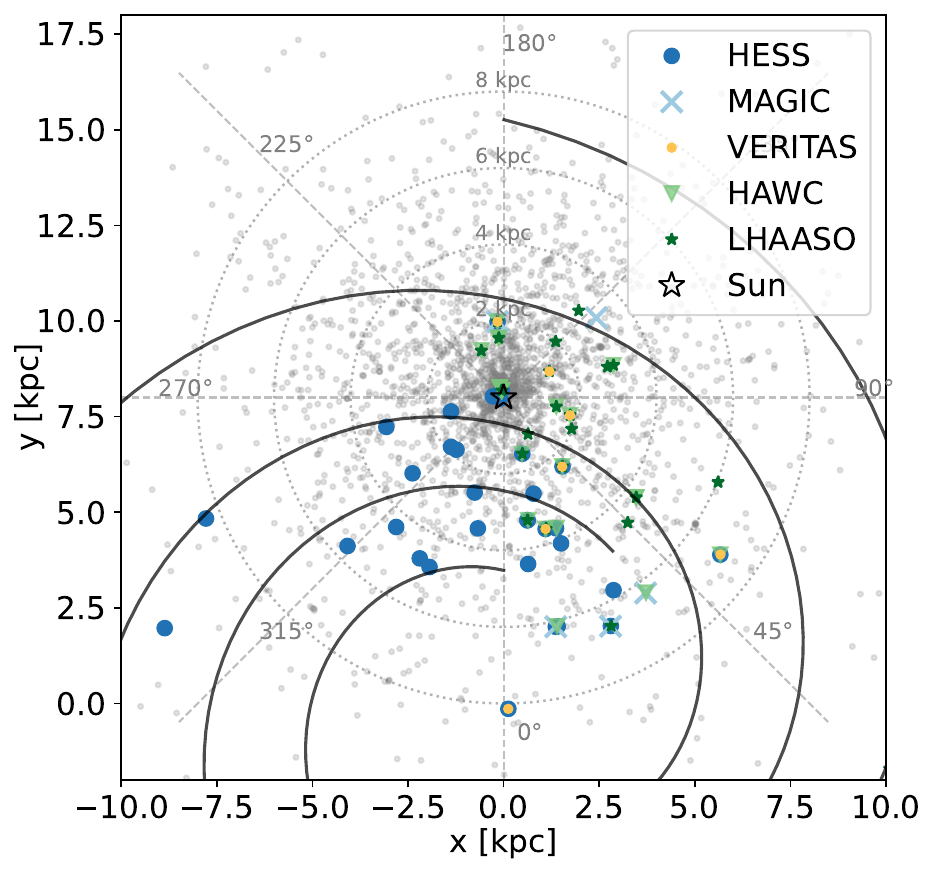}
\caption{Top-down view of the galaxy with a spiral-arm model presented in \citet{Faucher}. The pulsars that were identified to host $\gamma$-ray structures are over-plotted. The distance for each pulsar is taken from the ATNF \citep{ATNF}. Both TeV and ATNF pulsars are systematically found relatively close to the Sun.}
\label{fig:topdown}
\end{figure}

These sensitivity effects are also evident in the overall spatial coverage of current TeV observations. Figure~\ref{fig:topdown} presents the face-on Galactic distribution of all TeV PWN candidates in the catalogue, with the Sun and major spiral arms indicated following the model of \citet{Faucher}. The sources are broadly concentrated towards the inner Galaxy but do not trace individual spiral arms. This pattern primarily reflects the uneven exposure of current instruments, which achieve their highest sensitivity along the Galactic plane and within a few kiloparsecs of the Sun. Uncertainties in pulsar distance estimates further blur any potential association with specific spiral-arm segments.

\begin{table}
\caption{Subset of pulsar parameters from the catalogue that are used for a correlation study with the $\gamma$-ray properties collected in this work.}
\centering
\small
\begin{tabular}{ll}
\hline\hline
Parameter & Description \\ \hline
$\pi$ & Parallax (mas) \\
$P$ & Spin period (s) \\
$\dot{P}$ & Period derivative (s\,s$^{-1}$) \\
DM & Dispersion measure (pc\,cm$^{-3}$) \\
RM & Rotation measure (rad\,m$^{-2}$) \\
$d$ & Distance (kpc) \\
$L_{400}$ & Radio luminosity at 400 MHz (mJy\,kpc$^{2}$) \\
$\tau_c = P / (2\dot{P})$ & Characteristic age (kyr) \\
$B_s = 3.2\times10^{19}(P\dot{P})^{1/2}$ & Surface magnetic field (G) \\
$\dot{E}$ & Spin-down power (erg\,s$^{-1}$) \\
$\dot{E}/d^2$ & Spin-down flux (erg\,s$^{-1}$\,kpc$^{-2}$) \\
$\mu$ & Proper motion (mas\,yr$^{-1}$) \\
$B_{\mathrm{LC}}$ & Magnetic field at the light cylinder (G) \\
$F_{100}$ & Photon Index at 100 MeV () \\
$E_{\mathrm{peak}}$ & Energy at which the pulsar SED peaks (GeV)  \\
& \\
\hline
\end{tabular}
\label{tab:params_corr}
\tablefoot{$E_\mathrm{peak}$ denotes the energy at which the SED peaks, following Equation 21 in \citet{Fermi-LAT:2023zzt}}
\end{table}

Finally, we note that this catalogue is intended to provide a reference dataset against which theoretical models and evolutionary scenarios of pulsar environments can be tested. While the development of a comprehensive evolutionary model is beyond the scope of this work, the catalogue enables a systematic assessment of commonly invoked assumptions regarding the evolution of $\gamma$-ray emission across different stages of pulsar and nebula development.

\subsection{Correlations in the $P-\dot{P}$-plane}
\label{sec:corr}

To this end, we investigate correlations, or the lack thereof, between observed $\gamma$-ray emission parameters and several pulsar parameters. Since the spin period $P$ and its derivative $\dot P$ are the fundamental observables characterising pulsar spin-down, many commonly used pulsar parameters, such as characteristic age, spin-down power, and magnetic field strength, are derived quantities that depend algebraically on $P$ and $\dot P$. As a consequence, apparent correlations with these derived parameters are not independent and can be difficult to interpret when considered in isolation.
To account for this, we adopt an approach in which correlations are analysed directly in the $\log P$–$\log \dot P$ plane. For each $\gamma$-ray observable $X$, we fitted a relation of the form:
\begin{equation}
\label{eq:plane}
    \log(X) = a\log(P) + b\log(\dot{P}) \,.
\end{equation}
This formulation allows us to identify whether a given $\gamma$-ray property is primarily associated with variations in $P$, $\dot P$, or a combination of both, without introducing redundant dependencies through derived quantities. The regression coefficients between these parameters can be seen in Table \ref{tab:ppdot}. Uncertainties on the coefficients were estimated via Monte-Carlo resampling of each parameter within their respective measurement errors, so they naturally include the covariance between a and b.

We find a strong dependence of the flux normalisation at $1\,$TeV on both the pulsar spin period and its derivative, with brighter nebulae associated with shorter spin periods \mbox{($a = -1.33 \pm 0.44$)} and larger spin-down rates \mbox{($b = 0.91 \pm 0.18$)}. This behaviour is consistent with an increased particle injection power in young, energetic systems that are capable of sustaining bright inverse-Compton emission at TeV energies.
In contrast, the spectral index of the $\gamma$-ray nebula does not exhibit any measurable dependence on either $P$ or $\dot P$ ($a = 0.01 \pm 0.03$, $b = 0.00 \pm 0.01$), reflecting that the spectral shape is primarily governed by radiative cooling and environmental conditions, such as magnetic field strength and ambient photon fields, rather than by the instantaneous spin parameters of the pulsar.

We also observe strong dependencies between $P$ and $\dot{P}$ with the $\gamma$-ray luminosity ($a = -1.61 \pm 0.41$, $b = 0.28 \pm 0.20$) and $\gamma$-ray surface brightness ($a = -1.97 \pm 0.27$, $b = 0.99 \pm 0.14$). While the luminosity mainly depends on the pulsar energetics, suggesting that brighter nebulae are preferentially associated with faster-spinning pulsars, the surface brightness exhibits a significant dependence on both $P$ and $\dot P$, consistent with its sensitivity to both the injected particle power and the spatial evolution of the nebula.

Once the dependence in the $P$–$\dot P$ plane is established, the implied scaling with commonly used derived parameters follows directly. A dependence on spin-down power $\dot E \propto \dot{P} / P^{3}$ corresponds to an effective scaling exponent $\beta = b - 3a$, while the dependence on the characteristic age $\tau \propto P/ \dot{P}$ can be inferred as $\beta = a - b$. Dependencies on the surface magnetic field of the pulsar $B_\text{surf} \propto \sqrt{P\dot{P}}$ are given as $\beta = 0.5b + 0.5a$ and the light cylinder magnetic field $B_\text{LC} \propto \sqrt{P\dot{P}} / P^3$ as $\beta = 0.5b - 2.5a$. These inferred scalings should not be treated as independent correlations, but rather as convenient re-expressions of the underlying $P$–$\dot P$ dependence, avoiding over-interpretation of trends that arise purely from algebraic relationships among derived quantities. The resulting correlations can be seen in Table \ref{tab:inferred_corrs}

\paragraph{Spin-down Power}

The spin-down power of a pulsar provides a measure of the available energy injected into the pulsar wind and therefore sets an important scale for the energetics of the surrounding $\gamma$-ray nebula. Consistent with this picture, we infer strong positive scalings of the $\gamma$-ray luminosity ($\beta = 5.11 \pm 1.25$) and surface brightness ($\beta = 6.90 \pm 0.83$) with spin-down power, reflecting the increasing particle injection in more energetic systems.

The inferred positive scaling between nebular extension and spin-down power ($\beta = 1.76 \pm 0.53$) as well as the negative scaling of the nebular offset ($\beta = -1.18 \pm 0.69$), is at first glance counter-intuitive. In simple evolutionary scenarios, older, lower-$\dot E$ systems are expected to develop increasingly extended and displaced emission, eventually transitioning into halo-like morphologies.
An important factor that is not explicitly accounted for in this analysis is the role of surface-brightness sensitivity in defining the measured nebular extent. In particular for middle-aged systems, extended low-surface-brightness emission may fall below the detection threshold, leading to a systematic underestimation of the true nebular size. In contrast, more energetic systems remain detectable out to larger radii, even if their intrinsic nebulae are physically more compact.
This observational effect can naturally lead to an apparent positive correlation between measured extension and $\dot E$, while simultaneously producing a stronger and more robust scaling of surface brightness with spin-down power. The results therefore likely reflect a combination of intrinsic evolutionary trends and sensitivity-limited morphology measurements, rather than a direct causal relation between spin-down power and physical nebular size.

\paragraph{Characteristic Age}

The characteristic age, inferred from the pulsar period and its time derivative, is commonly employed as a proxy for the evolutionary state of pulsar systems and is widely used in population studies and theoretical models. However, $\tau_{\mathrm{c}}$ only approximates the true system age and does not capture the evolutionary history of the pulsar–nebula system, as shown in studies comparing the true age of pulsars to their characteristic ages \citep{age1, age2, age3}. This discrepancy directly maps onto the absence of a strong correlation between the characteristic age and $\gamma$-ray properties 
found in previous studies (e.g. \citep{hesspwnpop, halofrac}), which identify substantial scatter in the TeV luminosity and nebula size among systems with comparable characteristic ages. Using the power-law scaling inferred from the $P$–$\dot P$ dependence, we identify a trend suggesting that pulsars with younger characteristic ages tend to host brighter nebulae ($\beta = -2.96 \pm 0.31$). Nevertheless, we emphasise that additional factors, including environmental conditions, nebula magnetic field evolution, and pulsar-to-pulsar variations in particle injection efficiency, can decouple high-energy nebula properties from the spin-down timescale. Consequently, characteristic age should not be relied upon as a sole predictor of PWN evolution in population-level studies.

\paragraph{Surface magnetic field}

The correlations between surface magnetic field $B_\text{surf}$ and $\gamma$-ray properties are generally weak, with the exception of a moderate dependence on $\gamma$-ray luminosity ($\beta = -0.66 \pm 0.23$). This is consistent with expectations. While $B_\text{surf}$ can influence the initial pair creation and particle acceleration near the pulsar, it does not directly determine the fraction of spin-down energy injected into the nebula, nor does it govern the high-energy transport and radiative losses that shape the observed TeV emission. The absence of strong correlations with surface brightness, extension, or spectral index therefore supports the idea that surface magnetic field plays only an indirect role in PWN evolution as observed in $\gamma$ rays.

\paragraph{Light-cylinder magnetic field}

While $B_\mathrm{LC}$ characterises the pulsar magnetosphere on scales far smaller than the surrounding nebula and therefore cannot directly influence particle confinement in the PWN, it serves as a sensitive tracer of the pulsar’s energetic and evolutionary state. Pulsars with large $B_\mathrm{LC}$ are typically young and have high spin-down power, conditions that favour compact, high-surface-brightness nebulae through enhanced energy injection and reduced spatial dilution of relativistic particles. While we find a significant dependency of the luminosity ($\beta = 4.17 \pm 1.03$) and the surface brightness ($\beta = 5.42 \pm 0.69$) on $B_\text{LC}$, the observed connection includes a large scatter and is likely only indirect. Detailed magnetospheric models demonstrate that pair multiplicity depends sensitively on the magnetic field geometry and the location of the accelerating gaps, rather than on field strength alone \citep{multiplicity1}, making a direct correspondence unlikely. The observed correlations should thus be interpreted phenomenologically, reflecting the role of $B_\mathrm{LC}$ as a proxy for pulsar energetics rather than evidence for a direct causal influence on nebula confinement or particle production.

\subsection{Correlations with other pulsar parameters}

Aside for pulsar parameters derived from $P$ and $\dot{P}$, we also compute Spearman rank correlations between the $\gamma$-ray properties of the surrounding structures and other commonly used pulsar parameters. These pulsar parameters can be seen in Table \ref{tab:params_corr}. For each pair, we derive a Spearman correlation coefficient between the pulsar parameter and $\gamma$-ray emission parameter.

Since many pulsar and nebular properties span several orders of magnitude and are expected to follow non-linear or powerlaw-like trends, correlations are evaluated in log-log space for all strictly positive quantities. The use of the Spearman rank coefficient allows us to probe monotonic relationships without assuming an underlying linear dependence and reduces the impact of outliers and intrinsic scatter within the population. The obtained correlations are summarised in Table \ref{tab:corr_matrix}. For some specific parameters we additionally investigate alternative non-linear scalings (e.g. $d^2$). However, this did not reveal additional non-trivial correlations and is therefore not discussed further.

Some of the observed correlations with $\gamma$-ray nebula properties can be attributed to geometric or selection effects. The Luminosity of the $\gamma$-ray nebula shows a strong dependence on the distance and dispersion measure, as expected from Equation \ref{eq:lum}. A similar reasoning applies for the correlations between extension of the nebula or halo with distance to the pulsar. However, this test does point towards two interesting dependencies.

The mechanisms responsible for radio and $\gamma$-ray emission from the pulsar magnetosphere and the TeV emission from the surrounding nebula are physically and morphologically distinct. Radio waves are produced by coherent processes in the magnetosphere, while TeV emission arises from inverse-Compton scattering of relativistic electrons and positrons accelerated in the nebula. 

Nevertheless, we identify moderate positive correlations between the TeV luminosity and surface brightness of the nebula and both the pulsar radio luminosity at $400\,$MHz ($L_{400}$) ($p = 0.46$) and the pulsed GeV luminosity ($p = 0.61$). Since these correlations are also observed for the nebular surface brightness ($p = 0.57$ and  and $p = 0.50$), they are unlikely to arise solely from geometric distance effects. A plausible interpretation is that both the magnetospheric emission and the TeV nebular emission trace the overall energetic state of the system. Young pulsars with high spin-down power inject larger populations of relativistic leptons into their surrounding nebulae, enhancing the inverse-Compton emission, while simultaneously sustaining stronger non-thermal magnetospheric emission. The stronger correlation observed with pulsed GeV luminosity may indicate that both observables are more directly connected to the efficiency with which rotational energy is converted into high-energy particle populations. Fully understanding the connections between these mechanisms, however, requires a more thorough investigation that is beyond the scope of this study.

\section{Non-detections}

Due to the correlation between spin-down power and energy available for particle acceleration and gamma-ray production, almost all PWNe are observed around energetic pulsars, with less energetic, middle-aged pulsars often hosting a dimmer and more diffuse TeV halo. However, there are still powerful pulsars around which no PWN has been detected.
The most striking examples are the pulsars PSR~J1032$-$5804, PSR~J1400$-$6325, PSR~J1524$-$5625, PSR~J1638$-$4713, and PSR~J1935+2025, all with $\dot{E}/d^2 > 3.0 \times 10^{35}\,$erg\,s$^{-1}$\,kpc$^{-2}$. The relationship between $P$ and $\dot{P}$ for non-detected pulsars, compared to pulsars for which $\gamma$-ray emission was detected, can be seen in Figure \ref{fig:non-detect}. All of these pulsars are located in the southern hemisphere at a distance $d < 8\,$kpc and are covered by the HGPS \citep{HGPS}.  The non-detection of a PWN around these pulsars suggests either that the exposure at their respective positions was low, that the nebula is too extended to be picked up with the FoV of the H.E.S.S. telescopes, or that the formation of a PWN around the pulsar is inhibited due to some environmental effect.

\paragraph{PSR~J1032$-$5804} was only recently discovered using the Square Kilometre Array Pathfinder in Australia \citep{nodet1}. The pulsar is highly scattered and has a characteristic age of $\tau_c \sim 35$\,kyr  and a spin-down power of $\dot{E} = 2.9 \times 10^{36}\,$erg. Its located in the southern hemisphere at an estimated distance of $4.3\,$kpc, with no SNR or $\gamma$-ray source close by. X-ray observations conducted with the Swift X-ray Telescope \citep{swift} do not yield in a detection of a significant PWN, suggesting that PSR~J1032$-$5804 is a good candidate for studying the non-formation of a PWN around powerful pulsars.

\paragraph{PSR~J1400$-$6325} is one of the most energetic rotation-powered pulsars in the Galaxy, with a spin-down power of $\dot{E} = 5.1 \times 10^{37}\,$erg/s and a distance of $7.00\,$kpc. Its positional coincidence with the SNR~G310.6$-$1.6 and the characteristic age of $12.7\,$kyr suggests that the pulsar is still young and make it a prime candidate for the formation of a PWN. The pulsar features a bright X-ray PWN detected by INTEGRAL, however, no $\gamma$-ray emission at either GeV or TeV energies was reported from either the pulsar or the SNR \citep{nondet2}.

\paragraph{PSR~J1524$-$5625} has a spin-down power of $\dot{E} = 3.2 \times 10^{36}\,$erg/s and is located at a distance of $3.38\,$kpc and has a characteristic age of $\tau_c = 32\,$kyr. There is no evidence for a PWN detected around the pulsar in either radio or X-ray \citep{nondet_some}.

\paragraph{PSR~J1638$-$4713} powers the bow-shock PWN named Potoroo, with one of the longest projected radio tails observed for a PWN \citep{nodet41}. The pulsar has a spin-down power of $\dot{E} = 6.1 \times 10^{36}\,$erg/s and is located at a distance of $7.54\,$kpc. The characteristic age derived for the pulsar is $23.6\,$kyr. A X-ray PWN has been observed by Chandra and given the identifier CXOU~J163802.6$–$471358 \citep{nodet42}

\paragraph{PSR~J1935$+$2025} is located at a distance of $4.60\,$kpc and has a spin-down power of $\dot{E} = 4.6 \times 10^{36}\,$erg/s and an characteristic age of $21\,$kyr. $\gamma$-ray pulses have been detected from the pulsar by Fermi-\emph{LAT}, but no extended emission connected to a PWN has been reported. While the high spin-down power and close distance would suggest the presence of a PWN, a H.E.S.S. study of the region in the context of the PWN population study has also not yielded a significant detection of extended emission \citep{hesspwnpop}.

Understanding these non-detections is essential for interpreting the full population of TeV pulsar wind nebulae and for guiding future observational campaigns with next-generation instruments. To explore whether their non-detection can be understood in terms of an intrinsically low surface brightness rather than a lack of particle acceleration, we use the information gathered from the catalogue to predict the expected TeV surface brightness of a population typical PWN for these pulsars.

Given the heterogeneous and potentially non-linear relationships between pulsar properties and TeV emission characteristics, we adopt a data-driven approach based on supervised machine learning. Rather than attempting to model the detailed physical evolution of individual PWNe, this approach is intended to capture empirical trends present in the currently observed TeV PWN population and to extrapolate them to the undetected sample.

To this end, we train an ensemble gradient-boosted decision tree regressor (XGBoost; \citep{xgboost}) on the catalogue data. The model consists of an ensemble of 800 trees with a maximum depth of five and is trained using a learning rate of 0.05. To reduce over-fitting given the limited sample size and the dimensionality of the feature space, we apply stochastic subsampling both at the level of training examples and input features. The model is optimised using a squared-error loss function. For pulsar environments that have been observed with more than one instrument, we derive the median of the $\gamma$-ray properties between the instruments as training input. 

The dataset was randomly split into training and test subsets, with 80\% of the data used for training and 20\% reserved for validation. We chose model hyperparameters conservatively to limit over-fitting given the small sample size, and estimate feature relevance using permutation importance and SHAP (SHapley Additive exPlanations) values. Parameters that did not improve the prediction accuracy of the model following both estimations were iteratively removed. 

In order to parametrise the environment in which the nebulae or halos expand, we estimate the energy density and average black-body temperature of the IR and optical photon fields at the position of the pulsars using a model derived in \citet{rad_fields}. Both parameters were added to the input features. We find that the parameters with the largest influence on the surface brightness are the characteristic age and the first time derivative of barycentric rotation frequency (F1), the period derivative, and the energy density of the photon fields. The final parameters used to train the model can be found in Figure~\ref{fig:model_params}.

Several commonly invoked pulsar parameters do not appear as significant predictors in the empirical model. This does not imply that these parameters are physically irrelevant, but rather that their influence on the TeV surface brightness may be indirect, degenerate with other quantities, or masked by the large intrinsic scatter in the current sample. In this exercise, parameters derived from $P$ and $\dot{P}$ are kept, since their removal significantly decreases the prediction accuracy of the classifier.

While the limited size of the test sample prevents a detailed assessment of generalisation performance, its predictive performance can nevertheless be evaluated on the held-out test sample using the coefficient of determination ($R^2$), the mean absolute error (MAE), and the root mean square error (RMSE) in log space. The model achieves $R^2 = 0.58$, MAE = $0.41$, and RMSE = $0.45$, significantly outperforming a baseline model that predicts a constant mean surface brightness for all sources.
The residuals show no strong systematic trends with respect to the predicted surface brightness or the dominant input features. This suggests that the model captures at least part of the intrinsic variation in TeV surface brightness across the observed PWN population.

The trained model was subsequently applied to the sample of energetic pulsars for which no extended TeV emission has been detected to date. For each source, we obtain a predicted TeV $\gamma$-ray surface brightness together with an associated uncertainty, estimated via bootstrap resampling of the training data. These predictions are summarised in Table~\ref{tab:pred_vals}.

\begin{table*}
\caption{Predicted $\gamma$-ray emission properties for the PWNe or TeV halos around the five most powerful pulsars, without significant detection of extended TeV emission. 
} 
\label{tab:pred_vals} 
\centering    
\begin{tabular}{ c | c c c c c }       
\hline\hline  
\noalign{\smallskip}
 Pulsar Identifier & Extension [$^\circ$]  & $N_0$  & $N_\text{UL}$ & $S$ [$10^{30}$ erg s$^{-1}$ pc$^{-2}$]  \\    
\noalign{\smallskip}
\hline 
\noalign{\medskip}
PSR~J1032$-$5804 & $0.17^{+0.18}_{-0.09}$   & $1.22^{+0.26}_{-0.24}$  & $0.99$ &  $4.97^{+5.72}_{-2.58}$  \\ [0.3cm]
PSR~J1400$-$6325 & $0.11^{+0.13}_{-0.06}$  & $2.11^{+0.85}_{-0.84}$  & $1.10$  &  $20.84^{+15.44}_{-9.95}$   \\  [0.3cm]
PSR~J1524$-$5625 & $0.22^{+0.23}_{-0.11}$  & $2.20^{+0.48}_{-0.46}$  & $1.01$  &  $5.37^{+5.71}_{-2.78}$  \\ [0.3cm]
PSR~J1638$-$4713 & $0.10^{+0.10}_{-0.05}$  & $1.15^{+0.40}_{-0.38}$  & $1.21$  &  $13.58^{+14.39}_{-3.61}$   \\ [0.3cm]
PSR~J1935$+$2025 & $0.16^{+0.19}_{-0.08}$  & $2.90^{+0.81}_{-0.72}$  & $0.66$  &  $12.76^{+12.23}_{-7.07}$  \\ [0.3cm]
\noalign{\smallskip}
\hline   
\hline 
\end{tabular}
\tablefoot{Here $N_0$ is the normalisation at $1\,$TeV in $10^{-12}$ cm$^{-2}$ s$^{-1}$ TeV$^{-1}$, $N_\text{UL}$ is the upper limit derived from H.E.S.S. data in the same units and $S$ denotes the $\gamma$-ray surface brightness.}
\end{table*}

Given the limited number of pulsars currently known to host TeV PWNe or halos, the model was not extended to predict additional source parameters in a fully multivariate manner. Instead, we estimate the expected nebula extension and spectral index using linear regression relations derived from the observed population. The extension was predicted based on the pulsar's period and the derivative of the period. Since the spectral index of the nebulae shows no correlation with any pulsar parameter tested in this work, we adopt the mean of all PWNe observed by IACTs and use the spread of the measured $\gamma$-ray indices as the error on the predicted $\gamma$-ray index. We then combine these predicted parameters with the estimated surface brightness to derive the expected energy flux at $1\,\mathrm{TeV}$ for each source, and the parameters are likewise reported in Table~\ref{tab:pred_vals}.

\begin{figure}
\centering
\includegraphics[width=0.45\textwidth]{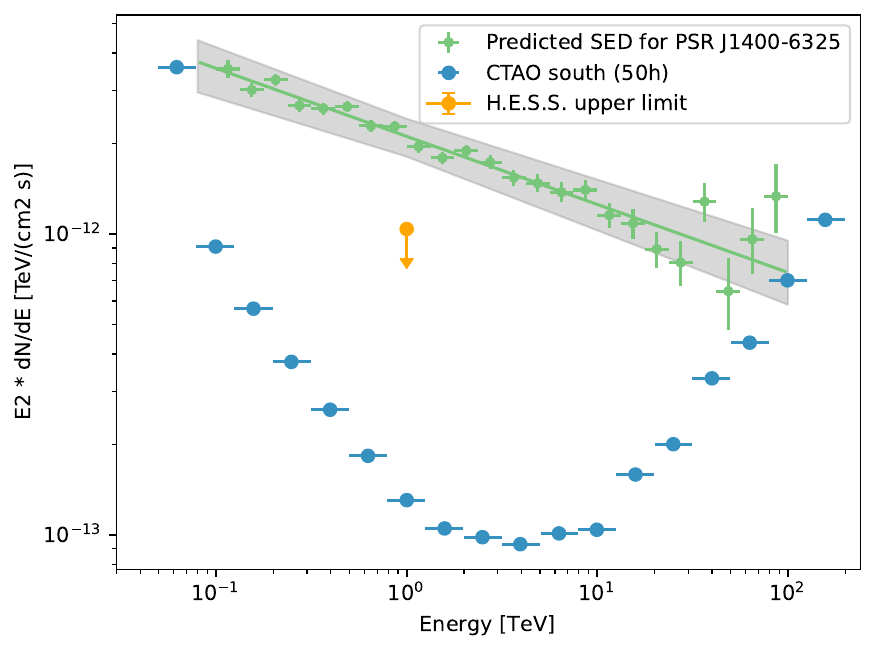}
\caption{Predicted SED for PSR~J1400$-$6325 compared to the 50 hour point-source sensitivity for CTAO south alpha configuration \citep{cta_sens}.}
\label{fig:predicted}%
\end{figure}

We use the resulting predictions to assess whether the inferred extended TeV emission would be detectable by next-generation $\gamma$-ray observatories. To this end, we employ Instrument Response Functions (IRFs) corresponding to 50 hours of observation with the Cherenkov Telescope Array Observatory (CTAO) \citep{cta_sens} to construct simulated datasets for each source. Using the CTAO IRFs, we generate forward-folded spectral simulations, which are then refitted, and the SEDs are extracted and compared to the published CTAO sensitivity curves. Figure~\ref{fig:predicted} shows the predicted SED for PSR~J1400$-$6325 compared to the CTAO sensitivity. While the uncertainties on the predicted $\gamma$-ray spectra are substantial, this test suggests that all five pulsars are expected to host extended TeV emission detectable by CTAO within a $50\,$hour exposure.

These results can also be compared to existing data. To this end, we derive conservative upper limits on the $\gamma$-ray flux normalisation at $1\,$TeV using integral flux maps from the HGPS \citep{HGPS}. These maps have been created with a kernel smoothing of $0.1^\circ$ to derive upper limits valid for the source extensions derived above, the upper limit is scaled to correspond to the $3\,\sigma$ containment region of the predicted $\gamma$-ray sources using a scaling factor:
\begin{equation}
    f = \sqrt{\frac{\sigma_\text{source}^2}{R_c^2} + 1} \,,
\end{equation}
with $R_c =0.1^\circ$ the kernel sizes used for the creation of the flux map and $\sigma_\text{source}$ the predicted $3\,\sigma$ containment radius of the source.

After extracting the integrated flux upper limit, we convert it into differential flux upper limits at a reference energy $E_{\rm ref} = 1\,$TeV assuming a power-law spectrum with photon index $\Gamma = 2.3$, which corresponds to the spectral index assumed in the HGPS flux map creation and is comparable to the index used for the simulations. These upper limits can be seen in Table \ref{tab:pred_vals}. 

The derived upper limit for the flux normalisation at $1\,$TeV at the position of PSR~J1032$-$5804 and PSR~J1638$-$4713 are comparable with the upper limits derived from the HGPS, while the predicted flux from the nebula powered by the other three nebulae exceeds the upper limits computed from H.E.S.S. data. This makes these three systems especially compelling targets for observations with CTAO. 

\section{Conclusion}

This work presents the first comprehensive catalogue of TeV pulsar environments, compiling information from all currently operating ground-based $\gamma$-ray instruments. The catalogue unifies morphological and spectral properties of pulsar wind nebulae, TeV halos, and related sources within a single, homogeneous framework and systematically links these observables to the physical characteristics of their powering pulsars. By consolidating previously heterogeneous datasets, it provides a coherent population-level view of pulsar-powered TeV emission across a wide range of evolutionary stages, observational conditions, and detection techniques.

The resulting population is dominated by young, energetic pulsars concentrated along the Galactic plane, reflecting both intrinsic luminosity effects and strong observational selection biases. These biases must be carefully accounted for when interpreting global trends and extrapolating to the broader pulsar population. As an initial demonstration of the diagnostic potential of the catalogue, this work examines correlations between TeV observables and pulsar parameters. The population trends revealed by this analysis demonstrate that the TeV properties of pulsar environments cannot be described by simple, one-to-one relationships with commonly used pulsar parameters and that characteristic age alone is a poor predictor of $\gamma$-ray morphology, spectral properties, or surface brightness. Instead, the observed diversity in TeV luminosities, extensions, and spectral indices points to a more complex evolutionary picture in which the present-day $\gamma$-ray emission reflects the cumulative particle injection history of the pulsar, the efficiency of particle confinement, and the influence of environmental conditions such as ambient density, magnetic turbulence, and pulsar motion. 

Correlations that do emerge, for example between surface brightness and the light-cylinder magnetic field or radio luminosity, indicate a tangible coupling between magnetospheric particle production and nebula emission, but these trends are often confined to specific evolutionary stages or observational subsamples. This highlights that population-wide correlations are easily diluted when heterogeneous systems and detector-dependent selection effects are considered together, and underscores the importance of homogeneous, multi-instrument datasets for interpreting the evolution of pulsar-powered $\gamma$-ray sources.

Finally, we use the catalogue to perform a targeted, data-driven assessment of the detectability of pulsar environments that have so far remained undetected at TeV energies. Focusing on five particularly powerful pulsars for which extended TeV emission would be expected based on their spin-down properties, but for which no $\gamma$-ray nebula has yet been reported, we employ a regression model trained on the observed population to predict their expected $\gamma$-ray surface brightness. Using these predicted spectra, we find that all five systems should be detectable by CTAO for realistic observation times of 50 hours, but three of the five predicted fluxes exceed upper limits placed in the region by H.E.S.S. data. A detection of PWNe around these systems would validate population-based scaling relations, while a non-detection would facilitate the identification of physical conditions that suppress or inhibit the formation of observable TeV pulsar wind nebulae in individual systems. This exercise demonstrates how the catalogue can be used not only to summarize the known TeV pulsar population, but also to prioritise future observations and guide targeted searches with next-generation $\gamma$-ray instruments.

\subsection*{Data availability}

All catalogue files are available in electronic form at the CDS via anonymous ftp to cdsarc.u-strasbg.fr (130.79.128.5) or via the CDS portal. 
\\
The catalogue data is also available and maintained on Zenodo via \url{https://doi.org/10.5281/zenodo.20705516}
\\
Details on the catalogue structure can be found in Appendix \ref{sec:layout}.

\begin{acknowledgements}
      Part of this work was supported by the German
      \emph{Deut\-sche For\-schungs\-ge\-mein\-schaft, DFG\/} project
      number 452934793. TL acknowledges support by the Swedish
Research Council under contract 2022- 04283. TL also acknowledges sabbatical support from the Wenner-Gren foundation under contract SSh2024-0037.
\end{acknowledgements}

\bibliographystyle{aa} 
\bibliography{references.bib} 

\begin{thebibliography}{86}
\expandafter\ifx\csname natexlab\endcsname\relax\def\natexlab#1{#1}\fi

\bibitem[{Abdalla {et~al.}(2018)Abdalla, Abramowski, Aharonian, Ait~Benkhali, Akhperjanian, Andersson, Angüner, Arrieta, Aubert, Backes, Balzer, Barnard, Becherini, Becker~Tjus, Berge, Bernhard, Bernlöhr, Blackwell, Böttcher, Boisson, Bolmont, Bordas, Bregeon, Brun, Brun, Bryan, Bulik, Capasso, Carr, Carrigan, Casanova, Cerruti, Chakraborty, Chalme-Calvet, Chaves, Chen, Chevalier, Chrétien, Colafrancesco, Cologna, Condon, Conrad, Couturier, Cui, Davids, Degrange, Deil, Devin, deWilt, Dirson, Djannati-Ataï, Domainko, Donath, Drury, Dubus, Dutson, Dyks, Edwards, Egberts, Eger, Ernenwein, Eschbach, Farnier, Fegan, Fernandes, Fiasson, Fontaine, Förster, Funk, Füßling, Gabici, Gajdus, Gallant, Garrigoux, Giavitto, Giebels, Glicenstein, Gottschall, Goyal, Grondin, Hadasch, Hahn, Haupt, Hawkes, Heinzelmann, Henri, Hermann, Hervet, Hillert, Hinton, Hofmann, Hoischen, Holler, Horns, Ivascenko, Jacholkowska, Jamrozy, Janiak, Jankowsky, Jankowsky, Jingo, Jogler, Jouvin, Jung-Richardt, Kastendieck, Katarzyński,
  Katz, Kerszberg, Khélifi, Kieffer, King, Klepser, Klochkov, Kluźniak, Kolitzus, Komin, Kosack, Krakau, Kraus, Krayzel, Krüger, Laffon, Lamanna, Lau, Lees, Lefaucheur, Lefranc, Lemière, Lemoine-Goumard, Lenain, Leser, Lohse, Lorentz, Liu, López-Coto, Lypova, Marandon, Marcowith, Mariaud, Marx, Maurin, Maxted, Mayer, Meintjes, Meyer, Mitchell, Moderski, Mohamed, Mohrmann, Morå, Moulin, Murach, de~Naurois, Niederwanger, Niemiec, Oakes, O’Brien, Odaka, Öttl, Ohm, de~Oña~Wilhelmi, Ostrowski, Oya, Padovani, Panter, Parsons, Paz~Arribas, Pekeur, Pelletier, Perennes, Petrucci, Peyaud, Pita, Poon, Prokhorov, Prokoph, Pühlhofer, Punch, Quirrenbach, Raab, Reimer, Reimer, Renaud, de~los Reyes, Rieger, Romoli, Rosier-Lees, Rowell, Rudak, Rulten, Sahakian, Salek, Sanchez, Santangelo, Sasaki, Schlickeiser, Schüssler, Schulz, Schwanke, Schwemmer, Settimo, Seyffert, Shafi, Shilon, Simoni, Sol, Spanier, Spengler, Spies, Stawarz, Steenkamp, Stegmann, Stinzing, Stycz, Sushch, Tavernet, Tavernier, Taylor, Terrier,
  Tibaldo, Tiziani, Tluczykont, Trichard, Tuffs, Uchiyama, Valerius, van~der Walt, van Eldik, van Soelen, Vasileiadis, Veh, Venter, Viana, Vincent, Vink, Voisin, Völk, Vuillaume, Wadiasingh, Wagner, Wagner, Wagner, White, Wierzcholska, Willmann, Wörnlein, Wouters, Yang, Zabalza, Zaborov, Zacharias, Zdziarski, Zech, Zefi, Ziegler, \& Żywucka}]{hesspwnpop}
Abdalla, H., Abramowski, A., Aharonian, F., {et~al.} 2018, Astronomy and Astrophysics, 612, A2

\bibitem[{{Abdo} {et~al.}(2007){Abdo}, {Allen}, {Berley}, {Casanova}, {Chen}, {Coyne}, {Dingus}, {Ellsworth}, {Fleysher}, {Fleysher}, {Gonzalez}, {Goodman}, {Hays}, {Hoffman}, {Hopper}, {H{\"u}ntemeyer}, {Kolterman}, {Lansdell}, {Linnemann}, {McEnery}, {Mincer}, {Nemethy}, {Noyes}, {Ryan}, {Saz Parkinson}, {Shoup}, {Sinnis}, {Smith}, {Sullivan}, {Vasileiou}, {Walker}, {Williams}, {Xu}, \& {Yodh}}]{veritas8_milagro}
{Abdo}, A.~A., {Allen}, B., {Berley}, D., {et~al.} 2007, \apjl, 664, L91

\bibitem[{{Abeysekara} {et~al.}(2023){Abeysekara}, {Albert}, {Alfaro}, {Alvarez}, {{\'A}lvarez}, {Araya}, {Arteaga-Vel{\'a}zquez}, {Arunbabu}, {Avila Rojas}, {Ayala Solares}, {Babu}, {Barber}, {Becerril}, {Belmont-Moreno}, {BenZvi}, {Blanco}, {Braun}, {Brisbois}, {Caballero-Mora}, {Cabrera Mart{\'\i}nez}, {Capistr{\'a}n}, {Carrami{\~n}ana}, {Casanova}, {Castillo}, {Chaparro-Amaro}, {Cotti}, {Cotzomi}, {Couti{\~n}o de Le{\'o}n}, {de la Fuente}, {de Le{\'o}n}, {De Young}, {Hernandez}, {Dingus}, {DuVernois}, {Durocher}, {D{\'\i}az-V{\'e}lez}, {Ellsworth}, {Engel}, {Espinoza}, {Fan}, {Fang}, {Fick}, {Fleischhack}, {Flores}, {Fraija}, {Garc{\'\i}a-Gonz{\'a}lez}, {Garcia-Torales}, {Garfias}, {Giacinti}, {Goksu}, {Gonz{\'a}lez}, {Gonz{\'a}lez-Mu{\~n}oz}, {Goodman}, {Harding}, {Hernandez}, {Hernandez}, {Hinton}, {Hona}, {Huang}, {Hueyotl-Zahuantitla}, {Hui}, {Humensky}, {H{\"u}ntemeyer}, {Iriarte}, {Imran}, {Jardin-Blicq}, {Joshi}, {Kaufmann}, {Kieda}, {Kunde}, {Lara}, {Lauer}, {Lee}, {Lennarz}, {Vargas},
  {Linnemann}, {Longinotti}, {Luis-Raya}, {Lundeen}, {Malone}, {Marandon}, {Marinelli}, {Martinez}, {Mart{\'\i}nez-Castellanos}, {Mart{\'\i}nez-Castro}, {Mart{\'\i}nez-Huerta}, {Matthews}, {Miranda-Romagnoli}, {Montaruli}, {Morales-Soto}, {Moreno}, {Mostaf{\'a}}, {Nayerhoda}, {Nellen}, {Newbold}, {Nisa}, {Noriega-Papaqui}, {Oceguera-Becerra}, {Olivera-Nieto}, {Omodei}, {Peisker}, {P{\'e}rez Araujo}, {P{\'e}rez-P{\'e}rez}, {Ponce}, {Pretz}, {Rho}, {Rosa-Gonz{\'a}lez}, {Ruiz-Velasco}, {Salazar}, {Salazar-Gallegos}, {Salesa Greus}, {Sandoval}, {Schneider}, {Schoorlemmer}, {Serna-Franco}, {Sinnis}, {Smith}, {Son}, {Sparks Woodle}, {Springer}, {Taboada}, {Tepe}, {Tibolla}, {Tollefson}, {Torres}, {Torres-Escobedo}, {Turner}, {Ure{\~n}a-Mena}, {Ukwatta}, {Varela}, {Vargas-Maga{\~n}a}, {Villase{\~n}or}, {Wang}, {Watson}, {Werner}, {Westerhoff}, {Willox}, {Wisher}, {Wood}, {Yodh}, {Zaborov}, {Zepeda}, {Zhou}, \& {HAWC Collaboration}}]{hawc}
{Abeysekara}, A.~U., {Albert}, A., {Alfaro}, R., {et~al.} 2023, Nuclear Instruments and Methods in Physics Research A, 1052, 168253

\bibitem[{{Abeysekara} {et~al.}(2017){Abeysekara}, {Albert}, {Alfaro}, {Alvarez}, {{\'A}lvarez}, {Arceo}, {Arteaga-Vel{\'a}zquez}, {Avila Rojas}, {Ayala Solares}, {Barber}, {Bautista-Elivar}, {Becerril}, {Belmont-Moreno}, {BenZvi}, {Berley}, {Bernal}, {Braun}, {Brisbois}, {Caballero-Mora}, {Capistr{\'a}n}, {Carrami{\~n}ana}, {Casanova}, {Castillo}, {Cotti}, {Cotzomi}, {Couti{\~n}o de Le{\'o}n}, {De Le{\'o}n}, {De la Fuente}, {Dingus}, {DuVernois}, {D{\'\i}az-V{\'e}lez}, {Ellsworth}, {Engel}, {Enr{\'\i}quez-Rivera}, {Fiorino}, {Fraija}, {Garc{\'\i}a-Gonz{\'a}lez}, {Garfias}, {Gerhardt}, {Gonz{\'a}lez Mu{\~n}oz}, {Gonz{\'a}lez}, {Goodman}, {Hampel-Arias}, {Harding}, {Hern{\'a}ndez}, {Hern{\'a}ndez-Almada}, {Hinton}, {Hona}, {Hui}, {H{\"u}ntemeyer}, {Iriarte}, {Jardin-Blicq}, {Joshi}, {Kaufmann}, {Kieda}, {Lara}, {Lauer}, {Lee}, {Lennarz}, {Vargas}, {Linnemann}, {Longinotti}, {Luis Raya}, {Luna-Garc{\'\i}a}, {L{\'o}pez-Coto}, {Malone}, {Marinelli}, {Martinez}, {Martinez-Castellanos}, {Mart{\'\i}nez-Castro},
  {Mart{\'\i}nez-Huerta}, {Matthews}, {Miranda-Romagnoli}, {Moreno}, {Mostaf{\'a}}, {Nellen}, {Newbold}, {Nisa}, {Noriega-Papaqui}, {Pelayo}, {Pretz}, {P{\'e}rez-P{\'e}rez}, {Ren}, {Rho}, {Rivi{\`e}re}, {Rosa-Gonz{\'a}lez}, {Rosenberg}, {Ruiz-Velasco}, {Salazar}, {Salesa Greus}, {Sandoval}, {Schneider}, {Schoorlemmer}, {Sinnis}, {Smith}, {Springer}, {Surajbali}, {Taboada}, {Tibolla}, {Tollefson}, {Torres}, {Ukwatta}, {Vianello}, {Weisgarber}, {Westerhoff}, {Wisher}, {Wood}, {Yapici}, {Yodh}, {Younk}, {Zepeda}, {Zhou}, {Guo}, {Hahn}, {Li}, \& {Zhang}}]{hawc_geminga}
{Abeysekara}, A.~U., {Albert}, A., {Alfaro}, R., {et~al.} 2017, Science, 358, 911

\bibitem[{{Abeysekara} {et~al.}(2018){Abeysekara}, {Archer}, {Benbow}, {Bird}, {Brose}, {Buchovecky}, {Buckley}, {Bugaev}, {Chromey}, {Connolly}, {Cui}, {Daniel}, {Falcone}, {Feng}, {Finley}, {Fortson}, {Furniss}, {H{\"u}tten}, {Hanna}, {Hervet}, {Holder}, {Hughes}, {Humensky}, {Johnson}, {Kaaret}, {Kar}, {Kertzman}, {Kieda}, {Krause}, {Krennrich}, {Kumar}, {Lang}, {Lin}, {McArthur}, {Moriarty}, {Mukherjee}, {O'Brien}, {Ong}, {Otte}, {Park}, {Petrashyk}, {Pohl}, {Pueschel}, {Quinn}, {Ragan}, {Reynolds}, {Richards}, {Roache}, {Rulten}, {Sadeh}, {Santander}, {Sembroski}, {Shahinyan}, {Sushch}, {Tyler}, {Wakely}, {Weinstein}, {Wells}, {Wilcox}, {Wilhelm}, {Williams}, {Williamson}, {Zitzer}, {VERITAS Collaboration}, {Abdollahi}, {Ajello}, {Baldini}, {Barbiellini}, {Bastieri}, {Bellazzini}, {Berenji}, {Bissaldi}, {Blandford}, {Bonino}, {Bottacini}, {Brandt}, {Bruel}, {Buehler}, {Cameron}, {Caputo}, {Caraveo}, {Castro}, {Cavazzuti}, {Charles}, {Chiaro}, {Ciprini}, {Cohen-Tanugi}, {Costantin}, {Cutini}, {D'Ammando},
  {de Palma}, {Di Lalla}, {Di Mauro}, {Di Venere}, {Dom{\'\i}nguez}, {Favuzzi}, {Fegan}, {Franckowiak}, {Fukazawa}, {Funk}, {Fusco}, {Gargano}, {Gasparrini}, {Giglietto}, {Giordano}, {Giroletti}, {Green}, {Grenier}, {Guillemot}, {Guiriec}, {Hays}, {Hewitt}, {Horan}, {J{\'o}hannesson}, {Kensei}, {Kuss}, {Larsson}, {Latronico}, {Lemoine-Goumard}, {Li}, {Longo}, {Loparco}, {Lovellette}, {Lubrano}, {Magill}, {Maldera}, {Mazziotta}, {McEnery}, {Michelson}, {Mitthumsiri}, {Mizuno}, {Monzani}, {Morselli}, {Moskalenko}, {Negro}, {Nuss}, {Ojha}, {Omodei}, {Orienti}, {Orlando}, {Palatiello}, {Paliya}, {Paneque}, {Perkins}, {Persic}, {Pesce-Rollins}, {Petrosian}, {Piron}, {Porter}, {Principe}, {Rain{\`o}}, {Rando}, {Rani}, {Razzano}, {Razzaque}, {Reimer}, {Reimer}, {Reposeur}, {Sgr{\`o}}, {Siskind}, {Spandre}, {Spinelli}, {Suson}, {Tajima}, {Thayer}, {Thompson}, {Torres}, {Tosti}, {Troja}, {Valverde}, {Vianello}, {Vogel}, {Wood}, {Yassine}, {Fermi-LAT Collaboration}, {Alfaro}, {{\'A}lvarez}, {{\'A}lvarez}, {Arceo},
  {Arteaga-Vel{\'a}zquez}, {Avila Rojas}, {Ayala Solares}, {Becerril}, {Belmont-Moreno}, {BenZvi}, {Bernal}, {Braun}, {Brisbois}, {Caballero-Mora}, {Capistr{\'a}n}, {Carrami{\~n}ana}, {Casanova}, {Castillo}, {Cotti}, {Cotzomi}, {Couti{\~n}o de Le{\'o}n}, {De Le{\'o}n}, {De la Fuente}, {Dichiara}, \& {Dingus}}]{Abeysekara_2018}
{Abeysekara}, A.~U., {Archer}, A., {Benbow}, W., {et~al.} 2018, \apj, 866, 24

\bibitem[{{Abeysekara} {et~al.}(2020){Abeysekara}, {Benbow}, {Bird}, {Brose}, {Christiansen}, {Chromey}, {Cui}, {Daniel}, {Falcone}, {Fortson}, {Hanna}, {Hassan}, {Hervet}, {Holder}, {Hughes}, {Humensky}, {Kaaret}, {Kar}, {Kelley-Hoskins}, {Kertzman}, {Kieda}, {Krause}, {Lang}, {Maier}, {Moriarty}, {Nieto}, {Nievas-Rosillo}, {Ong}, {Pandel}, {Pohl}, {Prado}, {Pueschel}, {Quinn}, {Ragan}, {Reynolds}, {Richards}, {Roache}, {Sadeh}, {Santander}, {Sembroski}, {Weinstein}, {Wilcox}, {Williams}, \& {Williamson}}]{vertias_source2}
{Abeysekara}, A.~U., {Benbow}, W., {Bird}, R., {et~al.} 2020, Astroparticle Physics, 117, 102403

\bibitem[{Abeysekara {et~al.}(2017{\natexlab{a}})}]{HAWC:2017kbo}
Abeysekara, A.~U. {et~al.} 2017{\natexlab{a}}, Science, 358, 911

\bibitem[{Abeysekara {et~al.}(2017{\natexlab{b}})}]{2hwc_catalog}
Abeysekara, A.~U. {et~al.} 2017{\natexlab{b}}, Astrophys. J., 843, 40

\bibitem[{{Acharyya} {et~al.}(2023){Acharyya}, {Adams}, {Archer}, {Bangale}, {Bartkoske}, {Batista}, {Benbow}, {Brill}, {Brose}, {Buckley}, {Capasso}, {Christiansen}, {Chromey}, {Daniel}, {Errando}, {Falcone}, {Farrell}, {Feng}, {Finley}, {Foote}, {Fortson}, {Furniss}, {Gallagher}, {Gent}, {Giuri}, {Gueta}, {Hanlon}, {Hanna}, {Hassan}, {Hervet}, {Hoang}, {Holder}, {Hughes}, {Humensky}, {Jin}, {Kaaret}, {Kertzman}, {Kieda}, {Kleiner}, {Korzoun}, {Krennrich}, {Kumar}, {Lang}, {Lundy}, {Maier}, {McGrath}, {Millard}, {Mooney}, {Moriarty}, {Mukherjee}, {Nieto}, {Nievas-Rosillo}, {O'Brien}, {Ong}, {Otte}, {Pandel}, {Park}, {Patel}, {Patel}, {Pfrang}, {Pichel}, {Pohl}, {Prado}, {Pueschel}, {Quinn}, {Ragan}, {Reynolds}, {Ribeiro}, {Richards}, {Roache}, {Rovero}, {Rulten}, {Ryan}, {Sadeh}, {Santander}, {Schlenstedt}, {Sembroski}, {Shang}, {Splettstoesser}, {Stevenson}, {Tak}, {Vassiliev}, {Wakely}, {Weinstein}, {Williams}, {Williamson}, {Angelini}, {Basu-Zych}, {Sabol}, \& {Smale}}]{vts_cat}
{Acharyya}, A., {Adams}, C.~B., {Archer}, A., {et~al.} 2023, Research Notes of the American Astronomical Society, 7, 6

\bibitem[{{Aharonian} {et~al.}(2024){Aharonian}, {Ait Benkhali}, {Aschersleben}, {Ashkar}, {Backes}, {Baktash}, {Barbosa Martins}, {Batzofin}, {Becherini}, {Berge}, {Bernl{\"o}hr}, {Bi}, {B{\"o}ttcher}, {Boisson}, {Bolmont}, {de Bony de Lavergne}, {Borowska}, {Bradascio}, {Breuhaus}, {Brose}, {Brown}, {Brun}, {Bruno}, {Bulik}, {Burger-Scheidlin}, {Bylund}, {Caroff}, {Casanova}, {Cecil}, {Celic}, {Cerruti}, {Chambery}, {Chand}, {Chandra}, {Chen}, {Chibueze}, {Chibueze}, {Cotter}, {Cristofari}, {Devin}, {Djannati-Ata{\"\i}}, {Djuvsland}, {Dmytriiev}, {Einecke}, {Ernenwein}, {Fegan}, {Feijen}, {Filipovi{\'c}}, {Fontaine}, {F{\"u}{\ss}ling}, {Funk}, {Gabici}, {Gallant}, {Giavitto}, {Glawion}, {Glicenstein}, {Glombitza}, {Goswami}, {Grolleron}, {Grondin}, {Haerer}, {Hinton}, {Hofmann}, {Holch}, {Holler}, {Horns}, {Jamrozy}, {Jankowsky}, {Joshi}, {Kasai}, {Katarzy{\'n}ski}, {Khatoon}, {Kh{\'e}lifi}, {Klu{\'z}niak}, {Komin}, {Kosack}, {Kostunin}, {Kundu}, {Lang}, {Le Stum}, {Leitl}, {Lemi{\`e}re}, {Lemoine-Goumard},
  {Lenain}, {Leuschner}, {Luashvili}, {Mackey}, {Malyshev}, {Malyshev}, {Marandon}, {Marinos}, {Mart{\'\i}-Devesa}, {Marx}, {Mehta}, {Meyer}, {Mitchell}, {Moderski}, {Mohrmann}, {Montanari}, {Moulin}, {Murach}, {de Naurois}, {Niemiec}, {O'Brien}, {Ohm}, {Olivera-Nieto}, {de Ona Wilhelmi}, {Ostrowski}, {Panny}, {Panter}, {Parsons}, {Peron}, {Prokhorov}, {P{\"u}hlhofer}, {Punch}, {Quirrenbach}, {Regeard}, {Reichherzer}, {Reimer}, {Reimer}, {Ren}, {Renaud}, {Reville}, {Rieger}, {Roellinghoff}, {Rudak}, {Sahakian}, {Salzmann}, {Sasaki}, {Sch{\"u}ssler}, {Schutte}, {Shapopi}, {Specovius}, {Spencer}, {Stawarz}, {Steenkamp}, {Steinmassl}, {Steppa}, {Streil}, {Sushch}, {Suzuki}, {Takahashi}, {Tanaka}, {Terrier}, {Tluczykont}, {Tsuji}, {Unbehaun}, {van Eldik}, {Vecchi}, {Veh}, {Venter}, {Vink}, {Wach}, {Wagner}, {Wierzcholska}, {Zacharias}, {Zargaryan}, {Zdziarski}, {Zech}, {Zouari}, {{\.Z}ywucka}, \& {Harding}}]{hess_crab2024}
{Aharonian}, F., {Ait Benkhali}, F., {Aschersleben}, J., {et~al.} 2024, \aap, 686, A308

\bibitem[{{Aharonian} {et~al.}(2009){Aharonian}, {Akhperjanian}, {Anton}, {Barres de Almeida}, {Bazer-Bachi}, {Becherini}, {Behera}, {Benbow}, {Bernl{\"o}hr}, {Boisson}, {Bochow}, {Borrel}, {Braun}, {Brion}, {Brucker}, {Brun}, {B{\"u}hler}, {Bulik}, {B{\"u}sching}, {Boutelier}, {Carrigan}, {Chadwick}, {Charbonnier}, {Chaves}, {Cheesebrough}, {Chounet}, {Clapson}, {Coignet}, {Dalton}, {Daniel}, {Degrange}, {Deil}, {Dickinson}, {Djannati-Ata{\"\i}}, {Domainko}, {O'C. Drury}, {Dubois}, {Dubus}, {Dyks}, {Dyrda}, {Egberts}, {Emmanoulopoulos}, {Espigat}, {Farnier}, {Feinstein}, {Fiasson}, {F{\"o}rster}, {Fontaine}, {F{\"u}{\ss}ling}, {Gabici}, {Gallant}, {G{\'e}rard}, {Giebels}, {Glicenstein}, {Gl{\"u}ck}, {Goret}, {Hauser}, {Hauser}, {Heinz}, {Heinzelmann}, {Henri}, {Hermann}, {Hinton}, {Hoffmann}, {Hofmann}, {Holleran}, {Hoppe}, {Horns}, {Jacholkowska}, {de Jager}, {Jung}, {Katarzy{\'n}ski}, {Katz}, {Kaufmann}, {Kendziorra}, {Kerschhaggl}, {Khangulyan}, {Kh{\'e}lifi}, {Keogh}, {Komin}, {Kosack}, {Lamanna},
  {Lenain}, {Lohse}, {Marandon}, {Martin}, {Martineau-Huynh}, {Marcowith}, {Maurin}, {McComb}, {Medina}, {Moderski}, {Moulin}, {Naumann-Godo}, {de Naurois}, {Nedbal}, {Nekrassov}, {Niemiec}, {Nolan}, {Ohm}, {Olive}, {de O{\~n}a Wilhelmi}, {Orford}, {Ostrowski}, {Panter}, {Paz Arribas}, {Pedaletti}, {Pelletier}, {Petrucci}, {Pita}, {P{\"u}hlhofer}, {Punch}, {Quirrenbach}, {Raubenheimer}, {Raue}, {Rayner}, {Renaud}, {Reimer}, {Rieger}, {Ripken}, {Rob}, {Rosier-Lees}, {Rowell}, {Rudak}, {Rulten}, {Ruppel}, {Sahakian}, {Santangelo}, {Schlickeiser}, {Sch{\"o}ck}, {Schr{\"o}der}, {Schwanke}, {Schwarzburg}, {Schwemmer}, {Shalchi}, {Skilton}, {Sol}, {Spangler}, {Stawarz}, {Steenkamp}, {Stegmann}, {Superina}, {Tam}, {Tavernet}, {Terrier}, {Tibolla}, {van Eldik}, {Vasileiadis}, {Venter}, {Venter}, {Vialle}, {Vincent}, {Vivier}, {V{\"o}lk}, {Volpe}, {Wagner}, {Ward}, {Zdziarski}, \& {Zech}}]{vertias_source3_hess}
{Aharonian}, F., {Akhperjanian}, A.~G., {Anton}, G., {et~al.} 2009, \aap, 499, 723

\bibitem[{{Aharonian} {et~al.}(2005){Aharonian}, {Akhperjanian}, {Aye}, {Bazer-Bachi}, {Beilicke}, {Benbow}, {Berge}, {Berghaus}, {Bernl{\"o}hr}, {Boisson}, {Bolz}, {Borgmeier}, {Braun}, {Breitling}, {Brown}, {Bussons Gordo}, {Chadwick}, {Chounet}, {Cornils}, {Costamante}, {Degrange}, {Djannati-Ata{\"\i}}, {O'C. Drury}, {Dubus}, {Ergin}, {Espigat}, {Feinstein}, {Fleury}, {Fontaine}, {Funk}, {Gallant}, {Giebels}, {Gillessen}, {Goret}, {Hadjichristidis}, {Hauser}, {Heinzelmann}, {Henri}, {Hermann}, {Hinton}, {Hofmann}, {Holleran}, {Horns}, {de Jager}, {Jung}, {Kh{\'e}lifi}, {Komin}, {Konopelko}, {Latham}, {Le Gallou}, {Lemi{\`e}re}, {Lemoine}, {Leroy}, {Lohse}, {Marcowith}, {Masterson}, {McComb}, {de Naurois}, {Nolan}, {Noutsos}, {Orford}, {Osborne}, {Ouchrif}, {Panter}, {Pelletier}, {Pita}, {P{\"u}hlhofer}, {Punch}, {Raubenheimer}, {Raue}, {Raux}, {Rayner}, {Redondo}, {Reimer}, {Reimer}, {Ripken}, {Rob}, {Rolland}, {Rowell}, {Sahakian}, {Saug{\'e}}, {Schlenker}, {Schlickeiser}, {Schuster}, {Schwanke},
  {Siewert}, {Sol}, {Steenkamp}, {Stegmann}, {Tavernet}, {Terrier}, {Th{\'e}oret}, {Tluczykont}, {Vasileiadis}, {Venter}, {Vincent}, {Visser}, {V{\"o}lk}, \& {Wagner}}]{veritas7_hess}
{Aharonian}, F., {Akhperjanian}, A.~G., {Aye}, K.-M., {et~al.} 2005, \aap, 432, L25

\bibitem[{{Aharonian} {et~al.}(2008){Aharonian}, {Akhperjanian}, {Barres de Almeida}, {Bazer-Bachi}, {Behera}, {Beilicke}, {Benbow}, {Bernl{\"o}hr}, {Boisson}, {Bolz}, {Borrel}, {Braun}, {Brion}, {Brown}, {B{\"u}hler}, {Bulik}, {B{\"u}sching}, {Boutelier}, {Carrigan}, {Chadwick}, {Chounet}, {Clapson}, {Coignet}, {Cornils}, {Costamante}, {Dalton}, {Degrange}, {Dickinson}, {Djannati-Ata{\"\i}}, {Domainko}, {Drury}, {Dubois}, {Dubus}, {Dyks}, {Egberts}, {Emmanoulopoulos}, {Espigat}, {Farnier}, {Feinstein}, {Fiasson}, {F{\"o}rster}, {Fontaine}, {Funk}, {F{\"u}{\ss}ling}, {Gallant}, {Giebels}, {Glicenstein}, {Gl{\"u}ck}, {Goret}, {Hadjichristidis}, {Hauser}, {Hauser}, {Heinzelmann}, {Henri}, {Hermann}, {Hinton}, {Hoffmann}, {Hofmann}, {Holleran}, {Hoppe}, {Horns}, {Jacholkowska}, {de Jager}, {Jung}, {Katarzy{\'n}ski}, {Kendziorra}, {Kerschhaggl}, {Kh{\'e}lifi}, {Keogh}, {Komin}, {Kosack}, {Lamanna}, {Latham}, {Lemi{\`e}re}, {Lemoine-Goumard}, {Lenain}, {Lohse}, {Martin}, {Martineau-Huynh}, {Marcowith},
  {Masterson}, {Maurin}, {Maurin}, {McComb}, {Moderski}, {Moulin}, {de Naurois}, {Nedbal}, {Nolan}, {Ohm}, {Olive}, {de O{\~n}a Wilhelmi}, {Orford}, {Osborne}, {Ostrowski}, {Panter}, {Pedaletti}, {Pelletier}, {Petrucci}, {Pita}, {P{\"u}hlhofer}, {Punch}, {Ranchon}, {Raubenheimer}, {Raue}, {Rayner}, {Renaud}, {Ripken}, {Rob}, {Rolland}, {Rosier-Lees}, {Rowell}, {Rudak}, {Ruppel}, {Sahakian}, {Santangelo}, {Schlickeiser}, {Sch{\"o}ck}, {Schr{\"o}der}, {Schwanke}, {Schwarzburg}, {Schwemmer}, {Shalchi}, {Sol}, {Spangler}, {Stawarz}, {Steenkamp}, {Stegmann}, {Superina}, {Tam}, {Tavernet}, {Terrier}, {van Eldik}, {Vasileiadis}, {Venter}, {Vialle}, {Vincent}, {Vivier}, {V{\"o}lk}, {Volpe}, {Wagner}, {Ward}, {Zdziarski}, \& {Zech}}]{magic4_hess}
{Aharonian}, F., {Akhperjanian}, A.~G., {Barres de Almeida}, U., {et~al.} 2008, \aap, 477, 353

\bibitem[{{Aharonian} {et~al.}(2006){Aharonian}, {Akhperjanian}, {Bazer-Bachi}, {Beilicke}, {Benbow}, {Berge}, {Bernl{\"o}hr}, {Boisson}, {Bolz}, {Borrel}, {Braun}, {Breitling}, {Brown}, {Chadwick}, {Chounet}, {Cornils}, {Costamante}, {Degrange}, {Dickinson}, {Djannati-Ata{\"\i}}, {Drury}, {Dubus}, {Emmanoulopoulos}, {Espigat}, {Feinstein}, {Fontaine}, {Fuchs}, {Funk}, {Gallant}, {Giebels}, {Gillessen}, {Glicenstein}, {Goret}, {Hadjichristidis}, {Hauser}, {Heinzelmann}, {Henri}, {Hermann}, {Hinton}, {Hofmann}, {Holleran}, {Horns}, {Jacholkowska}, {de Jager}, {Kh{\'e}lifi}, {Komin}, {Konopelko}, {Latham}, {Le Gallou}, {Lemi{\`e}re}, {Lemoine-Goumard}, {Leroy}, {Lohse}, {Martin}, {Martineau-Huynh}, {Marcowith}, {Masterson}, {McComb}, {de Naurois}, {Nolan}, {Noutsos}, {Orford}, {Osborne}, {Ouchrif}, {Panter}, {Pelletier}, {Pita}, {P{\"u}hlhofer}, {Punch}, {Raubenheimer}, {Raue}, {Raux}, {Rayner}, {Reimer}, {Reimer}, {Ripken}, {Rob}, {Rolland}, {Rowell}, {Sahakian}, {Saug{\'e}}, {Schlenker}, {Schlickeiser},
  {Schuster}, {Schwanke}, {Siewert}, {Sol}, {Spangler}, {Steenkamp}, {Stegmann}, {Tavernet}, {Terrier}, {Th{\'e}oret}, {Tluczykont}, {Vasileiadis}, {Venter}, {Vincent}, {V{\"o}lk}, \& {Wagner}}]{hess_survey1}
{Aharonian}, F., {Akhperjanian}, A.~G., {Bazer-Bachi}, A.~R., {et~al.} 2006, \apj, 636, 777

\bibitem[{{Aharonian} {et~al.}(2021){Aharonian}, {An}, {Axikegu}, {Bai}, {Bai}, {Bao}, {Bastieri}, {Bi}, {Bi}, {Cai}, {Cai}, {Cao}, {Cao}, {Chang}, {Chang}, {Chang}, {Chen}, {Chen}, {Chen}, {Chen}, {Chen}, {Chen}, {Chen}, {Chen}, {Chen}, {Chen}, {Chen}, {Chen}, {Chen}, {Cheng}, {Cheng}, {Cui}, {Cui}, {Cui}, {Dai}, {Dai}, {Dai}, {Danzengluobu}, {Della Volpe}, {Piazzoli}, {Dong}, {Fan}, {Fan}, {Fan}, {Fang}, {Fang}, {Feng}, {Feng}, {Feng}, {Feng}, {Gao}, {Gao}, {Gao}, {Gao}, {Ge}, {Geng}, {Gong}, {Gou}, {Gu}, {Guo}, {Guo}, {Guo}, {Guo}, {Han}, {He}, {He}, {He}, {He}, {He}, {He}, {Heller}, {Hor}, {Hou}, {Hou}, {Hu}, {Hu}, {Hu}, {Hu}, {Huang}, {Huang}, {Huang}, {Huang}, {Huang}, {Ji}, {Ji}, {Jia}, {Jiang}, {Jiang}, {Jin}, {Kuleshov}, {Levochkin}, {Li}, {Li}, {Li}, {Li}, {Li}, {Li}, {Li}, {Li}, {Li}, {Li}, {Li}, {Li}, {Li}, {Li}, {Li}, {Li}, {Li}, {Liang}, {Liang}, {Lin}, {Liu}, {Liu}, {Liu}, {Liu}, {Liu}, {Liu}, {Liu}, {Liu}, {Liu}, {Liu}, {Liu}, {Liu}, {Liu}, {Liu}, {Liu}, {Long}, {Lu}, {Lv}, {Ma}, {Ma}, {Ma},
  {Mao}, {Masood}, {Mitthumsiri}, {Montaruli}, {Nan}, {Pang}, {Pattarakijwanich}, {Pei}, {Qi}, {Qiao}, {Ruffolo}, {Rulev}, {S{\'a}iz}, {Shao}, {Shchegolev}, {Sheng}, {Shi}, {Song}, {Stenkin}, {Stepanov}, {Sun}, {Sun}, {Sun}, {Tam}, {Tang}, {Tian}, {Wang}, {Wang}, {Wang}, {Wang}, {Wang}, {Wang}, {Wang}, {Wang}, {Wang}, {Wang}, {Wang}, {Wang}, {Wang}, {Wang}, {Wang}, {Wang}, {Wang}, {Wang}, {Wang}, {Wang}, {Wang}, {Wei}, {Wei}, {Wei}, {Wen}, {Wu}, {Wu}, {Wu}, {Wu}, {Wu}, {Xi}, {Xia}, {Xia}, {Xiang}, {Xiao}, {Xiao}, {Xin}, {Xin}, {Xing}, {Xu}, {Xu}, \& {Xue}}]{wcda_perf}
{Aharonian}, F., {An}, Q., {Axikegu}, {et~al.} 2021, Chinese Physics C, 45, 085002

\bibitem[{Aharonian {et~al.}(1995)Aharonian, Atoyan, \& Volk}]{Aharonian:1995zz}
Aharonian, F.~A., Atoyan, A.~M., \& Volk, H.~J. 1995, Astron. Astrophys., 294, L41

\bibitem[{{Aharonian, F.} {et~al.}(2006){Aharonian, F.}, {Akhperjanian, A. G.}, {Bazer-Bachi, A. R.}, {Beilicke, M.}, {Benbow, W.}, {Berge, D.}, {Bernlöhr, K.}, {Boisson, C.}, {Bolz, O.}, {Borrel, V.}, {Braun, I.}, {Breitling, F.}, {Brown, A. M.}, {Bühler, R.}, {Büsching, I.}, {Carrigan, S.}, {Chadwick, P. M.}, {Chounet, L.-M.}, {Cornils, R.}, {Costamante, L.}, {Degrange, B.}, {Dickinson, H. J.}, {Djannati-Ataï, A.}, {Drury, L. O'C.}, {Dubus, G.}, {Egberts, K.}, {Emmanoulopoulos, D.}, {Espigat, P.}, {Feinstein, F.}, {Ferrero, E.}, {Fiasson, A.}, {Fontaine, G.}, {Funk, Seb.}, {Funk, S.}, {Gallant, Y. A.}, {Giebels, B.}, {Glicenstein, J. F.}, {Goret, P.}, {Hadjichristidis, C.}, {Hauser, D.}, {Hauser, M.}, {Heinzelmann, G.}, {Henri, G.}, {Hermann, G.}, {Hinton, J. A.}, {Hofmann, W.}, {Holleran, M.}, {Horns, D.}, {Jacholkowska, A.}, {de Jager, O. C.}, {Khélifi, B.}, {Komin, Nu.}, {Konopelko, A.}, {Kosack, K.}, {Latham, I. J.}, {Le Gallou, R.}, {Lemière, A.}, {Lemoine-Goumard, M.}, {Lohse, T.}, {Martin, J.
  M.}, {Martineau-Huynh, O.}, {Marcowith, A.}, {Masterson, C.}, {McComb, T. J. L.}, {de Naurois, M.}, {Nedbal, D.}, {Nolan, S. J.}, {Noutsos, A.}, {Orford, K. J.}, {Osborne, J. L.}, {Ouchrif, M.}, {Panter, M.}, {Pelletier, G.}, {Pita, S.}, {Pühlhofer, G.}, {Punch, M.}, {Raubenheimer, B. C.}, {Raue, M.}, {Rayner, S. M.}, {Reimer, A.}, {Reimer, O.}, {Ripken, J.}, {Rob, L.}, {Rolland, L.}, {Rowell, G.}, {Sahakian, V.}, {Saugé, L.}, {Schlenker, S.}, {Schlickeiser, R.}, {Schwanke, U.}, {Sol, H.}, {Spangler, D.}, {Spanier, F.}, {Steenkamp, R.}, {Stegmann, C.}, {Superina, G.}, {Tavernet, J.-P.}, {Terrier, R.}, {Théoret, C. G.}, {Tluczykont, M.}, {van Eldik, C.}, {Vasileiadis, G.}, {Venter, C.}, {Vincent, P.}, {Völk, H. J.}, {Wagner, S. J.}, \& {Ward, M.}}]{hess}
{Aharonian, F.}, {Akhperjanian, A. G.}, {Bazer-Bachi, A. R.}, {et~al.} 2006, A\&A, 457, 899

\bibitem[{{Albert} {et~al.}(2024{\natexlab{a}}){Albert}, {Alfaro}, {Alvarez}, {Andr{\'e}s}, {Arteaga-Vel{\'a}zquez}, {Avila Rojas}, {Ayala Solares}, {Babu}, {Belmont-Moreno}, {Bernal}, {Caballero-Mora}, {Capistr{\'a}n}, {Carrami{\~n}ana}, {Carre{\'o}n}, {Casanova}, {Cotti}, {Cotzomi}, {Couti{\~n}o de Le{\'o}n}, {De la Fuente}, {de Le{\'o}n}, {Depaoli}, {Di Lalla}, {D{\'\i}az Hern{\'a}ndez}, {Dingus}, {DuVernois}, {Engel}, {Ergin}, {Espinoza}, {Fan}, {Fang}, {Fraija}, {Fraija}, {Garc{\'\i}a-Gonz{\'a}lez}, {Garfias}, {Goksu}, {Gonz{\'a}lez}, {Goodman}, {Groetsch}, {Harding}, {Hern{\'a}ndez-Cadena}, {Herzog}, {Hinton}, {Huang}, {Hueyotl-Zahuantitla}, {H{\"u}ntemeyer}, {Iriarte}, {Kaufmann}, {Lara}, {Lee}, {Le{\'o}n Vargas}, {Linnemann}, {Longinotti}, {Luis-Raya}, {Malone}, {Mart{\'\i}nez-Castro}, {Matthews}, {Miranda-Romagnoli}, {Montes}, {Moreno}, {Mostaf{\'a}}, {Nellen}, {Nisa}, {Noriega-Papaqui}, {Olivera-Nieto}, {Omodei}, {Osorio-Archila}, {P{\'e}rez Araujo}, {P{\'e}rez-P{\'e}rez}, {Rho},
  {Rosa-Gonz{\'a}lez}, {Ruiz-Velasco}, {Salazar}, {Salazar-Gallegos}, {Sandoval}, {Schneider}, {Schwefer}, {Serna-Franco}, {Smith}, {Son}, {Springer}, {Tibolla}, {Tollefson}, {Torres}, {Torres-Escobedo}, {Turner}, {Ure{\~n}a-Mena}, {Varela}, {Wang}, {Watson}, {Whitaker}, {Willox}, {Wu}, {Yu}, {Yun-C{\'a}rcamo}, {Zhou}, \& {HAWC Collaboration}}]{hawc_perf}
{Albert}, A., {Alfaro}, R., {Alvarez}, C., {et~al.} 2024{\natexlab{a}}, \apj, 972, 144

\bibitem[{{Albert} {et~al.}(2024{\natexlab{b}}){Albert}, {Alfaro}, {Alvarez}, {Arteaga-Vel{\'a}zquez}, {Avila Rojas}, {Ayala Solares}, {Babu}, {Belmont-Moreno}, {Bernal}, {Caballero-Mora}, {Capistr{\'a}n}, {Carrami{\~n}ana}, {Casanova}, {Cotti}, {Cotzomi}, {Couti{\~n}o de Le{\'o}n}, {de la Fuente}, {Depaoli}, {Di Lalla}, {Diaz Hernandez}, {Dingus}, {DuVernois}, {Durocher}, {D{\'\i}az-V{\'e}lez}, {Engel}, {Espinoza}, {Fan}, {Fang}, {Fraija}, {Garc{\'\i}a-Gonz{\'a}lez}, {Garfias}, {Goksu}, {Gonz{\'a}lez}, {Goodman}, {Groetsch}, {Harding}, {Hern{\'a}ndez-Cadena}, {Herzog}, {H{\"u}ntemeyer}, {Huang}, {Hueyotl-Zahuantitla}, {Iriarte}, {Joshi}, {Kaufmann}, {Kieda}, {Lara}, {Lee}, {Lee}, {Le{\'o}n Vargas}, {Linnemann}, {Longinotti}, {Luis-Raya}, {Malone}, {Martinez}, {Mart{\'\i}nez-Castro}, {Matthews}, {Miranda-Romagnoli}, {Montes}, {Morales-Soto}, {Moreno}, {Mostaf{\'a}}, {Nayerhoda}, {Nellen}, {Noriega-Papaqui}, {Olivera-Nieto}, {Omodei}, {P{\'e}rez Araujo}, {P{\'e}rez-P{\'e}rez}, {Rho}, {Rosa-Gonz{\'a}lez},
  {Salazar}, {Salazar-Gallegos}, {Sandoval}, {Schneider}, {Schwefer}, {Serna-Franco}, {Son}, {Springer}, {Tibolla}, {Tollefson}, {Torres}, {Torres-Escobedo}, {Turner}, {Urea-Mena}, {Varela}, {Villase{\~n}or}, {Wang}, {Watson}, {Willox}, {Wu}, {Yun-C{\'a}rcamo}, {Zhou}, {de Le{\'o}n}, \& {Di Mauro}}]{geminga_hawc_new}
{Albert}, A., {Alfaro}, R., {Alvarez}, C., {et~al.} 2024{\natexlab{b}}, \apj, 974, 246

\bibitem[{{Albert} {et~al.}(2023{\natexlab{a}}){Albert}, {Alfaro}, {Alvarez}, {Arteaga-Vel{\'a}zquez}, {Avila Rojas}, {Ayala Solares}, {Babu}, {Belmont-Moreno}, {Brisbois}, {Caballero-Mora}, {Capistr{\'a}n}, {Carrami{\~n}ana}, {Casanova}, {Chaparro-Amaro}, {Cotti}, {Cotzomi}, {Couti{\~n}o de Le{\'o}n}, {De la Fuente}, {de Le{\'o}n}, {Diaz Hernandez}, {D{\'\i}az-V{\'e}lez}, {Dingus}, {DuVernois}, {Durocher}, {Engel}, {Espinoza}, {Fan}, {Fern{\'a}ndez Alonso}, {Fraija}, {Garc{\'\i}a-Gonz{\'a}lez}, {Garfias}, {Goksu}, {Gonz{\'a}lez}, {Goodman}, {Harding}, {Hernandez}, {Hinton}, {Hona}, {Huang}, {Hueyotl-Zahuantitla}, {H{\"u}ntemeyer}, {Iriarte}, {Jardin-Blicq}, {Joshi}, {Kaufmann}, {Kieda}, {Lee}, {Vargas}, {Linnemann}, {Longinotti}, {Luis-Raya}, {L{\'o}pez-Coto}, {Malone}, {Marandon}, {Martinez}, {Mart{\'\i}nez-Castro}, {Matthews}, {Miranda-Romagnoli}, {Morales-Soto}, {Moreno}, {Mostaf{\'a}}, {Nayerhoda}, {Nellen}, {Newbold}, {Nisa}, {Noriega-Papaqui}, {Olivera-Nieto}, {Omodei}, {Peisker}, {P{\'e}rez Araujo},
  {P{\'e}rez-P{\'e}rez}, {Rho}, {Rosa-Gonz{\'a}lez}, {Ruiz-Velasco}, {Salazar}, {Salazar-Gallegos}, {Greus}, {Sandoval}, {Schneider}, {Serna-Franco}, {Smith}, {Son}, {Springer}, {Tibolla}, {Tollefson}, {Torres}, {Torres-Escobedo}, {Turner}, {Ure{\~n}a-Mena}, {Villase{\~n}or}, {Wang}, {Werner}, {Willox}, {Zhou}, \& {HAWC Collaboration}}]{hawc_source2}
{Albert}, A., {Alfaro}, R., {Alvarez}, C., {et~al.} 2023{\natexlab{a}}, \apj, 942, 96

\bibitem[{{Albert} {et~al.}(2020){Albert}, {Alfaro}, {Alvarez}, {Camacho}, {Arteaga-Vel{\'a}zquez}, {Arunbabu}, {Avila Rojas}, {Ayala Solares}, {Baghmanyan}, {Belmont-Moreno}, {BenZvi}, {Brisbois}, {Caballero-Mora}, {Capistr{\'a}n}, {Carrami{\~n}ana}, {Casanova}, {Cotti}, {Couti{\~n}o de Le{\'o}n}, {De la Fuente}, {Diaz Hernandez}, {Diaz-Cruz}, {Dingus}, {DuVernois}, {Durocher}, {D{\'\i}az-V{\'e}lez}, {Ellsworth}, {Engel}, {Espinoza}, {Fan}, {Fang}, {Alonso}, {Fleischhack}, {Fraija}, {Galv{\'a}n-G{\'a}mez}, {Garcia}, {Garc{\'\i}a-Gonz{\'a}lez}, {Garfias}, {Giacinti}, {Gonz{\'a}lez}, {Goodman}, {Harding}, {Hernandez}, {Hinton}, {Hona}, {Huang}, {Hueyotl-Zahuantitla}, {H{\"u}ntemeyer}, {Iriarte}, {Jardin-Blicq}, {Joshi}, {Kieda}, {Lara}, {Lee}, {Le{\'o}n Vargas}, {Linnemann}, {Longinotti}, {Luis-Raya}, {Lundeen}, {L{\'o}pez-Coto}, {Malone}, {Marandon}, {Martinez}, {Martinez-Castellanos}, {Mart{\'\i}nez-Castro}, {Matthews}, {Miranda-Romagnoli}, {Morales-Soto}, {Moreno}, {Mostaf{\'a}}, {Nayerhoda}, {Nellen},
  {Newbold}, {Nisa}, {Noriega-Papaqui}, {Olivera-Nieto}, {Omodei}, {Peisker}, {P{\'e}rez Araujo}, {P{\'e}rez-P{\'e}rez}, {Ren}, {Rho}, {Rivi{\`e}re}, {Rosa-Gonz{\'a}lez}, {Ruiz-Velasco}, {Salazar}, {Salesa Greus}, {Sandoval}, {Schneider}, {Schoorlemmer}, {Serna}, {Sinnis}, {Smith}, {Springer}, {Surajbali}, {Tollefson}, {Torres}, {Torres-Escobedo}, {Ukwatta}, {Ure{\~n}a-Mena}, {Weisgarber}, {Werner}, {Willox}, {Zepeda}, {Zhou}, {de Le{\'o}n}, {{\'A}lvarez}, \& {HAWC Collaboration}}]{3hwc}
{Albert}, A., {Alfaro}, R., {Alvarez}, C., {et~al.} 2020, \apj, 905, 76

\bibitem[{{Albert} {et~al.}(2021){Albert}, {Alfaro}, {Alvarez}, {Camacho}, {Arteaga-Vel{\'a}zquez}, {Arunbabu}, {Rojas}, {Ayala Solares}, {Baghmanyan}, {Belmont-Moreno}, {BenZvi}, {Brisbois}, {Capistr{\'a}n}, {Carrami{\~n}ana}, {Casanova}, {Cotti}, {Cotzomi}, {Fuente}, {Hernandez}, {Dingus}, {DuVernois}, {Durocher}, {D{\'\i}az-V{\'e}lez}, {Engel}, {Espinoza}, {Fang}, {Fleischhack}, {Fraija}, {Galv{\'a}n-G{\'a}mez}, {Garcia}, {Garc{\'\i}a-Gonz{\'a}lez}, {Garfias}, {Giacinti}, {Gonz{\'a}lez}, {Goodman}, {Harding}, {Hona}, {Huang}, {Hueyotl-Zahuantitla}, {H{\"u}ntemeyer}, {Iriarte}, {Jardin-Blicq}, {Joshi}, {Kunde}, {Lara}, {Lee}, {Vargas}, {Linnemann}, {Longinotti}, {Luis-Raya}, {Lundeen}, {Malone}, {Marandon}, {Martinez}, {Mart{\'\i}nez-Castro}, {Matthews}, {Miranda-Romagnoli}, {Moreno}, {Mostaf{\'a}}, {Nayerhoda}, {Nellen}, {Newbold}, {Nisa}, {Noriega-Papaqui}, {Omodei}, {Peisker}, {Araujo}, {P{\'e}rez-P{\'e}rez}, {Rho}, {Rosa-Gonz{\'a}lez}, {Salazar}, {Greus}, {Sandoval}, {Schneider}, {Serna}, {Springer},
  {Tollefson}, {Torres}, {Torres-Escobedo}, {Ure{\~n}a-Mena}, {Villase{\~n}or}, {Willox}, {Zhou}, \& {Le{\'o}n}}]{J1826_HAWC}
{Albert}, A., {Alfaro}, R., {Alvarez}, C., {et~al.} 2021, \apjl, 907, L30

\bibitem[{{Albert} {et~al.}(2023{\natexlab{b}}){Albert}, {Alfaro}, {Arteaga-Vel{\'a}zquez}, {Ayala Solares}, {Belmont-Moreno}, {Capistr{\'a}n}, {Carrami{\~n}ana}, {Casanova}, {Cotzomi}, {Couti{\~n}o De Le{\'o}n}, {De la Fuente}, {de Le{\'o}n}, {Diaz Hernandez}, {DuVernois}, {D{\'\i}az-V{\'e}lez}, {Espinoza}, {Fan}, {Fraija}, {Fang}, {Garc{\'\i}a-Gonz{\'a}lez}, {Garfias}, {Jardin-Blicq}, {Gonz{\'a}lez}, {Goodman}, {Harding}, {Hernandez}, {Huang}, {Hueyotl-Zahuantitla}, {H{\"u}ntemeyer}, {Iriarte}, {Joshi}, {Lara}, {Lee}, {Le{\'o}n Vargas}, {Linnemann}, {Longinotti}, {Luis-Raya}, {Malone}, {Martinez}, {Mart{\'\i}nez-Castro}, {Matthews}, {Morales-Soto}, {Moreno}, {Mostaf{\'a}}, {Nayerhoda}, {Nellen}, {Newbold}, {Nisa}, {P{\'e}rez Araujo}, {Son}, {P{\'e}rez-P{\'e}rez}, {Rho}, {Rosa-Gonz{\'a}lez}, {Sandoval}, {Schneider}, {Serna-Franco}, {Smith}, {Springer}, {Tollefson}, {Torres}, {Torres-Escobedo}, {Wang}, {Whitaker}, {Willox}, {Zhou}, \& {HAWC Collaboration}}]{hawc_source1}
{Albert}, A., {Alfaro}, R., {Arteaga-Vel{\'a}zquez}, J.~C., {et~al.} 2023{\natexlab{b}}, \apjl, 944, L29

\bibitem[{{Albert} {et~al.}(2006){Albert}, {Aliu}, {Anderhub}, {Antoranz}, {Armada}, {Asensio}, {Baixeras}, {Barrio}, {Bartel}, {Bartko}, {Bastieri}, {Bavikadi}, {Bednarek}, {Berger}, {Bigongiari}, {Biland}, {Bisesi}, {Blanch}, {Bock}, {Bretz}, {Britvitch}, {Camara}, {Chilingarian}, {Ciprini}, {Coarasa}, {Commichau}, {Contreras}, {Cortina}, {Curtev}, {Danielyan}, {Dazzi}, {De Angelis}, {de los Reyes}, {De Lotto}, {Domingo-Santamaria}, {Dorner}, {Doro}, {Errando}, {Fagiolini}, {Ferenc}, {Fern{\'a}ndez}, {Firpo}, {Flix}, {Fonseca}, {Font}, {Galante}, {Garczarczyk}, {Gaug}, {Gebauer}, {Giller}, {Goebel}, {Hakobyan}, {Hayashida}, {Hengstebeck}, {H{\"o}hne}, {Hose}, {Jacon}, {Kalekin}, {Kranich}, {Laille}, {Lenisa}, {Liebing}, {Lindfors}, {Longo}, {L{\'o}pez}, {L{\'o}pez}, {Lorenz}, {Lucarelli}, {Majumdar}, {Maneva}, {Mannheim}, {Mariotti}, {Mart{\'\i}nez}, {Mase}, {Mazin}, {Merck}, {Merck}, {Meucci}, {Meyer}, {Miranda}, {Mirzoyan}, {Mizobuchi}, {Moralejo}, {Nilsson}, {O{\~n}a-Wilhelmi}, {Ordu{\~n}a}, {Otte},
  {Oya}, {Paneque}, {Paoletti}, {Pasanen}, {Pascoli}, {Pauss}, {Pavel}, {Pegna}, {Peruzzo}, {Piccioli}, {Prandini}, {Rico}, {Rhode}, {Riegel}, {Rissi}, {Robert}, {Rossato}, {R{\"u}gamer}, {Saggion}, {Sanchez}, {Sartori}, {Scalzotto}, {Schmitt}, {Schweizer}, {Shayduk}, {Shinozaki}, {Shore}, {Sidro}, {Sillanp{\"a}{\"a}}, {Sobczynska}, {Stamerra}, {Stark}, {Takalo}, {Temnikov}, {Tescaro}, {Teshima}, {Tonello}, {Torres}, {Torres}, {Turini}, {Vankov}, {Vitale}, {Wagner}, {Wibig}, {Wittek}, \& {Zapatero}}]{magic3}
{Albert}, J., {Aliu}, E., {Anderhub}, H., {et~al.} 2006, \apjl, 637, L41

\bibitem[{{Aleksi{\'c}} {et~al.}(2014){Aleksi{\'c}}, {Ansoldi}, {Antonelli}, {Antoranz}, {Babic}, {Bangale}, {Barrio}, {Becerra Gonz{\'a}lez}, {Bednarek}, {Bernardini}, {Biasuzzi}, {Biland}, {Blanch}, {Bonnefoy}, {Bonnoli}, {Borracci}, {Bretz}, {Carmona}, {Carosi}, {Colin}, {Colombo}, {Contreras}, {Cortina}, {Covino}, {Da Vela}, {Dazzi}, {De Angelis}, {De Caneva}, {De Lotto}, {de O{\~n}a Wilhelmi}, {Delgado Mendez}, {Dominis Prester}, {Dorner}, {Doro}, {Einecke}, {Eisenacher}, {Elsaesser}, {Fonseca}, {Font}, {Frantzen}, {Fruck}, {Galindo}, {Garc{\'\i}a L{\'o}pez}, {Garczarczyk}, {Garrido Terrats}, {Gaug}, {Godinovi{\'c}}, {Gonz{\'a}lez Mu{\~n}oz}, {Gozzini}, {Hadasch}, {Hanabata}, {Hayashida}, {Herrera}, {Hildebrand}, {Hose}, {Hrupec}, {Idec}, {Kadenius}, {Kellermann}, {Kodani}, {Konno}, {Krause}, {Kubo}, {Kushida}, {La Barbera}, {Lelas}, {Lewandowska}, {Lindfors}, {Lombardi}, {L{\'o}pez}, {L{\'o}pez-Coto}, {L{\'o}pez-Oramas}, {Lorenz}, {Lozano}, {Makariev}, {Mallot}, {Maneva}, {Mankuzhiyil}, {Mannheim},
  {Maraschi}, {Marcote}, {Mariotti}, {Mart{\'\i}nez}, {Mazin}, {Menzel}, {Miranda}, {Mirzoyan}, {Moralejo}, {Munar-Adrover}, {Nakajima}, {Niedzwiecki}, {Nilsson}, {Nishijima}, {Noda}, {Orito}, {Overkemping}, {Paiano}, {Palatiello}, {Paneque}, {Paoletti}, {Paredes}, {Paredes-Fortuny}, {Persic}, {Prada Moroni}, {Prandini}, {Puljak}, {Reinthal}, {Rhode}, {Rib{\'o}}, {Rico}, {Rodriguez Garcia}, {R{\"u}gamer}, {Saito}, {Saito}, {Satalecka}, {Scalzotto}, {Scapin}, {Schultz}, {Schweizer}, {Shore}, {Sillanp{\"a}{\"a}}, {Sitarek}, {Snidaric}, {Sobczynska}, {Spanier}, {Stamatescu}, {Stamerra}, {Steinbring}, {Storz}, {Strzys}, {Takalo}, {Takami}, {Tavecchio}, {Temnikov}, {Terzi{\'c}}, {Tescaro}, {Teshima}, {Thaele}, {Tibolla}, {Torres}, {Toyama}, {Treves}, {Uellenbeck}, {Vogler}, \& {Zanin}}]{magic_3C58}
{Aleksi{\'c}}, J., {Ansoldi}, S., {Antonelli}, L.~A., {et~al.} 2014, \aap, 567, L8

\bibitem[{Aleksić {et~al.}(2015)Aleksić, Ansoldi, Antonelli, Antoranz, Babic, Bangale, Barrio, {Becerra González}, Bednarek, Bernardini, Biasuzzi, Biland, Blanch, Bonnefoy, Bonnoli, Borracci, Bretz, Carmona, Carosi, Colin, Colombo, Contreras, Cortina, Covino, {Da Vela}, Dazzi, {De Angelis}, {De Caneva}, {De Lotto}, {de Oña Wilhelmi}, {Delgado Mendez}, Doert, {Dominis Prester}, Dorner, Doro, Einecke, Eisenacher, Elsaesser, Fonseca, Font, Frantzen, Fruck, Galindo, {García López}, Garczarczyk, {Garrido Terrats}, Gaug, Godinović, {González Muñoz}, Gozzini, Hadasch, Hanabata, Hayashida, Herrera, Hildebrand, Hose, Hrupec, Idec, Kadenius, Kellermann, Kodani, Konno, Krause, Kubo, Kushida, {La Barbera}, Lelas, Lewandowska, Lindfors, Lombardi, López, López-Coto, López-Oramas, Lorenz, Lozano, Makariev, Mallot, Maneva, Mankuzhiyil, Mannheim, Maraschi, Marcote, Mariotti, Martínez, Mazin, Menzel, Miranda, Mirzoyan, Moralejo, Munar-Adrover, Nakajima, Niedzwiecki, Nilsson, Nishijima, Noda, Nowak, Orito,
  Overkemping, Paiano, Palatiello, Paneque, Paoletti, Paredes, Paredes-Fortuny, Persic, {Prada Moroni}, Prandini, Preziuso, Puljak, Reinthal, Rhode, Ribó, Rico, {Rodriguez Garcia}, Rügamer, Saggion, Saito, Saito, Satalecka, Scalzotto, Scapin, Schultz, Schweizer, Shore, Sillanpää, Sitarek, Snidaric, Sobczynska, Spanier, Stamatescu, Stamerra, Steinbring, Storz, Strzys, Takalo, Takami, Tavecchio, Temnikov, Terzić, Tescaro, Teshima, Thaele, Tibolla, Torres, Toyama, Treves, Uellenbeck, Vogler, Wagner, Zanin, Horns, Martín, \& Meyer}]{magic_crab}
Aleksić, J., Ansoldi, S., Antonelli, L., {et~al.} 2015, Journal of High Energy Astrophysics, 5-6, 30

\bibitem[{{Aliu} {et~al.}(2014{\natexlab{a}}){Aliu}, {Archambault}, {Aune}, {Behera}, {Beilicke}, {Benbow}, {Berger}, {Bird}, {Buckley}, {Bugaev}, {Cardenzana}, {Cerruti}, {Chen}, {Ciupik}, {Collins-Hughes}, {Connolly}, {Cui}, {Dumm}, {Dwarkadas}, {Errando}, {Falcone}, {Federici}, {Feng}, {Finley}, {Fleischhack}, {Fortin}, {Fortson}, {Furniss}, {Galante}, {Gall}, {Gillanders}, {Griffin}, {Griffiths}, {Grube}, {Gyuk}, {Hanna}, {Holder}, {Hughes}, {Humensky}, {Kaaret}, {Kertzman}, {Khassen}, {Kieda}, {Krennrich}, {Kumar}, {Lang}, {Madhavan}, {Maier}, {McCann}, {Meagher}, {Millis}, {Moriarty}, {Mukherjee}, {Nieto}, {O'Faol{\'a}in de Bhr{\'o}ithe}, {Ong}, {Otte}, {Pandel}, {Park}, {Pohl}, {Popkow}, {Prokoph}, {Quinn}, {Ragan}, {Rajotte}, {Ratliff}, {Reyes}, {Reynolds}, {Richards}, {Roache}, {Rousselle}, {Sembroski}, {Shahinyan}, {Sheidaei}, {Smith}, {Staszak}, {Telezhinsky}, {Tsurusaki}, {Tucci}, {Tyler}, {Varlotta}, {Vassiliev}, {Vincent}, {Wakely}, {Ward}, {Weinstein}, {Welsing}, \& {Wilhelm}}]{vertias_source4}
{Aliu}, E., {Archambault}, S., {Aune}, T., {et~al.} 2014{\natexlab{a}}, \apj, 787, 166

\bibitem[{{Aliu} {et~al.}(2014{\natexlab{b}}){Aliu}, {Aune}, {Behera}, {Beilicke}, {Benbow}, {Berger}, {Bird}, {Bouvier}, {Buckley}, {Bugaev}, {Cerruti}, {Chen}, {Ciupik}, {Connolly}, {Cui}, {Dumm}, {Dwarkadas}, {Errando}, {Falcone}, {Federici}, {Feng}, {Finley}, {Fleischhack}, {Fortin}, {Fortson}, {Furniss}, {Galante}, {Gillanders}, {Gotthelf}, {Griffin}, {Griffiths}, {Grube}, {Gyuk}, {Hanna}, {Holder}, {Hughes}, {Humensky}, {Johnson}, {Kaaret}, {Kargaltsev}, {Kertzman}, {Khassen}, {Kieda}, {Krennrich}, {Lang}, {Madhavan}, {Maier}, {McArthur}, {McCann}, {Millis}, {Moriarty}, {Mukherjee}, {Nieto}, {O'Faol{\'a}in de Bhr{\'o}ithe}, {Ong}, {Otte}, {Pandel}, {Park}, {Pohl}, {Popkow}, {Prokoph}, {Quinn}, {Ragan}, {Rajotte}, {Reyes}, {Reynolds}, {Richards}, {Roache}, {Roberts}, {Sembroski}, {Shahinyan}, {Smith}, {Staszak}, {Telezhinsky}, {Tucci}, {Tyler}, {Vincent}, {Wakely}, {Weinstein}, {Welsing}, {Wilhelm}, {Williams}, \& {Zitzer}}]{veritas8}
{Aliu}, E., {Aune}, T., {Behera}, B., {et~al.} 2014{\natexlab{b}}, \apj, 788, 78

\bibitem[{{Ansoldi, S.} {et~al.}(2016){Ansoldi, S.}, {Antonelli, L. A.}, {Antoranz, P.}, {Babic, A.}, {Bangale, P.}, {Barres de Almeida, U.}, {Barrio, J. A.}, {Becerra González, J.}, {Bednarek, W.}, {Bernardini, E.}, {Biasuzzi, B.}, {Biland, A.}, {Blanch, O.}, {Bonnefoy, S.}, {Bonnoli, G.}, {Borracci, F.}, {Bretz, T.}, {Carmona, E.}, {Carosi, A.}, {Colin, P.}, {Colombo, E.}, {Contreras, J. L.}, {Cortina, J.}, {Covino, S.}, {Da Vela, P.}, {Dazzi, F.}, {De Angelis, A.}, {De Caneva, G.}, {De Lotto, B.}, {de Oña Wilhelmi, E.}, {Delgado Mendez, C.}, {Di Pierro, F.}, {Dominis Prester, D.}, {Dorner, D.}, {Doro, M.}, {Einecke, S.}, {Eisenacher Glawion, D.}, {Elsaesser, D.}, {Fernández-Barral, A.}, {Fidalgo, D.}, {Fonseca, M. V.}, {Font, L.}, {Frantzen, K.}, {Fruck, C.}, {Galindo, D.}, {García López, R. J.}, {Garczarczyk, M.}, {Garrido Terrats, D.}, {Gaug, M.}, {Godinović, N.}, {González Muñoz, A.}, {Gozzini, S. R.}, {Hanabata, Y.}, {Hayashida, M.}, {Herrera, J.}, {Hirotani, K.}, {Hose, J.}, {Hrupec, D.},
  {Hughes, G.}, {Idec, W.}, {Kellermann, H.}, {Knoetig, M. L.}, {Kodani, K.}, {Konno, Y.}, {Krause, J.}, {Kubo, H.}, {Kushida, J.}, {La Barbera, A.}, {Lelas, D.}, {Lewandowska, N.}, {Lindfors, E.}, {Lombardi, S.}, {Longo, F.}, {López, M.}, {López-Coto, R.}, {López-Oramas, A.}, {Lorenz, E.}, {Makariev, M.}, {Mallot, K.}, {Maneva, G.}, {Mannheim, K.}, {Maraschi, L.}, {Marcote, B.}, {Mariotti, M.}, {Martínez, M.}, {Mazin, D.}, {Menzel, U.}, {Miranda, J. M.}, {Mirzoyan, R.}, {Moralejo, A.}, {Munar-Adrover, P.}, {Nakajima, D.}, {Neustroev, V.}, {Niedzwiecki, A.}, {Nevas Rosillo, M.}, {Nilsson, K.}, {Nishijima, K.}, {Noda, K.}, {Orito, R.}, {Overkemping, A.}, {Paiano, S.}, {Palatiello, M.}, {Paneque, D.}, {Paoletti, R.}, {Paredes, J. M.}, {Paredes-Fortuny, X.}, {Persic, M.}, {Poutanen, J.}, {Prada Moroni, P. G.}, {Prandini, E.}, {Puljak, I.}, {Reinthal, R.}, {Rhode, W.}, {Ribó, M.}, {Rico, J.}, {Rodriguez Garcia, J.}, {Saito, T.}, {Saito, K.}, {Satalecka, K.}, {Scalzotto, V.}, {Scapin, V.}, {Schultz, C.},
  {Schweizer, T.}, {Shore, S. N.}, {Sillanpää, A.}, {Sitarek, J.}, {Snidaric, I.}, {Sobczynska, D.}, {Stamerra, A.}, {Steinbring, T.}, {Strzys, M.}, {Takalo, L.}, {Takami, H.}, {Tavecchio, F.}, {Temnikov, P.}, {Terzić, T.}, {Tescaro, D.}, {Teshima, M.}, {Thaele, J.}, {Torres, D. F.}, {Toyama, T.}, {Treves, A.}, {Ward, J.}, {Will, M.}, \& {Zanin, R.}}]{magiccrab}
{Ansoldi, S.}, {Antonelli, L. A.}, {Antoranz, P.}, {et~al.} 2016, aap, 585, A133

\bibitem[{{Archer} {et~al.}(2016){Archer}, {Benbow}, {Bird}, {Buchovecky}, {Buckley}, {Bugaev}, {Byrum}, {Cardenzana}, {Cerruti}, {Chen}, {Ciupik}, {Collins-Hughes}, {Connolly}, {Eisch}, {Falcone}, {Feng}, {Finley}, {Fleischhack}, {Flinders}, {Fortson}, {Furniss}, {Gillanders}, {Griffin}, {Grube}, {Gyuk}, {H{\r{a}}kansson}, {Hanna}, {Holder}, {Humensky}, {H{\"u}tten}, {Johnson}, {Kaaret}, {Kar}, {Kelley-Hoskins}, {Kertzman}, {Kieda}, {Krause}, {Krennrich}, {Kumar}, {Lang}, {McArthur}, {McCann}, {Meagher}, {Millis}, {Moriarty}, {Mukherjee}, {Nieto}, {Ong}, {Park}, {Pelassa}, {Pohl}, {Popkow}, {Pueschel}, {Quinn}, {Ragan}, {Ratliff}, {Reynolds}, {Richards}, {Roache}, {Rousselle}, {Santander}, {Sembroski}, {Shahinyan}, {Smith}, {Staszak}, {Telezhinsky}, {Tucci}, {Tyler}, {Vassiliev}, {Wakely}, {Weiner}, {Weinstein}, {Wilhelm}, {Williams}, {Zitzer}, \& {Yusef-Zadeh}}]{veritas7}
{Archer}, A., {Benbow}, W., {Bird}, R., {et~al.} 2016, \apj, 821, 129

\bibitem[{{Bandiera} {et~al.}(2023){Bandiera}, {Bucciantini}, {Olmi}, \& {Torres}}]{pwn_reverb}
{Bandiera}, R., {Bucciantini}, N., {Olmi}, B., \& {Torres}, D.~F. 2023, \mnras, 525, 2839

\bibitem[{{Burrows} {et~al.}(2005){Burrows}, {Hill}, {Nousek}, {Kennea}, {Wells}, {Osborne}, {Abbey}, {Beardmore}, {Mukerjee}, {Short}, {Chincarini}, {Campana}, {Citterio}, {Moretti}, {Pagani}, {Tagliaferri}, {Giommi}, {Capalbi}, {Tamburelli}, {Angelini}, {Cusumano}, {Br{\"a}uninger}, {Burkert}, \& {Hartner}}]{swift}
{Burrows}, D.~N., {Hill}, J.~E., {Nousek}, J.~A., {et~al.} 2005, \ssr, 120, 165

\bibitem[{{Camilo} {et~al.}(2004){Camilo}, {Gaensler}, {Gotthelf}, {Halpern}, \& {Manchester}}]{hess_source7_xray}
{Camilo}, F., {Gaensler}, B.~M., {Gotthelf}, E.~V., {Halpern}, J.~P., \& {Manchester}, R.~N. 2004, \apj, 616, 1118

\bibitem[{{Camilo} {et~al.}(2002){Camilo}, {Lorimer}, {Bhat}, {Gotthelf}, {Halpern}, {Wang}, {Lu}, \& {Mirabal}}]{veritas5_mw}
{Camilo}, F., {Lorimer}, D.~R., {Bhat}, N.~D.~R., {et~al.} 2002, \apjl, 574, L71

\bibitem[{Cao {et~al.}(2023)Cao, Aharonian, An, Axikegu, Bai, Bao, Bastieri, Bi, Bi, Cai, Cao, Cao, Cao, Chang, Chang, Chen, Chen, Chen, Chen, Chen, Chen, Chen, Chen, Chen, Chen, Chen, Chen, Cheng, Cheng, Cui, Cui, Cui, Cui, Dai, Dai, Dai, Danzengluobu, della Volpe, Dong, Duan, Fan, Fan, Fang, Fang, Feng, Feng, Feng, Feng, Feng, Gabici, Gao, Gao, Gao, Gao, Gao, Gao, Ge, Geng, Giacinti, Gong, Gou, Gu, Guo, Guo, Guo, Guo, Han, He, He, He, He, He, Heller, Hor, Hou, Hou, Hou, Hu, Hu, Hu, Huang, Huang, Huang, Huang, Huang, Huang, Huang, Ji, Jia, Jia, Jiang, Jiang, Jiang, Jin, Kang, Ke, Kuleshov, Kurinov, Li, Li, Li, Li, Li, Li, Li, Li, Li, Li, Li, Li, Li, Li, Li, Li, Li, Li, Li, Liang, Liang, Lin, Liu, Liu, Liu, Liu, Liu, Liu, Liu, Liu, Liu, Liu, Liu, Liu, Liu, Liu, Lu, Luo, Lv, Ma, Ma, Ma, Mao, Min, Mitthumsiri, Mu, Nan, Neronov, Ou, Pang, Pattarakijwanich, Pei, Qi, Qi, Qiao, Qin, Ruffolo, Sáiz, Semikoz, Shao, Shao, Shchegolev, Sheng, Shu, Song, Stenkin, Stepanov, Su, Sun, Sun, Sun, Tam, Tang, Tang, Tian, Wang,
  Wang, Wang, Wang, Wang, Wang, Wang, Wang, Wang, Wang, Wang, Wang, Wang, Wang, Wang, Wang, Wang, Wang, Wang, Wang, Wang, Wei, Wei, Wei, Wen, Wu, Wu, Wu, Wu, Wu, Xi, Xia, Xia, Xiang, Xiao, Xiao, Xin, Xin, Xing, Xiong, Xu, Xu, Xu, Xu, Xue, Yan, Yan, Yan, Yang, Yang, Yang, Yang, Yang, Yang, Yang, Yang, Yang, Yao, Yao, Ye, Yin, Yin, You, You, Yu, Yuan, Yue, Zeng, Zeng, Zeng, Zha, Zhang, Zhang, Zhang, Zhang, Zhang, Zhang, Zhang, Zhang, Zhang, Zhang, Zhang, Zhang, Zhang, Zhang, Zhang, Zhang, Zhang, Zhang, Zhao, Zhao, Zhao, Zhao, Zhao, Zheng, Zhou, Zhou, Zhou, Zhou, Zhou, Zhou, Zhou, Zhu, Zhu, Zhu, Zhu, \& Zuo.}]{cao2023lhaaso}
Cao, Z., Aharonian, F., An, Q., {et~al.} 2023, The First LHAASO Catalog of Gamma-Ray Sources

\bibitem[{{Cao} {et~al.}(2024){Cao}, {Aharonian}, {An}, {Axikegu}, {Bai}, {Bao}, {Bastieri}, {Bi}, {Bi}, {Cai}, {Cao}, {Cao}, {Cao}, {Chang}, {Chang}, {Chen}, {Chen}, {Chen}, {Chen}, {Chen}, {Chen}, {Chen}, {Chen}, {Chen}, {Chen}, {Chen}, {Chen}, {Cheng}, {Cheng}, {Cui}, {Cui}, {Cui}, {Cui}, {Dai}, {Dai}, {Dai}, {Danzengluobu}, {Della Volpe}, {Dong}, {Duan}, {Fan}, {Fan}, {Fang}, {Fang}, {Feng}, {Feng}, {Feng}, {Feng}, {Feng}, {Gabici}, {Gao}, {Gao}, {Gao}, {Gao}, {Gao}, {Gao}, {Ge}, {Geng}, {Giacinti}, {Gong}, {Gou}, {Gu}, {Guo}, {Guo}, {Guo}, {Guo}, {Han}, {He}, {He}, {He}, {He}, {He}, {Heller}, {Hor}, {Hou}, {Hou}, {Hou}, {Hu}, {Hu}, {Hu}, {Huang}, {Huang}, {Huang}, {Huang}, {Huang}, {Huang}, {Huang}, {Ji}, {Jia}, {Jia}, {Jiang}, {Jiang}, {Jiang}, {Jin}, {Kang}, {Ke}, {Kuleshov}, {Kurinov}, {Li}, {Li}, {Li}, {Li}, {Li}, {Li}, {Li}, {Li}, {Li}, {Li}, {Li}, {Li}, {Li}, {Li}, {Li}, {Li}, {Li}, {Li}, {Li}, {Liang}, {Liang}, {Lin}, {Liu}, {Liu}, {Liu}, {Liu}, {Liu}, {Liu}, {Liu}, {Liu}, {Liu}, {Liu}, {Liu},
  {Liu}, {Liu}, {Liu}, {Lu}, {Luo}, {Lv}, {Ma}, {Ma}, {Ma}, {Mao}, {Min}, {Mitthumsiri}, {Mu}, {Nan}, {Neronov}, {Ou}, {Pang}, {Pattarakijwanich}, {Pei}, {Qi}, {Qi}, {Qiao}, {Qin}, {Ruffolo}, {S{\'a}iz}, {Semikoz}, {Shao}, {Shao}, {Shchegolev}, {Sheng}, {Shu}, {Song}, {Stenkin}, {Stepanov}, {Su}, {Sun}, {Sun}, {Sun}, {Tam}, {Tang}, {Tang}, {Tian}, {Wang}, {Wang}, {Wang}, {Wang}, {Wang}, {Wang}, {Wang}, {Wang}, {Wang}, {Wang}, {Wang}, {Wang}, {Wang}, {Wang}, {Wang}, {Wang}, {Wang}, {Wang}, {Wang}, {Wang}, {Wang}, {Wei}, {Wei}, {Wei}, {Wen}, {Wu}, \& {Wu}}]{1LHAASO}
{Cao}, Z., {Aharonian}, F., {An}, Q., {et~al.} 2024, \apjs, 271, 25

\bibitem[{{Chen} \& {Guestrin}(2016)}]{xgboost}
{Chen}, T. \& {Guestrin}, C. 2016, arXiv e-prints, arXiv:1603.02754

\bibitem[{{Cherenkov Telescope Array Observatory} \& Consortium(2021)}]{cta_sens}
{Cherenkov Telescope Array Observatory} \& Consortium, C. T.~A. 2021, CTAO Instrument Response Functions - prod5 version v0.1

\bibitem[{{Da Vela} {et~al.}(2018){Da Vela}, {Stamerra}, {Neronov}, {Prandini}, {Konno}, \& {Sitarek}}]{IACT_perf}
{Da Vela}, P., {Stamerra}, A., {Neronov}, A., {et~al.} 2018, Astroparticle Physics, 98, 1

\bibitem[{{Faucher-Gigu{\`e}re} \& {Kaspi}(2006)}]{Faucher}
{Faucher-Gigu{\`e}re}, C.-A. \& {Kaspi}, V.~M. 2006, \apj, 643, 332

\bibitem[{{Fornasini} {et~al.}(2014){Fornasini}, {Tomsick}, {Bodaghee}, {Krivonos}, {An}, {Rahoui}, {Gotthelf}, {Bauer}, \& {Stern}}]{nodet42}
{Fornasini}, F.~M., {Tomsick}, J.~A., {Bodaghee}, A., {et~al.} 2014, \apj, 796, 105

\bibitem[{{Gaensler} \& {Frail}(2000)}]{age1}
{Gaensler}, B.~M. \& {Frail}, D.~A. 2000, \nat, 406, 158

\bibitem[{{Gaensler} {et~al.}(2003){Gaensler}, {Schulz}, {Kaspi}, {Pivovaroff}, \& {Becker}}]{veritas_source2_xray1}
{Gaensler}, B.~M., {Schulz}, N.~S., {Kaspi}, V.~M., {Pivovaroff}, M.~J., \& {Becker}, W.~E. 2003, \apj, 588, 441

\bibitem[{{Giacinti} {et~al.}(2020){Giacinti}, {Mitchell}, {L{\'o}pez-Coto}, {Joshi}, {Parsons}, \& {Hinton}}]{halofrac}
{Giacinti}, G., {Mitchell}, A.~M.~W., {L{\'o}pez-Coto}, R., {et~al.} 2020, \aap, 636, A113

\bibitem[{{Gotthelf} \& {Halpern}(2008)}]{magic5_pulsar}
{Gotthelf}, E.~V. \& {Halpern}, J.~P. 2008, \apj, 681, 515

\bibitem[{{H.E.S.S. Collaboration} {et~al.}(2018){H.E.S.S. Collaboration}, {Abdalla}, {Abramowski}, {Aharonian}, {Ait Benkhali}, {Ang{\"u}ner}, {Arakawa}, {Arrieta}, {Aubert}, {Backes}, {Balzer}, {Barnard}, {Becherini}, {Becker Tjus}, {Berge}, {Bernhard}, {Bernl{\"o}hr}, {Blackwell}, {B{\"o}ttcher}, {Boisson}, {Bolmont}, {Bonnefoy}, {Bordas}, {Bregeon}, {Brun}, {Brun}, {Bryan}, {B{\"u}chele}, {Bulik}, {Capasso}, {Carrigan}, {Caroff}, {Carosi}, {Casanova}, {Cerruti}, {Chakraborty}, {Chaves}, {Chen}, {Chevalier}, {Colafrancesco}, {Condon}, {Conrad}, {Davids}, {Decock}, {Deil}, {Devin}, {deWilt}, {Dirson}, {Djannati-Ata{\"\i}}, {Domainko}, {Donath}, {Drury}, {Dutson}, {Dyks}, {Edwards}, {Egberts}, {Eger}, {Emery}, {Ernenwein}, {Eschbach}, {Farnier}, {Fegan}, {Fernandes}, {Fiasson}, {Fontaine}, {F{\"o}rster}, {Funk}, {F{\"u}{\ss}ling}, {Gabici}, {Gallant}, {Garrigoux}, {Gast}, {Gat{\'e}}, {Giavitto}, {Giebels}, {Glawion}, {Glicenstein}, {Gottschall}, {Grondin}, {Hahn}, {Haupt}, {Hawkes}, {Heinzelmann}, {Henri},
  {Hermann}, {Hinton}, {Hofmann}, {Hoischen}, {Holch}, {Holler}, {Horns}, {Ivascenko}, {Iwasaki}, {Jacholkowska}, {Jamrozy}, {Jankowsky}, {Jankowsky}, {Jingo}, {Jouvin}, {Jung-Richardt}, {Kastendieck}, {Katarzy{\'n}ski}, {Katsuragawa}, {Katz}, {Kerszberg}, {Khangulyan}, {Kh{\'e}lifi}, {King}, {Klepser}, {Klochkov}, {Klu{\'z}niak}, {Komin}, {Kosack}, {Krakau}, {Kraus}, {Kr{\"u}ger}, {Laffon}, {Lamanna}, {Lau}, {Lees}, {Lefaucheur}, {Lemi{\`e}re}, {Lemoine-Goumard}, {Lenain}, {Leser}, {Lohse}, {Lorentz}, {Liu}, {L{\'o}pez-Coto}, {Lypova}, {Marandon}, {Malyshev}, {Marcowith}, {Mariaud}, {Marx}, {Maurin}, {Maxted}, {Mayer}, {Meintjes}, {Meyer}, {Mitchell}, {Moderski}, {Mohamed}, {Mohrmann}, {Mor{\r{a}}}, {Moulin}, {Murach}, {Nakashima}, {de Naurois}, {Ndiyavala}, {Niederwanger}, {Niemiec}, {Oakes}, {O'Brien}, {Odaka}, {Ohm}, {Ostrowski}, {Oya}, {Padovani}, {Panter}, {Parsons}, {Paz Arribas}, {Pekeur}, {Pelletier}, {Perennes}, {Petrucci}, {Peyaud}, {Piel}, {Pita}, {Poireau}, {Poon}, {Prokhorov}, {Prokoph},
  {P{\"u}hlhofer}, {Punch}, {Quirrenbach}, {Raab}, {Rauth}, {Reimer}, {Reimer}, {Renaud}, {de los Reyes}, {Rieger}, {Rinchiuso}, {Romoli}, {Rowell}, {Rudak}, {Rulten}, {Safi-Harb}, {Sahakian}, {Saito}, {Sanchez}, {Santangelo}, {Sasaki}, {Schandri}, {Schlickeiser}, {Sch{\"u}ssler}, {Schulz}, {Schwanke}, \& {Schwemmer}}]{HGPS}
{H.E.S.S. Collaboration}, {Abdalla}, H., {Abramowski}, A., {et~al.} 2018, \aap, 612, A1

\bibitem[{{H.E.S.S. Collaboration} {et~al.}(2020){H.E.S.S. Collaboration}, {Abdalla}, {Adam}, {Aharonian}, {Ait Benkhali}, {Ang{\"u}ner}, {Arcaro}, {Armand}, {Armstrong}, {Ashkar}, {Backes}, {Baghmanyan}, {Barbosa Martins}, {Barnacka}, {Barnard}, {Becherini}, {Berge}, {Bernl{\"o}hr}, {Bi}, {B{\"o}ttcher}, {Boisson}, {Bolmont}, {de Bony de Lavergne}, {Bordas}, {Breuhaus}, {Brun}, {Brun}, {Bryan}, {B{\"u}chele}, {Bulik}, {Bylund}, {Caroff}, {Carosi}, {Casanova}, {Chand}, {Chandra}, {Chen}, {Cotter}, {Cury{\l}o}, {Damascene Mbarubucyeye}, {Davids}, {Davies}, {Deil}, {Devin}, {deWilt}, {Dirson}, {Djannati-Ata{\"\i}}, {Dmytriiev}, {Donath}, {Doroshenko}, {Duffy}, {Dyks}, {Egberts}, {Eichhorn}, {Einecke}, {Emery}, {Ernenwein}, {Feijen}, {Fegan}, {Fiasson}, {Fichet de Clairfontaine}, {Fontaine}, {Funk}, {F{\"u}{\ss}ling}, {Gabici}, {Gallant}, {Giavitto}, {Giunti}, {Glawion}, {Glicenstein}, {Gottschall}, {Grondin}, {Hahn}, {Haupt}, {Hermann}, {Hinton}, {Hofmann}, {Hoischen}, {Holch}, {Holler}, {H{\"o}rbe}, {Horns},
  {Huber}, {Jamrozy}, {Jankowsky}, {Jankowsky}, {Jardin-Blicq}, {Joshi}, {Jung-Richardt}, {Kasai}, {Kastendieck}, {Katarzy{\'n}ski}, {Katz}, {Khangulyan}, {Kh{\'e}lifi}, {Klepser}, {Klu{\'z}niak}, {Komin}, {Konno}, {Kosack}, {Kostunin}, {Kreter}, {Lamanna}, {Lemi{\`e}re}, {Lemoine-Goumard}, {Lenain}, {Levy}, {Lohse}, {Lypova}, {Mackey}, {Majumdar}, {Malyshev}, {Malyshev}, {Marandon}, {Marchegiani}, {Marcowith}, {Mares}, {Mart{\'\i}-Devesa}, {Marx}, {Maurin}, {Meintjes}, {Meyer}, {Mitchell}, {Moderski}, {Mohamed}, {Mohrmann}, {Montanari}, {Moore}, {Morris}, {Moulin}, {Muller}, {Murach}, {Nakashima}, {Nayerhoda}, {de Naurois}, {Ndiyavala}, {Niederwanger}, {Niemiec}, {Oakes}, {O'Brien}, {Odaka}, {Ohm}, {Olivera-Nieto}, {de Ona Wilhelmi}, {Ostrowski}, {Oya}, {Panter}, {Panny}, {Parsons}, {Peron}, {Peyaud}, {Piel}, {Pita}, {Poireau}, {Priyana Noel}, {Prokhorov}, {Prokoph}, {P{\"u}hlhofer}, {Punch}, {Quirrenbach}, {Raab}, {Rauth}, {Reichherzer}, {Reimer}, {Reimer}, {Remy}, {Renaud}, {Rieger}, {Rinchiuso}, {Romoli},
  {Rowell}, {Rudak}, {Ruiz-Velasco}, {Sahakian}, {Sailer}, {Sanchez}, {Santangelo}, {Sasaki}, {Scalici}, {Sch{\"u}ssler}, {Schutte}, {Schwanke}, {Schwemmer}, {Seglar-Arroyo}, {Senniappan}, {Seyffert}, {Shafi}, {Shiningayamwe}, {Simoni}, {Sinha}, {Sol}, {Specovius}, {Spencer}, {Spir-Jacob}, {Stawarz}, {Sun}, {Steenkamp}, {Stegmann}, {Steinmassl}, \& {Steppa}}]{hess_source5}
{H.E.S.S. Collaboration}, {Abdalla}, H., {Adam}, R., {et~al.} 2020, \aap, 644, A112

\bibitem[{{H.E.S.S. Collaboration} {et~al.}(2019){H.E.S.S. Collaboration}, {Abdalla}, {Aharonian}, {Ait Benkhali}, {Ang{\"u}ner}, {Arakawa}, {Arcaro}, {Armand}, {Arrieta}, {Backes}, {Barnard}, {Becherini}, {Becker Tjus}, {Berge}, {Bernl{\"o}hr}, {Blackwell}, {B{\"o}ttcher}, {Boisson}, {Bolmont}, {Bonnefoy}, {Bordas}, {Bregeon}, {Brun}, {Brun}, {Bryan}, {B{\"u}chele}, {Bulik}, {Bylund}, {Capasso}, {Caroff}, {Carosi}, {Casanova}, {Cerruti}, {Chakraborty}, {Chand}, {Chandra}, {Chaves}, {Chen}, {Colafrancesco}, {Condon}, {Davids}, {Deil}, {Devin}, {deWilt}, {Dirson}, {Djannati-Ata{\"\i}}, {Dmytriiev}, {Donath}, {Doroshenko}, {Drury}, {Dyks}, {Egberts}, {Emery}, {Ernenwein}, {Eschbach}, {Fegan}, {Fiasson}, {Fontaine}, {Funk}, {F{\"u}{\ss}ling}, {Gabici}, {Gallant}, {Gat{\'e}}, {Giavitto}, {Glawion}, {Glicenstein}, {Gottschall}, {Grondin}, {Hahn}, {Haupt}, {Heinzelmann}, {Henri}, {Hermann}, {Hinton}, {Hofmann}, {Hoischen}, {Holch}, {Holler}, {Horns}, {Huber}, {Iwasaki}, {Jacholkowska}, {Jamrozy}, {Jankowsky},
  {Jankowsky}, {Jouvin}, {Jung-Richardt}, {Kastendieck}, {Katarzy{\'n}ski}, {Katsuragawa}, {Katz}, {Kerszberg}, {Khangulyan}, {Kh{\'e}lifi}, {King}, {Klepser}, {Klu{\'z}niak}, {Komin}, {Kosack}, {Kraus}, {Lamanna}, {Lau}, {Lefaucheur}, {Lemi{\`e}re}, {Lemoine-Goumard}, {Lenain}, {Leser}, {Lohse}, {L{\'o}pez-Coto}, {Lypova}, {Malyshev}, {Marandon}, {Marcowith}, {Mariaud}, {Mart{\'\i}-Devesa}, {Marx}, {Maurin}, {Meintjes}, {Mitchell}, {Moderski}, {Mohamed}, {Mohrmann}, {Moore}, {Moulin}, {Murach}, {Nakashima}, {de Naurois}, {Ndiyavala}, {Niederwanger}, {Niemiec}, {Oakes}, {O'Brien}, {Odaka}, {Ohm}, {Ostrowski}, {Oya}, {Panter}, {Parsons}, {Perennes}, {Petrucci}, {Peyaud}, {Piel}, {Pita}, {Poireau}, {Priyana Noel}, {Prokhorov}, {Prokoph}, {P{\"u}hlhofer}, {Punch}, {Quirrenbach}, {Raab}, {Rauth}, {Reimer}, {Reimer}, {Renaud}, {Rieger}, {Rinchiuso}, {Romoli}, {Rowell}, {Rudak}, {Ruiz-Velasco}, {Sahakian}, {Saito}, {Sanchez}, {Santangelo}, {Sasaki}, {Schlickeiser}, {Sch{\"u}ssler}, {Schulz}, {Schutte}, {Schwanke},
  {Schwemmer}, {Seglar-Arroyo}, {Senniappan}, {Seyffert}, {Shafi}, {Shilon}, {Shiningayamwe}, {Simoni}, {Sinha}, {Sol}, {Specovius}, {Spir-Jacob}, {Stawarz}, {Steenkamp}, {Stegmann}, {Steppa}, {Takahashi}, {Tavernet}, {Tavernier}, {Taylor}, {Terrier}, {Tibaldo}, {Tiziani}, {Tluczykont}, {Trichard}, {Tsirou}, {Tsuji}, {Tuffs}, \& {Uchiyama}}]{hess_source6}
{H.E.S.S. Collaboration}, {Abdalla}, H., {Aharonian}, F., {et~al.} 2019, \aap, 621, A116

\bibitem[{{H.E.S.S. Collaboration} {et~al.}(2012{\natexlab{a}}){H.E.S.S. Collaboration}, {Abramowski}, {Acero}, {Aharonian}, {Akhperjanian}, {Anton}, {Balenderan}, {Balzer}, {Barnacka}, {Becherini}, {Becker}, {Bernl{\"o}hr}, {Birsin}, {Biteau}, {Bochow}, {Boisson}, {Bolmont}, {Bordas}, {Brucker}, {Brun}, {Brun}, {Bulik}, {Carrigan}, {Casanova}, {Cerruti}, {Chadwick}, {Charbonnier}, {Chaves}, {Cheesebrough}, {Cologna}, {Conrad}, {Couturier}, {Dalton}, {Daniel}, {Davids}, {Degrange}, {Deil}, {Dickinson}, {Djannati-At{\"a}{\i}}, {Domainko}, {Drury}, {Dubus}, {Dutson}, {Dyks}, {Dyrda}, {Egberts}, {Eger}, {Espigat}, {Fallon}, {Farnier}, {Fegan}, {Feinstein}, {Fernandes}, {Fernandez}, {Fiasson}, {Fontaine}, {F{\"o}rster}, {F{\"u}{\ss}ling}, {Gajdus}, {Gallant}, {Garrigoux}, {Gast}, {G{\textasciiacute}rard}, {Giebels}, {Glicenstein}, {Gl{\"u}ck}, {G{\"o}ring}, {Grondin}, {H{\"a}ffner}, {Hague}, {Hahn}, {Hampf}, {Harris}, {Hauser}, {Heinz}, {Heinzelmann}, {Henri}, {Hermann}, {Hillert}, {Hinton}, {Hofmann},
  {Hofverberg}, {Holler}, {Horns}, {Jacholkowska}, {de Jager}, {Jahn}, {Jamrozy}, {Jung}, {Kastendieck}, {K{\textasciiacute}ski}, {Katz}, {Kaufmann}, {K{\textasciiacute}lifi}, {Klochkov}, {K{\textasciiacute}niak}, {Kneiske}, {Komin}, {Kosack}, {Kossakowski}, {Krayzel}, {Laffon}, {Lamanna}, {Lenain}, {Lennarz}, {Lohse}, {Lopatin}, {Lu}, {Marandon}, {Marcowith}, {Masbou}, {Maurin}, {Maxted}, {Mayer}, {McComb}, {Medina}, {M{\textasciiacute}hault}, {Menzler}, {Moderski}, {Mohamed}, {Moulin}, {Naumann}, {Naumann-Godo}, {de Naurois}, {Nedbal}, {Nguyen}, {Nicholas}, {Niemiec}, {Nolan}, {Ohm}, {de O{\~n}a Wilhelmi}, {Opitz}, {Ostrowski}, {Oya}, {Panter}, {Paz Arribas}, {Pekeur}, {Pelletier}, {Perez}, {Petrucci}, {Peyaud}, {Pita}, {P{\"u}hlhofer}, {Punch}, {Quirrenbach}, {Raue}, {Reimer}, {Reimer}, {Renaud}, {de los Reyes}, {Rieger}, {Ripken}, {Rob}, {Rosier-Lees}, {Rowell}, {Rudak}, {Rulten}, {Sahakian}, {Sanchez}, {Santangelo}, {Schlickeiser}, {Schulz}, {Schwanke}, {Schwarzburg}, {Schwemmer}, {Sheidaei}, {Skilton},
  {Sol}, {Spengler}, {Stawarz}, {.}, {Steenkamp}, {Stegmann}, {Stinzing}, {Stycz}, {Sushch}, {Szostek}, {Tavernet}, {Terrier}, {Tluczykont}, {Valerius}, {van Eldik}, {Vasileiadis}, {Venter}, {Viana}, {Vincent}, {V{\"o}lk}, {Volpe}, {Vorobiov}, {Vorster}, {Wagner}, {Ward}, {White}, {Wierzcholska}, {Zacharias}, {Zajczyk}, {Zdziarski}, {Zech}, \& {Zechlin}}]{hess_source1}
{H.E.S.S. Collaboration}, {Abramowski}, A., {Acero}, F., {et~al.} 2012{\natexlab{a}}, \aap, 545, L2

\bibitem[{{H.E.S.S. Collaboration} {et~al.}(2012{\natexlab{b}}){H.E.S.S. Collaboration}, {Abramowski}, {Acero}, {Aharonian}, {Akhperjanian}, {Anton}, {Balzer}, {Barnacka}, {Becherini}, {Becker}, {Bernl{\"o}h}, {Birsin}, {Biteau}, {Bochow}, {Boisson}, {Bolmont}, {Bordas}, {Brucker}, {Brun}, {Brun}, {Bulik}, {B{\"u}sching}, {Carrigan}, {Casanova}, {Cerruti}, {Chadwick}, {Charbonnier}, {Chaves}, {Cheesebrough}, {Cologna}, {Conrad}, {Dalton}, {Daniel}, {Davids}, {Degrange}, {Deil}, {Dickinson}, {Djannati-Ata{\"\i}}, {Domainko}, {Drury}, {Dubus}, {Dutson}, {Dyks}, {Dyrda}, {Egberts}, {Eger}, {Espigat}, {Fallon}, {Fegan}, {Feinstein}, {Fernandes}, {Fiasson}, {Fontaine}, {F{\"o}rster}, {F{\"u}{\ss}ling}, {Gallant}, {Gast}, {G{\'e}rard}, {Gerbig}, {Giebels}, {Glicenstein}, {Gl{\"u}ck}, {G{\"o}ring}, {H{\"a}ffner}, {Hague}, {Hahn}, {Hampf}, {Harris}, {Hauser}, {Heinz}, {Heinzelmann}, {Henri}, {Hermann}, {Hillert}, {Hinton}, {Hofmann}, {Hofverberg}, {Holler}, {Horns}, {Jacholkowska}, {de Jager}, {Jahn}, {Jamrozy},
  {Jung}, {Kastendieck}, {Katarzy{\'n}ski}, {Katz}, {Kaufmann}, {Keogh}, {Kh{\'e}lifi}, {Klochkov}, {Klu{\.z}niak}, {Kneiske}, {Komin}, {Kosack}, {Kossakowski}, {Krayzel}, {Laffon}, {Lamanna}, {Lenain}, {Lennarz}, {Lohse}, {Lopatin}, {Lu}, {Marandon}, {Marcowith}, {Masbou}, {Maxted}, {Mayer}, {McComb}, {Medina}, {M{\'e}hault}, {Moderski}, {Mohamed}, {Moulin}, {Naumann}, {Naumann-Godo}, {de Naurois}, {Nedbal}, {Nekrassov}, {Nguyen}, {Nicholas}, {Niemiec}, {Nolan}, {Ohm}, {de O{\~n}a Wilhelmi}, {Opitz}, {Ostrowski}, {Oya}, {Panter}, {Paz Arribas}, {Pekeur}, {Pelletier}, {Perez}, {Petrucci}, {Peyaud}, {Pita}, {P{\"u}hlhofer}, {Punch}, {Quirrenbach}, {Raue}, {Rayner}, {Reimer}, {Reimer}, {Renaud}, {de los Reyes}, {Rieger}, {Ripken}, {Rob}, {Rosier-Lees}, {Rowell}, {Rudak}, {Rulten}, {Sahakian}, {Sanchez}, {Santangelo}, {Schlickeiser}, {Schulz}, {Schwanke}, {Schwarzburg}, {Schwemmer}, {Sheidaei}, {Skilton}, {Sol}, {Spengler}, {Stawarz}, {Steenkamp}, {Stegmann}, {Stinzing}, {Stycz}, {Sushch}, {Szostek}, {Tavernet},
  {Terrier}, {Tluczykont}, {Valerius}, {van Eldik}, {Vasileiadis}, {Venter}, {Viana}, {Vincent}, {V{\"o}lk}, {Volpe}, {Vorobiov}, {Vorster}, {Wagner}, {Ward}, {White}, {Wierzcholska}, {Zacharias}, {Zajczyk}, {Zdziarski}, {Zech}, \& {Zechlin}}]{hess_source7}
{H.E.S.S. Collaboration}, {Abramowski}, A., {Acero}, F., {et~al.} 2012{\natexlab{b}}, \aap, 541, A5

\bibitem[{{H.E.S.S. Collaboration} {et~al.}(2024){H.E.S.S. Collaboration}, {Aharonian}, {Ait Benkhali}, {Aschersleben}, {Ashkar}, {Backes}, {Baktash}, {Barbosa Martins}, {Barnard}, {Batzofin}, {Becherini}, {Berge}, {Bernl{\"o}hr}, {Bi}, {B{\"o}ttcher}, {Boisson}, {Bolmont}, {de Bony de Lavergne}, {Borowska}, {Bouyahiaoui}, {Breuhaus}, {Brose}, {Brun}, {Bruno}, {Bulik}, {Burger-Scheidlin}, {Caroff}, {Casanova}, {Cecil}, {Celic}, {Cerruti}, {Chambery}, {Chand}, {Chen}, {Chibueze}, {Chibueze}, {Cotter}, {Damascene Mbarubucyeye}, {Djannati-Ata{\"\i}}, {Dmytriiev}, {Doroshenko}, {Einecke}, {Ernenwein}, {Feijen}, {Filipovic}, {Fontaine}, {F{\"u}{\ss}ling}, {Funk}, {Gabici}, {Gallant}, {Ghafourizadeh}, {Giavitto}, {Glawion}, {Glicenstein}, {Goswami}, {Grolleron}, {Grondin}, {Hinton}, {Hofmann}, {Holch}, {Holler}, {Jamrozy}, {Jankowsky}, {Joshi}, {Jung-Richardt}, {Katarzy{\'n}ski}, {Khatoon}, {Kh{\'e}lifi}, {Klepser}, {Klu{\'z}niak}, {Komin}, {Kosack}, {Kostunin}, {Kundu}, {Lang}, {Le Stum}, {Leitl}, {Lemi{\`e}re},
  {Lemoine-Goumard}, {Lenain}, {Leuschner}, {Mackey}, {Malyshev}, {Malyshev}, {Marandon}, {Marinos}, {Mart{\'\i}-Devesa}, {Marx}, {Mehta}, {Mitchell}, {Moderski}, {Mohrmann}, {Montanari}, {Moulin}, {Murach}, {de Naurois}, {Niemiec}, {Priyana Noel}, {O'Brien}, {Ohm}, {Olivera-Nieto}, {de Ona Wilhelmi}, {Ostrowski}, {Panny}, {Panter}, {Parsons}, {Prokhorov}, {P{\"u}hlhofer}, {Punch}, {Quirrenbach}, {Regeard}, {Reichherzer}, {Reimer}, {Reimer}, {Ren}, {Renaud}, {Reville}, {Rieger}, {Roellinghoff}, {Rudak}, {Sahakian}, {Salzmann}, {Sasaki}, {Sch{\"u}ssler}, {Schutte}, {Shapopi}, {Specovius}, {Spencer}, {Steenkamp}, {Steinmassl}, {Steppa}, {Sushch}, {Suzuki}, {Takahashi}, {Tanaka}, {Terrier}, {Tluczykont}, {Tsuji}, {Unbehaun}, {van Eldik}, {Vecchi}, {Veh}, {Venter}, {Vink}, {Wach}, {Wagner}, {Wierzcholska}, {Zacharias}, {Zargaryan}, {Zdziarski}, {Zouari}, \& {{\.Z}ywucka}}]{hess_source3}
{H.E.S.S. Collaboration}, {Aharonian}, F., {Ait Benkhali}, F., {et~al.} 2024, \aap, 686, A149

\bibitem[{{H.E.S.S. Collaboration} {et~al.}(2023{\natexlab{a}}){H.E.S.S. Collaboration}, {Aharonian}, {Ait Benkhali}, {Aschersleben}, {Ashkar}, {Backes}, {Barbosa Martins}, {Batzofin}, {Becherini}, {Berge}, {Bernl{\"o}hr}, {Bi}, {B{\"o}ttcher}, {Boisson}, {Bolmont}, {Borowska}, {Bouyahiaoui}, {Bradascio}, {Brose}, {Brun}, {Bruno}, {Bulik}, {Burger-Scheidlin}, {Cangemi}, {Caroff}, {Casanova}, {Celic}, {Cerruti}, {Chambery}, {Chand}, {Chandra}, {Chen}, {Chibueze}, {Chibueze}, {Cotter}, {Mbarubucyeye}, {Devin}, {Djannati-Ata{\"\i}}, {Dmytriiev}, {Egberts}, {Einecke}, {Ernenwein}, {Feijen}, {Fichet de Clairfontaine}, {Filipovic}, {Fontaine}, {F{\"u}{\ss}ling}, {Funk}, {Gabici}, {Gallant}, {Ghafourizadeh}, {Giavitto}, {Giunti}, {Glawion}, {Glicenstein}, {Goswami}, {Grolleron}, {Grondin}, {Haerer}, {Haupt}, {Hermann}, {Hinton}, {Hofmann}, {Holch}, {Holler}, {Horns}, {Huang}, {Jamrozy}, {Jankowsky}, {Joshi}, {Jung-Richardt}, {Kasai}, {Katarzy{\'n}ski}, {Kh{\'e}lifi}, {Klu{\'z}niak}, {Komin}, {Kosack}, {Kostunin},
  {Lang}, {Le Stum}, {Leitl}, {Lemi{\`e}re}, {Lemoine-Goumard}, {Lenain}, {Leuschner}, {Lohse}, {Luashvili}, {Lypova}, {Mackey}, {Malyshev}, {Marandon}, {Marchegiani}, {Marcowith}, {Marinos}, {Mart{\'\i}-Devesa}, {Marx}, {Maurin}, {Meintjes}, {Meyer}, {Mitchell}, {Moderski}, {Mohrmann}, {Montanari}, {Moulin}, {Muller}, {Nakashima}, {de Naurois}, {Niemiec}, {Noel}, {O'Brien}, {Ohm}, {Olivera-Nieto}, {de Ona Wilhelmi}, {Ostrowski}, {Panny}, {Panter}, {Parsons}, {Peron}, {Prokhorov}, {P{\"u}hlhofer}, {Quirrenbach}, {Reimer}, {Reimer}, {Renaud}, {Reville}, {Rieger}, {Rowell}, {Rudak}, {Ricarte}, {Ruiz-Velasco}, {Sahakian}, {Salzmann}, {Santangelo}, {Sasaki}, {Sch{\"u}ssler}, {Schutte}, {Schwanke}, {Shapopi}, {Sinha}, {Sol}, {Specovius}, {Spencer}, {Stawarz}, {Steinmassl}, {Sushch}, {Suzuki}, {Takahashi}, {Tanaka}, {Tavernier}, {Taylor}, {Terrier}, {Thorpe-Morgan}, {Tsirou}, {Tsuji}, {Vecchi}, {Venter}, {Vink}, {Wagner}, {White}, {Wierzcholska}, {Wong}, {Zacharias}, {Zargaryan}, {Zdziarski}, {Zech}, {Zouari}, \&
  {{\.Z}ywucka}}]{hess_source2}
{H.E.S.S. Collaboration}, {Aharonian}, F., {Ait Benkhali}, F., {et~al.} 2023{\natexlab{a}}, \aap, 673, A148

\bibitem[{{H.E.S.S. Collaboration} {et~al.}(2023{\natexlab{b}}){H.E.S.S. Collaboration}, {Aharonian}, {Ait Benkhali}, {Aschersleben}, {Ashkar}, {Backes}, {Barbosa Martins}, {Batzofin}, {Becherini}, {Berge}, {B{\"o}ttcher}, {Boisson}, {Bolmont}, {Borowska}, {Bouyahiaoui}, {Bradascio}, {Breuhaus}, {Brose}, {Brun}, {Bruno}, {Bulik}, {Burger-Scheidlin}, {Bylund}, {Caroff}, {Casanova}, {Celic}, {Cerruti}, {Chambery}, {Chand}, {Chen}, {Chibueze}, {Chibueze}, {Damascene Mbarubucyeye}, {Djannati-Ata{\"\i}}, {Dmytriiev}, {Einecke}, {Ernenwein}, {Feijen}, {Filipovic}, {Fontaine}, {F{\"u}{\ss}ling}, {Funk}, {Gabici}, {Gallant}, {Ghafourizadeh}, {Giavitto}, {Giunti}, {Glawion}, {Goswami}, {Grolleron}, {Grondin}, {Haerer}, {Hinton}, {Hofmann}, {Holch}, {Holler}, {Horns}, {Huang}, {Jamrozy}, {Jankowsky}, {Joshi}, {Jung-Richardt}, {Kasai}, {Katarzy{\'n}ski}, {Kh{\'e}lifi}, {Klu{\'z}niak}, {Komin}, {Kosack}, {Kostunin}, {Lang}, {Le Stum}, {Leitl}, {Lemi{\`e}re}, {Lemoine-Goumard}, {Lenain}, {Leuschner}, {Lohse}, {Luashvili},
  {Lypova}, {Mackey}, {Malyshev}, {Malyshev}, {Marandon}, {Marchegiani}, {Marcowith}, {Marinos}, {Mart{\'\i}-Devesa}, {Marx}, {Mitchell}, {Moderski}, {Mohrmann}, {Montanari}, {Moulin}, {Muller}, {Nakashima}, {de Naurois}, {Niemiec}, {Priyana Noel}, {Ohm}, {Olivera-Nieto}, {de Ona Wilhelmi}, {Ostrowski}, {Panny}, {Panter}, {Parsons}, {Prokhorov}, {P{\"u}hlhofer}, {Punch}, {Quirrenbach}, {Reichherzer}, {Reimer}, {Reimer}, {Renaud}, {Reville}, {Rieger}, {Rowell}, {Rudak}, {Sahakian}, {Santangelo}, {Sasaki}, {Schutte}, {Schwanke}, {Shapopi}, {Sol}, {Specovius}, {Spencer}, {Stawarz}, {Steenkamp}, {Steinmassl}, {Sushch}, {Suzuki}, {Takahashi}, {Tanaka}, {Terrier}, {Thorpe-Morgan}, {Tsirou}, {Tsuji}, {Uchiyama}, {van Eldik}, {Vecchi}, {Veh}, {Venter}, {Vink}, {Wach}, {Wagner}, {White}, {Wierzcholska}, {Wong}, {Zacharias}, {Zargaryan}, {Zdziarski}, {Zech}, {Zouari}, \& {{\.Z}ywucka}}]{hess_source4}
{H.E.S.S. Collaboration}, {Aharonian}, F., {Ait Benkhali}, F., {et~al.} 2023{\natexlab{b}}, \aap, 672, A103

\bibitem[{{Holler} {et~al.}(2020){Holler}, {Lenain}, {de Naurois}, {Rauth}, \& {Sanchez}}]{runwise}
{Holler}, M., {Lenain}, J.-P., {de Naurois}, M., {Rauth}, R., \& {Sanchez}, D.~A. 2020, Astroparticle Physics, 123, 102491

\bibitem[{{Karpova} {et~al.}(2017){Karpova}, {Shternin}, {Zyuzin}, {Danilenko}, \& {Shibanov}}]{hawc_source2_xray}
{Karpova}, A., {Shternin}, P., {Zyuzin}, D., {Danilenko}, A., \& {Shibanov}, Y. 2017, \mnras, 466, 1757

\bibitem[{Kirk {et~al.}(2009)Kirk, Lyubarsky, \& Petri}]{pwn_theory}
Kirk, J.~G., Lyubarsky, Y., \& Petri, J. 2009, The Theory of Pulsar Winds and Nebulae, ed. W.~Becker (Berlin, Heidelberg: Springer Berlin Heidelberg), 421--450

\bibitem[{{Konopelko} {et~al.}(2005){Konopelko}, {Chadwick}, {Eifert}, {Lohse}, {Noutsos}, {Rayner}, {Schmidt}, {Schwanke}, \& {Stegmann}}]{nondet_some}
{Konopelko}, A., {Chadwick}, P., {Eifert}, T., {et~al.} 2005, in International Cosmic Ray Conference, Vol.~4, 29th International Cosmic Ray Conference (ICRC29), Volume 4, ed. B.~S. {Acharya}, S.~{Gupta}, P.~{Jagadeesan}, A.~{Jain}, S.~{Karthikeyan}, S.~{Morris}, \& S.~{Tonwar}, 139

\bibitem[{{Kramer} {et~al.}(2003){Kramer}, {Lyne}, {Hobbs}, {L{\"o}hmer}, {Carr}, {Jordan}, \& {Wolszczan}}]{age2}
{Kramer}, M., {Lyne}, A.~G., {Hobbs}, G., {et~al.} 2003, \apjl, 593, L31

\bibitem[{{Lazarevi{\'c}} {et~al.}(2024){Lazarevi{\'c}}, {Filipovi{\'c}}, {Dai}, {Kothes}, {Ahmad}, {Alsaberi}, {Balzan}, {Barnes}, {Cotton}, {Edwards}, {Gordon}, {Haberl}, {Hopkins}, {Koribalski}, {Leahy}, {Maitra}, {Mi{\'c}i{\'c}}, {Rowell}, {Sasaki}, {Tothill}, {Umana}, \& {Velovi{\'c}}}]{nodet41}
{Lazarevi{\'c}}, S., {Filipovi{\'c}}, M.~D., {Dai}, S., {et~al.} 2024, \pasa, 41, e032

\bibitem[{Linden {et~al.}(2017)Linden, Auchettl, Bramante, Cholis, Fang, Hooper, Karwal, \& Li}]{Linden:2017vvb}
Linden, T., Auchettl, K., Bramante, J., {et~al.} 2017, Phys. Rev. D, 96, 103016

\bibitem[{{MAGIC Collaboration} {et~al.}(2019){MAGIC Collaboration}, {Acciari}, {Ansoldi}, {Antonelli}, {Arbet Engels}, {Arcaro}, {Baack}, {Babi{\'c}}, {}, {Banerjee}, {Bangale}, {de Almeida}, {Barrio}, {Becerra Gonz{\'a}lez}, {Bednarek}, {Bernardini}, {Berti}, {Besenrieder}, {Bhattacharyya}, {Bigongiari}, {Biland}, {Blanch}, {Bonnoli}, {Carosi}, {Ceribella}, {Chatterjee}, {Colak}, {Colin}, {Colombo}, {Contreras}, {Cortina}, {Covino}, {Cumani}, {D'Elia}, {da Vela}, {Dazzi}, {de Angelis}, {de Lotto}, {Delfino}, {Delgado}, {di Pierro}, {Dom{\'\i}nguez}, {Dominis Prester}, {Dorner}, {Doro}, {Einecke}, {Elsaesser}, {Fallah Ramazani}, {Fattorini}, {Fern{\'a}ndez-Barral}, {Ferrara}, {Fidalgo}, {Foffano}, {Fonseca}, {Font}, {Fruck}, {Galindo}, {Gallozzi}, {Garc{\'\i}a L{\'o}pez}, {Garczarczyk}, {Gaug}, {Giammaria}, {Godinovi{\'c}}, {}, {Guberman}, {Hadasch}, {Hahn}, {Hassan}, {Herrera}, {Hoang}, {Hrupec}, {Inoue}, {Ishio}, {Iwamura}, {Kubo}, {Kushida}, {Kuve{\v{z}}di{\'c}}, {}, {Lamastra}, {Lelas}, {Leone},
  {Lindfors}, {Lombardi}, {Longo}, {L{\'o}pez}, {L{\'o}pez-Oramas}, {Maggio}, {Majumdar}, {Makariev}, {Maneva}, {Manganaro}, {Mannheim}, {Maraschi}, {Mariotti}, {Mart{\'\i}nez}, {Masuda}, {Mazin}, {Minev}, {Miranda}, {Mirzoyan}, {Molina}, {Moralejo}, {Moreno}, {Moretti}, {Neustroev}, {Niedzwiecki}, {Nievas Rosillo}, {Nigro}, {Nilsson}, {Ninci}, {Nishijima}, {Noda}, {Nogu{\'e}s}, {Paiano}, {Palacio}, {Paneque}, {Paoletti}, {Paredes}, {Pedaletti}, {Pe{\~n}il}, {Peresano}, {Persic}, {Prada Moroni}, {Prandini}, {Puljak}, {Garcia}, {Rhode}, {Rib{\'o}}, {Rico}, {Righi}, {Rugliancich}, {Saha}, {Saito}, {Satalecka}, {Schweizer}, {Sitarek}, {{\v{S}}nidari{\'c}}, {}, {Sobczynska}, {Somero}, {Stamerra}, {Strzys}, {Suri{\'c}}, {}, {Tavecchio}, {Temnikov}, {Terzi{\'c}}, {}, {Teshima}, {Torres-Alb{\`a}}, {Tsujimoto}, {Vanzo}, {Vazquez Acosta}, {Vovk}, {Ward}, {Will}, {Zari{\'c}}, {}, {de O{\~n}a Wilhelmi}, {Torres}, \& {Zanin}}]{magic5}
{MAGIC Collaboration}, {Acciari}, V.~A., {Ansoldi}, S., {et~al.} 2019, \mnras, 483, 4578

\bibitem[{{MAGIC Collaboration} {et~al.}(2014){MAGIC Collaboration}, {Aleksi{\'c}}, {Ansoldi}, {Antonelli}, {Antoranz}, {Babic}, {Bangale}, {Barres de Almeida}, {Barrio}, {Becerra Gonz{\'a}lez}, {Bednarek}, {Bernardini}, {Biland}, {Blanch}, {Bonnefoy}, {Bonnoli}, {Borracci}, {Bretz}, {Carmona}, {Carosi}, {Carreto Fidalgo}, {Colin}, {Colombo}, {Contreras}, {Cortina}, {Covino}, {Da Vela}, {Dazzi}, {De Angelis}, {De Caneva}, {De Lotto}, {Delgado Mendez}, {Doert}, {Dom{\'\i}nguez}, {Dominis Prester}, {Dorner}, {Doro}, {Einecke}, {Eisenacher}, {Elsaesser}, {Farina}, {Ferenc}, {Fonseca}, {Font}, {Frantzen}, {Fruck}, {Garc{\'\i}a L{\'o}pez}, {Garczarczyk}, {Garrido Terrats}, {Gaug}, {Godinovi{\'c}}, {Gonz{\'a}lez Mu{\~n}oz}, {Gozzini}, {Hadasch}, {Hayashida}, {Herrera}, {Herrero}, {Hildebrand}, {Hose}, {Hrupec}, {Idec}, {Kadenius}, {Kellermann}, {Klepser}, {Kodani}, {Konno}, {Krause}, {Kubo}, {Kushida}, {La Barbera}, {Lelas}, {Lewandowska}, {Lindfors}, {Lombardi}, {L{\'o}pez}, {L{\'o}pez-Coto}, {L{\'o}pez-Oramas},
  {Lorenz}, {Lozano}, {Makariev}, {Mallot}, {Maneva}, {Mankuzhiyil}, {Mannheim}, {Maraschi}, {Marcote}, {Mariotti}, {Mart{\'\i}nez}, {Mazin}, {Menzel}, {Meucci}, {Miranda}, {Mirzoyan}, {Moralejo}, {Munar-Adrover}, {Nakajima}, {Niedzwiecki}, {Nilsson}, {Nishijima}, {Noda}, {Nowak}, {de O{\~n}a Wilhelmi}, {Orito}, {Overkemping}, {Paiano}, {Palatiello}, {Paneque}, {Paoletti}, {Paredes}, {Paredes-Fortuny}, {Partini}, {Persic}, {Prada}, {Prada Moroni}, {Prandini}, {Preziuso}, {Puljak}, {Reinthal}, {Rhode}, {Rib{\'o}}, {Rico}, {Rodriguez Garcia}, {R{\"u}gamer}, {Saggion}, {Saito}, {Saito}, {Satalecka}, {Scalzotto}, {Scapin}, {Schultz}, {Schweizer}, {Shore}, {Sillanp{\"a}{\"a}}, {Sitarek}, {Snidaric}, {Sobczynska}, {Spanier}, {Stamatescu}, {Stamerra}, {Steinbring}, {Storz}, {Strzys}, {Sun}, {Suri{\'c}}, {Takalo}, {Takami}, {Tavecchio}, {Temnikov}, {Terzi{\'c}}, {Tescaro}, {Teshima}, {Thaele}, {Tibolla}, {Torres}, {Toyama}, {Treves}, {Uellenbeck}, {Vogler}, {Wagner}, {Zandanel}, \& {Zanin}}]{magic4_magic}
{MAGIC Collaboration}, {Aleksi{\'c}}, J., {Ansoldi}, S., {et~al.} 2014, \aap, 571, A96

\bibitem[{{Manchester} {et~al.}(2005){Manchester}, {Hobbs}, {Teoh}, \& {Hobbs}}]{ATNF}
{Manchester}, R.~N., {Hobbs}, G.~B., {Teoh}, A., \& {Hobbs}, M. 2005, \aj, 129, 1993

\bibitem[{{Meagher} \& {VERITAS Collaboration}(2015)}]{vertias_source3}
{Meagher}, K. \& {VERITAS Collaboration}. 2015, in International Cosmic Ray Conference, Vol.~34, 34th International Cosmic Ray Conference (ICRC2015), 792

\bibitem[{{Mitchell} \& {Spencer}(2026)}]{PWN_hadrons}
{Mitchell}, A. M.~W. \& {Spencer}, S.~T. 2026, Universe, 12, 85

\bibitem[{{Mizuno} {et~al.}(2017){Mizuno}, {Tanaka}, {Takahashi}, {Katsuta}, {Hayashi}, \& {Yamazaki}}]{veritas8_xray}
{Mizuno}, T., {Tanaka}, N., {Takahashi}, H., {et~al.} 2017, \apj, 841, 104

\bibitem[{{Mohrmann} {et~al.}(2019){Mohrmann}, {Specovius}, {Tiziani}, {Funk}, {Malyshev}, {Nakashima}, \& {van Eldik}}]{bkg1_hess}
{Mohrmann}, L., {Specovius}, A., {Tiziani}, D., {et~al.} 2019, \aap, 632, A72

\bibitem[{{Noutsos} {et~al.}(2013){Noutsos}, {Schnitzeler}, {Keane}, {Kramer}, \& {Johnston}}]{age3}
{Noutsos}, A., {Schnitzeler}, D.~H.~F.~M., {Keane}, E.~F., {Kramer}, M., \& {Johnston}, S. 2013, \mnras, 430, 2281

\bibitem[{{Park} \& {VERITAS Collaboration}(2012)}]{veritas6}
{Park}, N. \& {VERITAS Collaboration}. 2012, in American Institute of Physics Conference Series, Vol. 1505, High Energy Gamma-Ray Astronomy: 5th International Meeting on High Energy Gamma-Ray Astronomy, ed. F.~A. {Aharonian}, W.~{Hofmann}, \& F.~M. {Rieger} (AIP), 354--357

\bibitem[{{Parsons} \& {Hinton}(2014)}]{impact}
{Parsons}, R.~D. \& {Hinton}, J.~A. 2014, Astroparticle Physics, 56, 26

\bibitem[{{Pavlov} {et~al.}(2008){Pavlov}, {Kargaltsev}, \& {Brisken}}]{veritas_source2_xray2}
{Pavlov}, G.~G., {Kargaltsev}, O., \& {Brisken}, W.~F. 2008, \apj, 675, 683

\bibitem[{{Petriella} {et~al.}(2021){Petriella}, {Duvidovich}, \& {Giacani}}]{magic4_radio}
{Petriella}, A., {Duvidovich}, L., \& {Giacani}, E. 2021, \aap, 652, A142

\bibitem[{{Popescu} \& {Tuffs}(2013)}]{rad_fields}
{Popescu}, C.~C. \& {Tuffs}, R.~J. 2013, \mnras, 436, 1302

\bibitem[{{Renaud} {et~al.}(2010){Renaud}, {Marandon}, {Gotthelf}, {Rodriguez}, {Terrier}, {Mattana}, {Lebrun}, {Tomsick}, \& {Manchester}}]{nondet2}
{Renaud}, M., {Marandon}, V., {Gotthelf}, E.~V., {et~al.} 2010, \apj, 716, 663

\bibitem[{{Roberts} {et~al.}(2007){Roberts}, {Gotthelf}, {Halpern}, {Brogan}, \& {Ransom}}]{hess_source5_xray}
{Roberts}, M. S.~E., {Gotthelf}, E.~V., {Halpern}, J.~P., {Brogan}, C.~L., \& {Ransom}, S.~M. 2007, in WE-Heraeus Seminar on Neutron Stars and Pulsars 40 years after the Discovery, ed. W.~{Becker} \& H.~H. {Huang}, 24

\bibitem[{{Saha}(2016)}]{veritas8_mw}
{Saha}, L. 2016, \mnras, 460, 3563

\bibitem[{Smith {et~al.}(2023)}]{Fermi-LAT:2023zzt}
Smith, D.~A. {et~al.} 2023, Astrophys. J., 958, 191

\bibitem[{{Timokhin} \& {Harding}(2019)}]{multiplicity1}
{Timokhin}, A.~N. \& {Harding}, A.~K. 2019, \apj, 871, 12

\bibitem[{Torres~Escobedo(2025)}]{magic4_hawc}
Torres~Escobedo, R. 2025, PoS, ICRC2025, 867

\bibitem[{{Wach} {et~al.}(2024){Wach}, {Mitchell}, \& {Mohrmann}}]{bkg2_hess}
{Wach}, T., {Mitchell}, A., \& {Mohrmann}, L. 2024, \aap, 690, A250

\bibitem[{Wach \& Mitchell(2025)}]{hess_source8}
Wach, T. \& Mitchell, A. M.~W. 2025, in Proceedings of the 39th International Cosmic Ray Conference (ICRC 2025), PoS(ICRC2025), 874

\bibitem[{{Wang} {et~al.}(2001){Wang}, {Gotthelf}, {Chu}, \& {Dickel}}]{hess_source1_xray}
{Wang}, Q.~D., {Gotthelf}, E.~V., {Chu}, Y.~H., \& {Dickel}, J.~R. 2001, \apj, 559, 275

\bibitem[{{Wang} {et~al.}(2024){Wang}, {Kaplan}, {Sengar}, {Lenc}, {Zic}, {Anumarlapudi}, {Gaensler}, {Hurley-Walker}, {Murphy}, \& {Wang}}]{nodet1}
{Wang}, Z., {Kaplan}, D.~L., {Sengar}, R., {et~al.} 2024, \apj, 961, 175

\bibitem[{{Yodh}(1996)}]{MILAGRO}
{Yodh}, G.~B. 1996, \ssr, 75, 199

\bibitem[{{Zari{\'c}} {et~al.}(2022){Zari{\'c}}, {Green}, {Strzys}, {Vovk}, {The MAGIC Collaboration}, {Acciari}, {Ansoldi}, {Antonelli}, {Arbet Engels}, {Artero}, {Asano}, {Baack}, {Babic}, {Baquero}, {Barres de Almeida}, {Barrio}, {Batkovi{\'c}}, {Becerra Gonzalez}, {Bednarek}, {Bellizzi}, {Bernardini}, {Bernardos}, {Berti}, {Besenrieder}, {Bhattacharyya}, {Bigongiari}, {Biland}, {Blanch}, {B{\"o}kenkamp}, {Bonnoli}, {Bosnjak}, {Busetto}, {Carosi}, {Ceribella}, {Cerruti}, {Chai}, {Chilingarian}, {Cikota}, {Colak}, {Colombo}, {Contreras}, {Cortina}, {Covino}, {D'Amico}, {D'Elia}, {da Vela}, {Dazzi}, {de Angelis}, {de Lotto}, {Delfino}, {Delgado}, {Delgado Mendez}, {Depaoli}, {di Pierro}, {di Venere}, {Do Souto Espi{\~n}eira}, {Dominis Prester}, {Donini}, {Dorner}, {Doro}, {Elsaesser}, {Fallah Ramazani}, {Fattorini}, {Fonseca}, {Font}, {Fruck}, {Fukami}, {Fukazawa}, {Garc{\'\i}a L{\'o}pez}, {Garczarczyk}, {Gasparyan}, {Gaug}, {Giglietto}, {Giordano}, {Gliwny}, {Godinovic}, {Green}, {Hadasch}, {Hahn},
  {Heckmann}, {Herrera}, {Hoang}, {Hrupec}, {H{\"u}tten}, {Inada}, {Ishio}, {Iwamura}, {Jim{\'e}nez Mart{\'\i}nez}, {Jormanainen}, {Jouvin}, {Karjalainen}, {Kerszberg}, {Kobayashi}, {Kubo}, {Kushida}, {Lamastra}, {Lelas}, {Leone}, {Lindfors}, {Linhoff}, {Lombardi}, {Longo}, {Lopez-Coto}, {L{\'o}pez-Moya}, {L{\'o}pez-Oramas}, {Loporchio}, {Machado de Oliveira Fraga}, {Maggio}, {Majumdar}, {Makariev}, {Mallamaci}, {Maneva}, {Manganaro}, {Mannheim}, {Maraschi}, {Mariotti}, {Martinez}, {Mazin}, {Menchiari}, {Mender}, {Mi{\'c}anovi{\'c}}, {Miceli}, {Miener}, {Miranda}, {Mirzoyan}, {Molina}, {Moralejo}, {Morcuende}, {Moreno}, {Moretti}, {Nakamori}, {Nava}, {Neustroev}, {Nigro}, {Nilsson}, {Nishijima}, {Noda}, {Nozaki}, {Ohtani}, {Oka}, {Otero-Santos}, {Paiano}, {Palatiello}, {Paneque}, {Paoletti}, {Paredes}, {Pavleti{\'c}}, {Pe{\~n}il}, {Persic}, {Pihet}, {Prada Moroni}, {Prandini}, {Priyadarshi}, {Puljak}, {Rhode}, {Rib{\'o}}, {Rico}, {Righi}, {Rugliancich}, {Sahakyan}, {Saito}, {Sakurai}, {Satalecka}, {Saturni},
  {Schleicher}, {Schmidt}, {Schweizer}, {Sitarek}, {{\v{S}}nidari{\'c}}, {Sobczy{\'n}ska}, {Spolon}, {Stamerra}, {Stri{\v{s}}kovi{\'c}}, {Strom}, {Suda}, {Suri{\'c}}, {Takahashi}, {Takeishi}, {Tavecchio}, {Temnikov}, {Terzic}, {Teshima}, {Tosti}, {Truzzi}, {Tutone}, {Ubach}, {van Scherpenberg}, {Vanzo}, {Vazquez Acosta}, {Ventura}, {Verguilov}, {Vigorito}, {Vitale}, {Will}, {Wunderlich}, \& {Yamamoto}}]{magic_1809}
{Zari{\'c}}, D., {Green}, D., {Strzys}, M., {et~al.} 2022, in 37th International Cosmic Ray Conference, 818

\bibitem[{{Zyuzin} {et~al.}(2018){Zyuzin}, {Karpova}, \& {Shibanov}}]{hawc_source1_xray}
{Zyuzin}, D.~A., {Karpova}, A.~V., \& {Shibanov}, Y.~A. 2018, \mnras, 476, 2177

\end{thebibliography}

\begin{appendix}

\clearpage

\section{Correlated parameters}

To identify first-order trends in the population between the $\gamma$-ray emission observed in pulsar environments and the properties of the pulsar, correlations in the $P- \dot{P}$ plane are analysed following Equation \ref{eq:plane}. The correlations can be seen in Table \ref{tab:ppdot}. From these dependencies, we can then infer correlations with commonly used pulsar parameters depending on $P$ and $\dot{P}$. These are given in Table \ref{tab:inferred_corrs}. Table \ref{tab:corr_matrix} shows the Spearman correlation coefficients derived between the $\gamma$-ray properties and the pulsar properties not derived from $P$ and $\dot{P}$.

\begin{table}[h]
\caption{Values for fit parameters $a$ and $b$ for a relation between the $\gamma$-ray properties $X$ of pulsar environments and their spin periods $P$ and $\dot{P}$, following Equation \ref{eq:plane}.} 
\label{tab:ppdot} 
\centering    
\begin{tabular}{ c | c c }       
\hline\hline  
\noalign{\smallskip}
 Gamma-ray property & a & b  \\    
\noalign{\smallskip}
\hline 
\noalign{\medskip}
$r_{39}$     & $0.37 \pm 0.11 $    & $-0.13 \pm 0.04 $    \\
$N_0$        & $-1.33 \pm 0.44 $    & $0.91 \pm 0.18 $    \\
$\Gamma$     & $-0.01 \pm 0.03 $    & $0.00 \pm 0.01 $    \\
Offset       & $0.31 \pm 0.23 $    & $-0.25 \pm 0.09 $    \\
Extension    & $-0.55 \pm 0.18 $    & $0.12 \pm 0.07 $    \\
$L_\gamma$   & $-1.61 \pm 0.41 $    & $0.28 \pm 0.20 $    \\
$S$          & $-1.97 \pm 0.27 $    & $0.99 \pm 0.14 $    \\
\noalign{\smallskip}
\hline   
\hline 
\end{tabular}
\end{table}

\begin{table}[h]
\caption{Values for scaling parameter $\beta$ linking the the $\gamma$-ray properties of pulsar environments to their pulsar properties derived from $P$ and $\dot{P}$. For the definitions of $\beta$ for each derived parameter, see section \ref{sec:corr}.} 
\label{tab:inferred_corrs} 
\centering    
\begin{tabular}{ c | c c c c }       
\hline\hline  
\noalign{\smallskip}
 Gamma-ray property & $\dot{E}$ & $\tau_c$ & $B_\text{surf}$ & $B_\text{LC}$  \\    
\noalign{\smallskip}
\hline 
\noalign{\medskip}
$r_{39}$     & $-1.24 \pm 0.33 $    & $ 0.50 \pm 0.12 $ & $ 0.12 \pm 0.24 $    & $-0.99 \pm 0.28 $    \\
$N_0$        & $ 4.90 \pm 1.32 $    & $-2.24 \pm 0.47 $ & $-0.21 \pm 0.02 $    & $ 3.78 \pm 1.10 $    \\
$\Gamma$     & $ 0.02 \pm 0.09 $    & $-0.01 \pm 0.04 $ & $-0.01 \pm 0.12 $    & $ 0.02 \pm 0.08 $    \\
Offset       & $-1.19 \pm 0.69 $    & $ 0.57 \pm 0.25 $ & $ 0.30 \pm 0.09 $    & $-0.90 \pm 0.57 $    \\
Extension    & $ 1.76 \pm 0.53 $    & $-0.67 \pm 0.19 $ & $-0.21 \pm 0.17 $    & $ 1.42 \pm 0.44 $    \\
$L_\gamma$   & $ 5.11 \pm 1.25 $    & $-1.89 \pm 0.46 $ & $-0.66 \pm 0.23 $    & $-4.17 \pm 1.03 $    \\
$S$          & $ 6.90 \pm 0.83 $    & $-2.96 \pm 0.31 $ & $-0.48 \pm 0.15 $    & $ 5.42 \pm 0.69 $    \\
\noalign{\smallskip}
\hline   
\hline 
\end{tabular}
\end{table}

\begin{table}[h]
\caption{Spearman correlation coefficients $p$ between gamma-ray and pulsar properties.}
\label{tab:corr_matrix}
\centering
\begin{tabular}{ c | c c c c c c  c c c }
\hline\hline
\noalign{\smallskip}
& $\pi$ & DM & RM & $d$ & $L_{400}$ & $\mu$ & $\Gamma_{100}$ & $E_\text{peak}$ & $L_\text{pulse}$ \\
\noalign{\smallskip}
\hline
\noalign{\medskip}

$r_{39}$
& 0.33 & 0.05 & 0.11 & -0.03 & -0.37 & 0.25  & -0.25 & 0.29 & -0.18 \\

$N_0$
& -0.15 & 0.28 & -0.03 & 0.32  & 0.33 & -0.17  & 0.22 & -0.12 & 0.36 \\

$\Gamma$ 
& -0.14 & -0.05 & -0.03 & 0.04  & 0.22 & -0.02  & -0.03 & 0.13 & -0.05 \\

Offset 
& 0.24 & 0.32 & 0.02 & 0.41 & -0.38 & 0.04  & -0.06 & 0.09 & -0.01 \\

Extension 
& -0.41 & 0.52 & 0.17 & 0.67  & -0.07 & -0.21  & 0.08 & 0.02 & 0.17 \\

$\text{L}_\gamma$
& -0.76 & 0.65 & 0.15 & 0.81  & 0.46 & -0.46 & 0.34 & -0.25 & 0.61 \\

S
& -0.40 & 0.25 & -0.00 & 0.30  & 0.57 & -0.34 & 0.15 & -0.17 & 0.50 \\

\noalign{\smallskip}
\hline
\hline
\end{tabular}
\end{table}


\clearpage

\section{Predictions of surface brightness for non-detected PWNe}

\begin{figure}[h]
\centering
\includegraphics[width=0.45\textwidth]{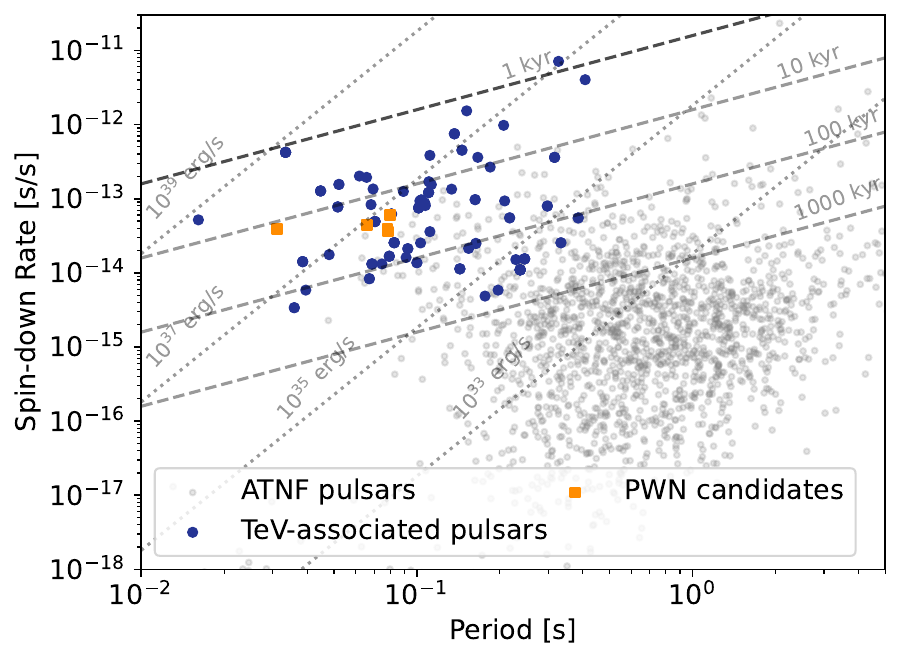}
\caption{$P$-$\dot{P}$-diagram showing all detected pulsars with a distance below $8\,$kpc. Powerful pulsars without any detected $\gamma$-ray emission are indicated in orange.}
\label{fig:non-detect}%
\end{figure}

$\gamma$-ray PWNe are detected around most powerful pulsars located in close vicinity from earth. However, for some of these powerful pulsars no extended $\gamma$-ray emission has been detected yet. The position of the most striking examples of the pulsars with such elusive nebulae in a diagram relating spin-down power corrected for pulsar distance squared to characteristic age can be seen in Figure \ref{fig:non-detect}.

To assess whether CTAO, a future $\gamma$-ray Observatory, can resolve the question of whether this non-detection is caused by an anomaly in the formation of the PWNe or a surface brightness to low to result in a significant detection, the PWNe catalogue is used to train an ensemble gradient-boosted decision tree regressor and predict the surface brightness of these nebulae, assuming their evolution is influenced only by the parent pulsar and the photon field density the nebula expands in. To assess the relative influence of the input parameters used by the model, SHAP (SHapley Additive exPlanations) values, which provide a model-agnostic measure of feature importance, are employed. These values, quantifying the contribution of each parameter to the predicted TeV $\gamma$-ray surface brightness, in difference to the simple Pearson correlation test reported in Section \ref{sec:corr}, allow for an interpretation of non-linear model behaviour.

\begin{figure}
\centering
\includegraphics[width=0.45\textwidth]{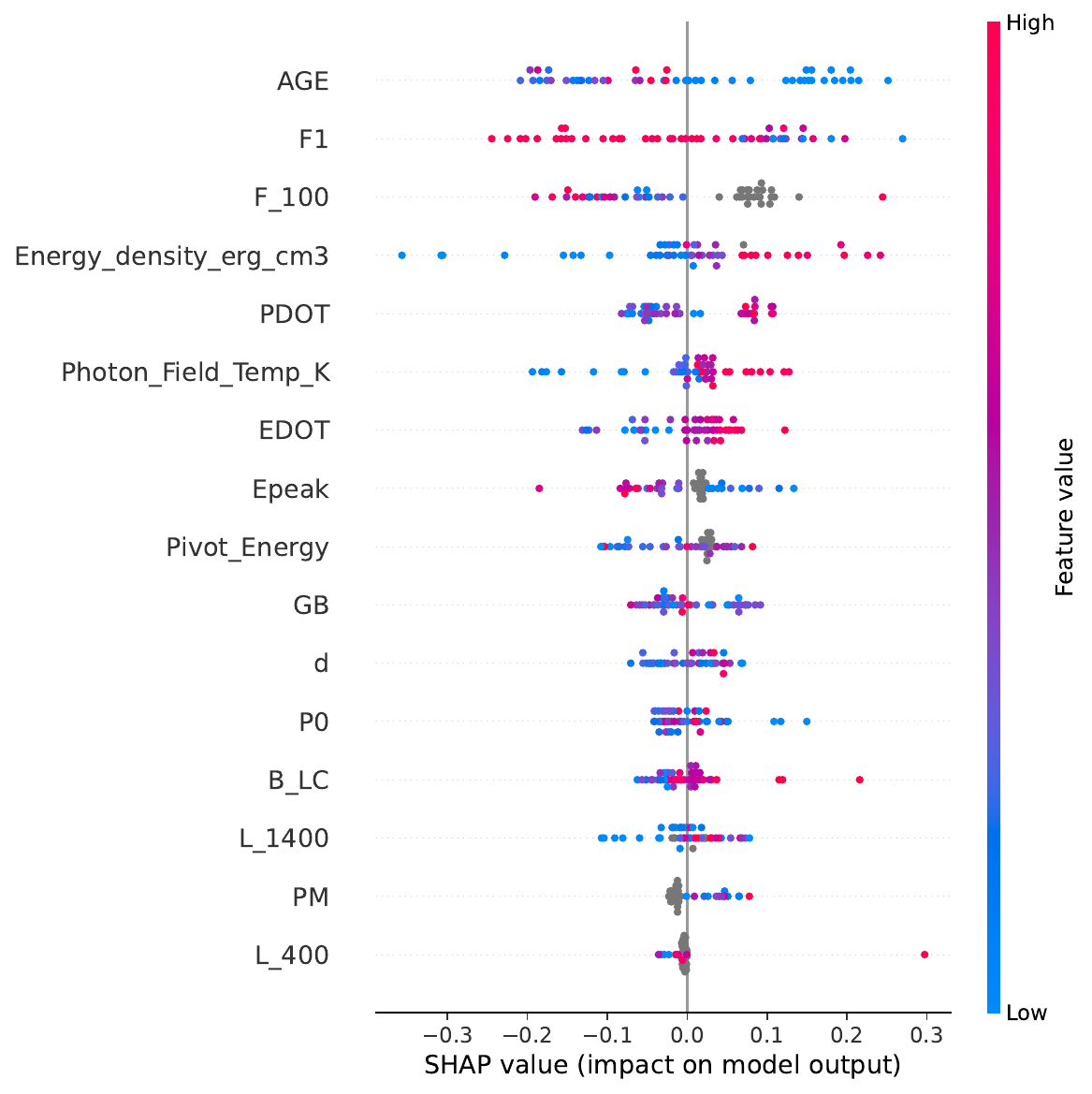}
\caption{SHAP summary plot for the machine-learning model used to predict the TeV gamma-ray surface brightness. The plot shows the distribution of SHAP values for each input parameter across the training sample, with points colour-coded by the corresponding parameter value. Parameters are ordered by their mean absolute SHAP value.}
\label{fig:model_params}%
\end{figure}

Figure~\ref{fig:model_params} shows the SHAP summary plot for the trained model. The horizontal spread of SHAP values reflects the strength of each parameter’s impact on the model predictions, while the colour coding indicates how different parameter values contribute across the pulsar population. Parameters with larger mean absolute SHAP values exert a stronger overall influence on the predicted surface brightness. Additionally to the parameters discussed in Table \ref{tab:params_corr}, the Time derivative of barycentric rotation frequency (F1), as well as the energy density of the photon fields at the pulsar position  (Energy\_density\_erg\_cm3) and the average black-body temperature of the photon fields (Photon\_field\_Temp\_K), as predicted by a stellar radiation field model derived in \citet{rad_fields}, are used.

\FloatBarrier
\section{Catalogue layout}
\label{sec:layout}

The catalogue can be downloaded from the CDS. It is given as nine csv files. The file `pulsar\_associations.csv' serves as a Table of navigation. It combines the pulsar identifier with the respective source names of all $\gamma$-ray instruments that observed emission connected to the pulsar. The available columns of this file can be found in Table \ref{tab:id_pulsars}. Then there are six instrument specific tables, detailing all available information from the $\gamma$-ray observations. The columns available in these tables are equal across instruments and their keys can be found in Table \ref{tab:id_tables}. Lastly the file `pulsar\_properties.csv' and `fermi\_properties.csv' detail the properties of the pulsar collected from the ATNF catalogue \citep{ATNF} and the third Fermi-\emph{LAT} pulsar catalog. All nine files share the column `JNAME', which can be used to unify the files for different investigations. The catalogue tables are available in electronic form at the CDS. 

The layout of the individual files can be found in Table \ref{tab:id_pulsars} and Table \ref{tab:id_tables}.

\begin{table}
\caption{Columns available in the `pulsar\_associations.csv' file}
\centering
\small
\begin{tabular}{ll}
\hline\hline
Column & Description \\ \hline
JNAME & Pulsar Identifier \\
LHAASO Assoc & Source Name given by LHAASO \\
HAWC Assoc & Source Name given by HAWC \\
HESS Assoc & Source Name given by HESS \\
MAGIC Assoc & Source Name given by MAGIC \\
VERITAS Assoc & Source Name given by VERITAS \\
\hline
\end{tabular}
\label{tab:id_pulsars}%
\end{table}

\begin{table*}
\caption{Column names for all available $\gamma$-ray properties for the observed TeV PWNe or haloes, as they are stored in the respective .csv file per instrument.}
\centering
\small
\begin{tabular}{lll}
\hline\hline
Parameter & Description & Format\\ \hline
Source Name & Identifier of the $\gamma$-ray source & str \\
JNAME & Identifier of the pulsar connected to the $\gamma$-ray emission & str \\
Spatial Model & Spatial Model used for the parameter extraction (Point, Disk or Gauss) & str \\
RA [deg] & Right Ascension ($^\circ$) & float64 \\
RA error & Error on the Right Ascension ($^\circ$) & float64 \\
Dec [deg] & Declination ($^\circ$) & float64 \\
Dec error & Error on the Declination ($^\circ$) & float64 \\
r\_{39} [deg] & $1\,\sigma$ containment radius ($^\circ$) & float64 \\
r\_{39} error & Error on the containment radius ($^\circ$) & float64 \\
$e$ & Eccentricity & float64 \\
$e$ error & Error on the eccentricity  & float64 \\
phi & Position Angle ($^\circ$) & float64 \\
phi error & Error on the Position Angle ($^\circ$) & float64 \\
E\_0 [TeV] & Reference Energy (TeV) & float64 \\
TS & Test statistic of the model & float64 \\
Spectral Model & Spectral Model used in the analysis (PL, BPL or ECPL) & str \\
N\_0 [10\textasciicircum-x TeV-1cm-2s-1]& Energy flux, the normalization varies for different detector files & float64 \\
N\_0 error & Error on the spectral normalization & float64 \\
Gamma & Spectral Index & float64 \\
Gamma error & Error on the spectral index  & float64\\
beta & Spectral curvature  & float64\\
beta error & Error on the spectral curvature & float64 \\
E\_cut [TeV] & Cutoff Energy (TeV) & float64 \\
E\_cut error & Error on the cutoff Energy & float64 \\
Catalogue & Parameters obtained from a catalogue & str \\
Publication & URL of the publication parameters are taken from & str \\
Offset [pc] & Between pulsar and emission centre (pc) & float64 \\
Offset error & Error on the offset & float64 \\
Extension [pc] & Physical extension (pc) & float64 \\
Extension error & Error on the extension & float64 \\
Luminosity [erg/s] & $\gamma$-ray Luminosity (erg/s) & float64 \\
Luminosity error & Error on the $\gamma$-ray Luminosity & float64 \\
Surface\_brightness [erg/s/pc2] & Surface brightness (erg/s/pc$^2$) & float64 \\
Surface\_brightness error & Error on the surface brightness & float64 \\
\hline
\end{tabular}
\label{tab:id_tables}%
\end{table*}

\end{appendix}

\end{document}